\documentclass[12pt]{article}

\usepackage{cite}
\usepackage{axodraw}
\usepackage[dvips]{graphicx}
\usepackage{epsfig}
\usepackage{amssymb}
\usepackage{rotating}

\def\theequation{\arabic{section}.\arabic{equation}}

\begin{document}
\begin{flushright}
{\tt CERN-PH-TH/2006-100}\\[-2pt]
{\tt MAN/HEP/2006/17}\\[-2pt]
{\tt hep-ph/0605264}
\end{flushright}
\bigskip

\begin{center}
{\LARGE {\bf Anatomy of {\boldmath $F_D$}-Term Hybrid Inflation}}\\[1.5cm]
{\sc Bj\"orn Garbrecht$^{\, a}$, Constantinos Pallis$^{\, a}$ and
Apostolos Pilaftsis$^{\, a,b}$}\\[0.5cm]
{\em $^a$School of Physics and Astronomy, University of Manchester,}\\
{\em Manchester M13 9PL, United Kingdom}\\[0.3cm]
{\em $^b$CERN, Physics Department, Theory Division, CH-1211 Geneva 23,
Switzerland}
\end{center}

\vspace{1.cm} \centerline{\bf ABSTRACT}

\noindent
{\small
We analyze the cosmological  implications of $F$-term hybrid inflation
with   a  subdominant   Fayet--Iliopoulos   $D$-term  whose   presence
explicitly breaks a $D$-parity in the inflaton-waterfall sector.  This
scenario  of inflation, which  is called  $F_D$-term hybrid  model for
brevity,  can  naturally  predict   lepton  number  violation  at  the
electroweak  scale, by  tying the  $\mu$-parameter of  the MSSM  to an
SO(3)-symmetric Majorana mass $m_N$,  via the vacuum expectation value
of the inflaton field. We show how a negative Hubble-induced mass term
in a  next-to-minimal extension  of supergravity helps  to accommodate
the present CMB data and considerably weaken the strict constraints on
the theoretical  parameters, resulting  from cosmic string  effects on
the power  spectrum $P_{\cal  R}$.  The usual  gravitino overabundance
constraint  may  be significantly  relaxed  in  this  model, once  the
enormous  entropy  release from  the  late  decays  of the  ultraheavy
waterfall  gauge particles  is properly  considered.  As  the Universe
enters  a second  thermalization  phase involving  a  very low  reheat
temperature,  which  might  be   as  low  as  about  0.3~TeV,  thermal
electroweak-scale  resonant leptogenesis  provides a  viable mechanism
for  successful baryogenesis,  while  thermal right-handed  sneutrinos
emerge as  new possible  candidates for solving  the cold  dark matter
problem.  In~addition, we discuss grand unified theory realizations of
$F_D$-term hybrid  inflation devoid  of cosmic strings  and monopoles,
based  on  the  complete  breaking  of  an  SU(2)$_X$  subgroup.   The
$F_D$-term  hybrid model  offers rich  particle-physics phenomenology,
which  could  be  probed  at  high-energy colliders,  as  well  as  in
low-energy experiments of lepton flavour or number violation. }

\thispagestyle{empty}

\noindent


\tableofcontents
\thispagestyle{empty}

\newpage
\pagestyle{plain}
\setcounter{page}{1}

\setcounter{equation}{0}
\section{Introduction}

Standard  big-bang cosmology faces  severe difficulties  in accounting
for  the observed  flatness  and  enormity of  the  causal horizon  of
today's Universe. It also leaves  unexplained the origin of the nearly
scale-invariant cosmic  microwave background (CMB), as was  found by a
number        of        observations        over       the        last
decade~\cite{COBE,WMAP,MT,WMAP3,Lyman}.   All these  pressing problems
can be successfully addressed  within the field-theoretic framework of
inflation~\cite{review}.   As a  source  of inflation,  it is  usually
considered to be a scalar field, the inflaton, which is displaced from
its  minimum and  whose  slow-roll dynamics  leads  to an  accelerated
expansion  of  the  early  Universe.   In this  phase  of  accelerated
expansion or inflation, the quantum fluctuations of the inflaton field
are  stretched on  large scales  and eventually  get frozen  when they
become much bigger than the Hubble radius.  These quantum fluctuations
get imprinted  in the form  of density perturbations, when  the former
are crossing  back inside the  Hubble radius long after  inflation has
ended. In this  way, inflation provides a causal  mechanism to explain
the observed nearly-scale invariant CMB spectrum.

A complete description  of the CMB spectrum involves  about a dozen of
cosmological parameters, such as  the power spectrum $P_{{\cal R}}$ of
curvature perturbations,  the spectral index $n_{\rm  s}$, the running
spectral index $dn_{\rm s}/d\ln  k$, the ratio $r$ of tensor-to-scalar
perturbations,   the  baryon-to-photon   ratio  of   number  densities
$\eta_B$, the fractions of  relic abundance $\Omega_{\rm DM}$ and dark
energy  $\Omega_{\rm   \Lambda}$  and  a  few   others.   Recent  WMAP
data~\cite{WMAP,WMAP3},     along      with     other     astronomical
observations~\cite{MT}, have improved upon the precision of almost all
of  the above  cosmological  observables. In  particular, the  precise
values of these cosmological  observables set stringent constraints on
the model-building  of successful models  of inflation. To  ensure the
slow-roll  dynamics of  the inflaton,  for example,  one would  need a
scalar potential, which  is almost flat.  Moreover, one  has to assure
that the  flatness of the inflaton  potential does not  get spoiled by
large quantum corrections that  depend quadratically on the cut-off of
the theory.   In this context, supersymmetry (SUSY),  softly broken at
the TeV scale,  emerges almost as a compelling  ingredient not only in
the model-building of inflationary  scenarios, but also for addressing
technically the so-called gauge-hierarchy problem.

One  of the  most  predictive and  potentially  testable scenarios  of
inflation   is  the  model   of  hybrid   inflation~\cite{Linde}.   An
advantageous feature  of this  model is that  the inflaton  $\phi$ may
start its  slow-roll from field  values well below the  reduced Planck
mass  $m_{\rm  Pl}  =   2.4\times  10^{18}$~GeV.   As  a  consequence,
cosmological observables,  such as $P_{{\cal R}}$ and  $n_{\rm s}$, do
not  generically  receive   significant  contributions  from  possible
higher-dimensional   non-renormalizable   operators,   as  these   are
suppressed  by inverse  powers of  $1/m_{\rm Pl}$.   Thus,  the hybrid
model becomes very predictive and possibly testable, in the sense that
the  inflaton dynamics  is  mainly governed  by  a few  renormalizable
operators  which  might have  observable  implications for  laboratory
experiments.  In  the hybrid  model, inflation terminates  through the
so-called waterfall mechanism.  This  mechanism is triggered, when the
inflaton field  $\phi$ passes below  some critical value  $\phi_c$. In
this case,  another field $X$  different from $\phi$, which  is called
the waterfall field and is  held fixed at origin initially, develops a
tachyonic  instability  and rolls  rapidly  down  to  its true  vacuum
expectation value~(VEV).

Hybrid  inflation can be  realized in  supersymmetric theories  in two
forms.   In the  first form,  the  hybrid potential  results from  the
$F$-terms of  a superpotential, where  the slope of the  potential may
come either from  supergravity (SUGRA) corrections~\cite{CLLSW} and/or
from   radiative   effects~\cite{DSS}.    The  second   supersymmetric
realization~\cite{Halyo}   of  hybrid   inflation   uses  a   dominant
Fayet--Iliopoulos~(FI) $D$-term~\cite{FI}, which may originate from an
anomalous  local  U(1)$_Q$  symmetry  within  the  context  of  string
theories.

All models  of inflation embedded in  SUGRA have to  address a serious
problem. This  is the  so-called gravitino overabundance  problem.  If
abundantly produced in the early Universe, gravitinos may disrupt, via
their late gravitationally-mediated decays, the nucleosynthesis of the
light elements.   In order to prevent this  from happening, gravitinos
$\widetilde{G}$   must   have    a   rather   low   abundance   today,
i.e.~$Y_{\widetilde{G}}  = n_{\widetilde{G}}/s\ \stackrel{<}{{}_\sim}\
10^{-12}$--$10^{-15}$, where $n_{\widetilde{G}}$ is the number density
of  gravitinos and  $s$ is  the entropy  density. The  upper  bound on
$Y_{\widetilde{G}}$  depends on  the properties  of the  gravitino and
becomes tighter,  if gravitinos  decay appreciably to  hadronic modes.
These  considerations set a  strict upper  bound on  Universe's reheat
temperature  $T_{\rm  reh}$, generically  implying  that $T_{\rm  reh}
\stackrel{<}{{}_\sim}  10^{10}$--$10^7$~GeV~\cite{kohri,oliveg}.  This
upper  limit on  $T_{\rm  reh}$  severely restricts  the  size of  any
renormalizable superpotential coupling of the inflaton to particles of
the  Standard   Model~(SM).   All  these  couplings   must  be  rather
suppressed.    Typically,  they   have  to   be  smaller   than  about
$10^{-5}$~\cite{SS}.

The aforementioned  gravitino constraint may  be considerably relaxed,
if there is  a mechanism that could cause late  entropy release in the
evolution of the  early Universe.  Such a mechanism  could then dilute
the gravitinos to a level that would not upset the limits derived from
Big  Bang nucleosynthesis  (BBN).  This possibility  might arise  even
within the context  of $F$-term hybrid inflation, if  a subdominant FI
$D$-tadpole associated with the  gauge group U(1)$_X$ of the waterfall
sector  were  added  to  the  model.  Such  a  scenario  was  recently
discussed in~\cite{GP}.  It  has been observed that the  presence of a
FI $D$-term breaks explicitly an exact discrete symmetry acting on the
gauged waterfall  sector, i.e.~a kind of $D$-parity,  which would have
remained  otherwise  unbroken  even  after  the  spontaneous  symmetry
breaking  (SSB)  of  U(1)$_X$.    As  a  consequence,  the  ultraheavy
U(1)$_X$-gauge-sector  bosons  and  fermions,  which would  have  been
otherwise stable, can  now decay with rates controlled  by the size of
the FI  $D$-term. Since these  particles could be  abundantly produced
during the  preheating epoch, their late  decays could give  rise to a
second reheat phase in the evolution of the early Universe.  Depending
on the actual size of  the FI $D$-term, this second reheat temperature
may be as low as 0.5--1~TeV, resulting in an enormous entropy release.
This could be sufficient to render the gravitinos underabundant, which
might  be  copiously produced  during  the  first  reheating from  the
perturbative inflaton decays.

In  this paper  we  present  a detailed  analysis  of $F$-term  hybrid
inflation with  a subdominant FI $D$-tadpole. As  mentioned above, the
presence of the FI $D$-tadpole is essential for explicitly breaking an
exact discrete symmetry, a $D$-parity,  which was acting on the gauged
waterfall sector.  In~\cite{GP}, we termed this inflationary scenario,
in  short,  $F_D$-term  hybrid  inflation.   As  the  inflaton  chiral
superfield   $\widehat{S}$  couples   to   the  Higgs-doublet   chiral
superfields        $\widehat{H}_{u,d}$,       through       $\lambda\,
\widehat{S}\widehat{H}_u   \widehat{H}_d$,  the  model   generates  an
effective  $\mu$-parameter  for  the Minimal  Supersymmetric  Standard
Model~(MSSM), through the VEV $\langle S \rangle$~\cite{DLS}. The same
mechanism may also generate an  effective Majorana mass matrix for the
singlet  neutrino  superfields $\widehat{N}_{1,2,3}$~\cite{Francesca},
through   the    operator   $\frac{1}{2}\,   \rho_{ij}\,   \widehat{S}
\widehat{N}_i  \widehat{N}_j$~\cite{PU2,GP}.  Assuming that  this last
operator  is  SO(3)-symmetric or  very  close  to it,  i.e.~$\rho_{ij}
\approx  \rho\,  {\bf  1}_3$,  the  resulting  lepton-number-violating
Majorana mass, $m_N = \rho\,  \langle S \rangle$, will be closely tied
to  the $\mu$-parameter  of the  MSSM.   If $\lambda  \sim \rho$,  the
$F_D$-term hybrid  model will  then give rise  to 3  nearly degenerate
heavy  Majorana neutrinos $\nu_{1,2,3\,R}$,  as well  as to  3 complex
right-handed       sneutrinos       $\widetilde{N}_{1,2,3}$,      with
electroweak-scale  masses.    Such  a  mass  spectrum   opens  up  the
possibility  to explain  the baryon  asymmetry in  the  Universe (BAU)
$\eta_B$~\cite{FY,BAUpapers}  by  thermal  electroweak-scale  resonant
leptogenesis~\cite{APRD,APRL,PU2}, almost independently of the initial
baryon-number composition  of the primordial  plasma.  Moreover, since
the  $F_D$-term  hybrid   model  conserves  $R$-parity~\cite{GP},  the
lightest supersymmetric  particle (LSP)  is stable.  Here,  we examine
the  possibility  that   {\em  thermal}  right-handed  sneutrinos  are
responsible  for solving  the cold  dark matter  (CDM) problem  of the
Universe.

In  this   paper  we  also  improve   an  earlier  approach~\cite{GP},
concerning  the production of  the quasi-stable  U(1)$_X$ gauge-sector
particles  during the  preheating epoch.   In addition,  we  present a
numerical  analysis  that properly  takes  into  account the  combined
effect on the  reheat temperature $T_{\rm reh}$ from  the inflaton and
gauge-sector particle  decays and  their annihilations.  We  call this
two-states'  mechanism   of  reheating  the   Universe,  {\em  coupled
reheating}.   After   solving  numerically  a   network  of  Boltzmann
equations~(BEs) that appropriately  treat coupled reheating, we obtain
estimates for the present abundance of gravitinos in the Universe.  We
show  explicitly, how  a small  breaking  of $D$-parity  sourced by  a
subdominant  FI  $D$-tadpole  helps  to  relax  the  strict  gravitino
overproduction constraint.

In addition to gravitinos, one  might have to worry that topologically
stable cosmic strings do not contribute significantly to the CMB power
spectrum~$P_{\cal   R}$.   Cosmic  strings,   global  or   local,  are
topological defects and  usually form after the SSB  of some global or
local  U(1)  symmetry~\cite{NielsenOlesen,Vilenkin,HK}.  According  to
recent analyses~\cite{strings}, cosmic strings, if any, should make up
no more than about 10\% of the power spectrum $P_{\cal R}$.  This last
requirement  puts severe  limits  on the  allowed  parameter space  of
models    of    inflation.     There    have   already    been    some
suggestions~\cite{JKLS} on how to get  rid of cosmic strings, based on
modified versions  of hybrid inflation.   Here, we follow  a different
approach to solving  this problem.  We consider models,  for which the
waterfall sector  possesses an  SU(2)$_X$ gauge symmetry  which breaks
completely,  i.e.~SU(2)$_X  \to {\bf  I}$,  such  that neither  cosmic
strings nor monopoles  are produced at the end  of inflation.  In this
case,  gauge   invariance  forbids  the  existence   of  an  SU(2)$_X$
$D$-tadpole $D^a$.  However, Planck-mass suppressed non-renormalizable
operators  that  originate from  the  superpotential  or the  K\"ahler
potential  can give  rise  to explicit  breaking  of $D$-parity.   The
latter   may  manifest   itself   by  the   generation  of   effective
$D^a$-tadpole terms  that arise after  the SSB of~SU(2)$_X$.   In this
way, all the SU(2)$_X$ gauge-sector particles can be made unstable.

The organization of the paper is as follows: in Section~\ref{FDmodel},
we  describe the  $F_D$-term  hybrid model  and  calculate the  1-loop
effective potential  relevant to  inflation.  In addition,  we discuss
the  possible cosmological  consequences of  radiative effects  on the
flat directions  in the MSSM.   We conclude this section  by outlining
how the  $F_D$-term hybrid  model could generally  be embedded  into a
grand unified  theory~(GUT), including possible realizations  of a GUT
without  cosmic strings and  monopoles.  Technical  details concerning
mechanisms  of  explicit $D$-parity  breaking  in  SUGRA, e.g.~via  an
effective subdominant  $D$-tadpole or non-renormalizable  operators in
K\"ahler    potential,   are   given    in   Appendix~\ref{Dappendix}.
Section~\ref{inflation}  analyzes the  constraints on  the theoretical
parameters, which are mainly  derived from considerations of the power
spectrum $P_{\cal R}$ and a strongly red-tilted spectral index $n_{\rm
s}$,  with $n_{\rm  s} \approx  0.95$, as  observed most  recently by
WMAP~\cite{WMAP3,Lyman}.  We  show how a  negative Hubble-induced mass
term in  a next-to-minimal extension of supergravity  helps to account
for  the present  CMB data,  as well  as to  substantially  weaken the
strict constraints  on the  model parameters, originating  from cosmic
string effects on  $P_{\cal R}$, within a U(1)$_X$  realization of the
$F_D$-term hybrid model.

In  Section~\ref{Preheat},  we  analyze   the  mass  spectrum  of  the
inflaton-waterfall  sector in  the post-inflationary  era  and present
naive estimates  of the reheat  temperature $T_{\rm reh}$  as obtained
from perturbative  inflaton decays.  We  then make use of  an improved
approach  to  preheating  and   compute  the  energy  density  of  the
quasi-stable waterfall  gauge particles.  In  Section~\ref{reheat}, we
solve numerically  the BEs relevant  to coupled reheating  and present
estimates  for the gravitino  abundance in  the present  Universe.  In
Section~\ref{BAU},  we   demonstrate,  how  thermal  electroweak-scale
resonant  leptogenesis can  be realized  within the  $F_D$-term hybrid
model  and discuss  the possibility  of  solving the  CDM problem,  if
thermal right-handed sneutrinos  are considered to be the  LSPs in the
spectrum.  In  Section~\ref{conclusions}, we present  our conclusions,
including a  summary of possible particle-physics  implications of the
$F_D$-term hybrid  model for high-energy colliders  and for low-energy
experiments of lepton flavour and/or number violation.

\setcounter{equation}{0}
\section{General Setup}\label{FDmodel}

In this section  we first present the general  setup of the $F_D$-term
hybrid  model  within the  minimal  SUGRA  framework  and compute  the
renormalized  1-loop effective  potential relevant  to  inflation.  We
then discuss the cosmological implications of radiative effects on the
MSSM  flat directions  for $F_D$-term  hybrid inflation  and  for SUSY
inflationary models  in general. Finally, we analyze  the prospects of
embedding the $F_D$-term hybrid model into a GUT.

\subsection{The Model}\label{FD}

The renormalizable  superpotential of  the $F_D$-term hybrid  model is
given by
\begin{eqnarray}
  \label{Wmodel}
 W & =& \kappa\, \widehat{S}\, \Big( \widehat{X}_1
\widehat{X}_2\:  -\: M^2\Big)\ +\ \lambda\, \widehat{S} \widehat{H}_u
\widehat{H}_d\ +\ \frac{\rho_{ij}}{2}\, \widehat{S}\, \widehat{N}_i
\widehat{N}_j\ +\ h^{\nu}_{ij} \widehat{L}_i \widehat{H}_u
\widehat{N}_j\nonumber\\ &&+\ W_{\rm MSSM}^{(\mu = 0)}\; ,
\end{eqnarray}
where $W_{\rm  MSSM}^{(\mu =  0)}$ denotes the  MSSM superpotential
without the $\mu$-term:
\begin{equation} W_{\rm MSSM}^{(\mu = 0)}\ =\
  h^u_{ij}\,\widehat{Q}_i\widehat{H}_u\widehat{U}_j\: +\:
h^d_{ij}\,\widehat{H}_d\widehat{Q}_i\widehat{D}_j\: +\:
  h_l\, \widehat{H}_d\widehat{L}_l\widehat{E}_l \; .
\end{equation}
The first term in~(\ref{Wmodel}) describes the inflaton-waterfall (IW)
sector.   Specifically,  $\widehat{S}$   is  the  SM-singlet  inflaton
superfield, and $\widehat{X}_{1,2}$ is  a chiral multiplet pair of the
waterfall fields with opposite charges under the U(1)$_X$ gauge group,
i.e.~$Q (\widehat{X}_1) = - Q  (\widehat{X}_2) = 1$.  In addition, the
corresponding   inflationary   soft   SUSY-breaking  sector   obtained
from~(\ref{Wmodel}) reads:
\begin{equation}
  \label{Lsoft}
-\, {\cal L}_{\rm soft}\ =\ M^2_S S^*S\: +\: \Big(
\kappa A_\kappa\, S X_1X_2\: +\: \lambda A_\lambda S H_u H_d\: \: +\:
\frac{\rho}{2}\, A_\rho\, S \widetilde{N}_i\widetilde{N}_i\:
-\: \kappa a_S M^2 S \: \ +\ {\rm  H.c.}\,\Big)\,,
\end{equation}
where   $M_S$,   $A_{\kappa,\lambda,\rho}$    and   $a_S$   are   soft
SUSY-breaking mass parameters of order $M_{\rm SUSY} \sim 1$~TeV.

The    second   term    in~(\ref{Wmodel}),    $\lambda\,   \widehat{S}
\widehat{H}_u  \widehat{H}_d$, induces  an  effective $\mu$-parameter,
when the scalar component of $\widehat{S}$, $S$, acquires a VEV, i.e.
\begin{equation}
  \label{mu}
\mu\ =\  \lambda\, \langle S  \rangle\ \approx\ \frac{\lambda}{2\kappa}\,
|A_\kappa - a_S|\ .
\end{equation}
In obtaining the last approximate equality in~(\ref{mu}), we neglected
the VEVs  of $H_{u,d}$ and  considered the fact  that the VEVs  of the
waterfall fields $X_{1,2}$ after inflation are: $\langle X_{1,2}\rangle
=   M$~\cite{DLS}.    For  $\lambda   \sim   \kappa$,   the  size   of
$\mu$-parameter turns out to be of the order of the soft-SUSY breaking
scale $M_{\rm  SUSY}$, as required for a  successful electroweak Higgs
mechanism.    By   analogy,    the   third   term   in~(\ref{Wmodel}),
$\frac{1}{2}\,\rho_{ij}\,  \widehat{S}\, \widehat{N}_i \widehat{N}_j$,
gives  rise  to  an  effective lepton-number-violating  Majorana  mass
matrix, i.e.~$M_S  = \rho_{ij}\,  v_S$.  Assuming that  $\rho_{ij}$ is
approximately  SO(3) symmetric,  viz.~$\rho_{ij}  \approx \rho\;  {\bf
1}_3$,  one   obtains  3  nearly   degenerate  right-handed  neutrinos
$\nu_{1,2,3\,R}$, with mass
\begin{equation}
  \label{mN}
m_N\ =\ \rho\, v_S\ .
\end{equation}
If  $\lambda$  and  $\rho$  are  comparable  in  magnitude,  then  the
$\mu$-parameter and  the SO(3)-symmetric Majorana mass  $m_N$ are tied
together, i.e.~$m_N  \sim \mu$, thus  leading to a scenario  where the
singlet   neutrinos  $\nu_{1,2,3\,R}$  can   naturally  have   TeV  or
electroweak-scale masses~\cite{PU2,GP}.

The renormalizable  superpotential~(\ref{Wmodel}) of the  model may be
uniquely determined by imposing the continuous $R$ symmetry:
\begin{equation}
  \label{RFD}
\widehat{S}\  \to\ e^{i\alpha}\,\widehat{S}\, ,\qquad
\widehat{L}\  \to\  e^{i\alpha}\, \widehat{L}\,, \qquad
\widehat{Q}\ \to\ e^{i\alpha}\, \widehat{Q}\; ,
\end{equation}
with $W \to e^{i\alpha} W$,  whereas all other fields remain invariant
under an $R$ transformation.  Notice that the $R$ symmetry~(\ref{RFD})
forbids  the  presence of  higher-dimensional  operators  of the  form
$\widehat{X}_1 \widehat{X}_2  \widehat{N}_i \widehat{N}_j/m_{\rm Pl}$.
This fact ensures that  the electroweak-scale Majorana mass $m_N$ does
not get destabilized by Planck-scale SUGRA effects.

One  may  now  observe   that  the  superpotential  (\ref{Wmodel})  is
symmetric   under   the   permutation   of   the   waterfall   fields,
i.e.~$\widehat{X}_1 \leftrightarrow  \widehat{X}_2$.  This permutation
symmetry persists,  even after the  SSB of U(1)$_X$, since  the ground
state, $\langle X_1  \rangle = \langle X_2 \rangle  = M$, is invariant
under the  same symmetry  as well. Hence,  there is an  exact discrete
symmetry acting on the gauged  waterfall sector, a kind of $D$-parity.
As a consequence of  $D$-parity conservation, the ultraheavy particles
of mass  $g M$, which  are related to  the U(1)$_X$ gauge  sector, are
stable.  Such a possibility is not very desirable, as these particles,
if abundantly produced, may overclose  the Universe at late times.  In
order to  break this unwanted  $D$-parity, a subdominant  FI $D$-term,
$-\frac{1}{2} g\, m^2_{\rm FI}\,  D$, is added to the model~\cite{GP},
giving rise  to the $D$-term potential~\footnote{The  $D$-parity is an
accidental discrete  symmetry and it  should not be confused  with the
${\rm   U(1)}_X$   charge  conjugation   symmetry   realized  by   the
transformations: $X_1 \leftrightarrow  X_1^*$ and $X_2 \leftrightarrow
X_2^*$. Although  both discrete symmetries  have the same  effect when
acting on the  ${\rm U(1)}_X$ scalar current $j_X^\mu={\rm  i} ( X_1^*
\stackrel{\leftrightarrow}{\partial^\mu}       X_1       -       X^*_2
\stackrel{\leftrightarrow}{\partial^\mu}      X_2)$,     i.e.~$j_X^\mu
\leftrightarrow  -  j_X^\mu$,  they  crucially differ  when  they  are
applied on  the FI $D$-term:  $-\,\frac{g}{2} m^2_{\rm FI}  (|X_1|^2 -
|X_2|^2)$.  This term is even  under charge conjugation, but odd under
a $D$-parity conjugation.}
\begin{equation}
  \label{Dterm}
V_D\ =\ \frac{g^2}{8}\ \Big( |X_1|^2\, -\, |X_2|^2\, -\, m^2_{\rm
  FI}\,\Big)^2\; .
\end{equation}
The FI $D$-term will not  affect the inflationary dynamics, as long as
$g m_{\rm FI}  \ll \kappa M$. Technically, a  subdominant $D$-term can
be  generated  radiatively after  integrating  out Planck-scale  heavy
degrees    of    freedom.    Further    discussion    is   given    in
Section~\ref{postinfl} and in  Appendix~\ref{Dappendix}, where we also
discuss   the  possibility  of   breaking  explicitly   $D$-parity  by
non-renormalizable  K\"ahler potential  terms.   The post-inflationary
implications  of  the  FI  $D$-term,  $m_{\rm  FI}$,  for  the  reheat
temperature    $T_{\rm    reh}$    and   the    gravitino    abundance
$Y_{\widetilde{G}}$ will be analyzed in Section~\ref{reheat}.

The inflationary potential $V_{\rm inf}$ may be represented by the sum
\begin{equation}
  \label{Vinf}
V_{\rm inf}\ =\ V^{(0)}_{\inf}\: +\:  V^{(1)}_{\inf}\: +\: V_{\rm SUGRA}\ ,
\end{equation}
where  $V^{(0)}_{\inf}$   and  $V^{(1)}_{\inf}$  are   the  tree-level
potential and the 1-loop effective potential, respectively and $V_{\rm
SUGRA}$ contains the  SUGRA contribution. Including soft-SUSY breaking
terms related to $S$,  the tree-level contribution to the inflationary
potential is
\begin{equation}
  \label{V0inf}
V^{(0)}_{\inf}\ =\ {\cal Z}_S\, \kappa^2 M^4\: +\:
M^2_S\, S^* S\: -\: \Big( \kappa a_S M^2 S\: +\: {\rm H.c.}\Big)\ ,
\end{equation}
where ${\cal  Z}^{1/2}_S$ is the wave-function  renormalization of the
inflaton  field which is  needed to  renormalize the  1-loop effective
potential  given  below  in  the   SUSY  limit  of  the  theory.   The
counter-term,  $\delta  {\cal Z}_S  =  {\cal Z}_S  -  1$,  due to  $S$
wave-function  renormalization  may  be  obtained  from  the  inflaton
self-energy $\Pi_{SS}(p^2)$, through the relation
\begin{equation}
\delta  {\cal Z}_S\ =\ -\, \frac{d\,{\rm Re}\,
  \Pi_{SS}(p^2)}{dp^2}\Bigg|_{p^2 = 0}\ .
\end{equation}
Calculating the  UV part  of $\delta {\cal  Z}_S$ from this  very last
relation, we find
\begin{equation}
  \label{dZs}
\delta  {\cal   Z}_S\ =\ -\, \frac{1}{32 \pi^2}\,
\Bigg[\, 2{\cal N}\kappa^2\, \ln\Bigg(\frac{\kappa^2 M^2}{Q^2}\Bigg)\: +\:
4\lambda^2\, \ln\Bigg(\frac{\lambda^2 M^2}{Q^2}\Bigg)\: +\:
3\rho^2\,\ln\Bigg(\frac{\rho^2 M^2}{Q^2}\Bigg)\,\Bigg]\; ,
\end{equation}
where $Q^2$ is the renormalization  scale and the inflaton field value
$|S_R|  = M$  is taken  as a  common mass  renormalization  point.  In
addition,  the  parameter  ${\cal  N}$ in~(\ref{dZs})  represents  the
dimensionality of the waterfall sector.   For example, it is ${\cal N}
= 1$ for an U(1)$_X$ waterfall sector, whilst it is ${\cal N} = N$, if
$\widehat{X}_1$   ($\widehat{X}_2$)   belongs   to   the   fundamental
(anti-fundamental)  representation  of  an SU($N$)  theory.   Observe,
finally,  that  only  the  fermionic components  of  the  superfields,
$\widehat{X}_{1,2}$,    $\widehat{H}_{u,d}$,    $\widehat{N}_{1,2,3}$,
contribute to $\delta {\cal Z}_S$.

Ignoring  soft  SUSY-breaking terms,  the  1-loop effective  potential
relevant to inflation is calculated to be
\begin{eqnarray}
  \label{V1loop}
V^{(1)}_{\rm inf} \!\!&=&\!\! \frac{1}{32\pi^2}\, \Bigg\{
{\cal N}\kappa^4\, \Bigg[ |S^2 + M^2|^2\,
\ln\Bigg(\frac{\kappa^2 (|S|^2 +M^2)}{Q^2}\Bigg) +\,
|S^2 - M^2|^2\,
\ln\Bigg(\frac{\kappa^2 (|S|^2 - M^2)}{Q^2}\Bigg)\Bigg]\nonumber\\
&&\hspace{-0.17cm} +\, 2\lambda^4\, \Bigg[
|S^2 + {\textstyle \frac{\kappa}{\lambda}}\, M^2|^2\,
\ln\Bigg(\frac{\lambda^2 (|S|^2 + {\textstyle \frac{\kappa}{\lambda}}
M^2)}{Q^2}\Bigg)\, +\,
|S^2 - {\textstyle \frac{\kappa}{\lambda}} M^2|^2\,
\ln\Bigg(\frac{\lambda^2 (|S|^2 - {\textstyle \frac{\kappa}{\lambda}}
M^2)}{Q^2}\Bigg)\Bigg]\nonumber\\
&&\hspace{-0.17cm} +\, \frac{3\rho^4}{2}\, \Bigg[
|S^2 + {\textstyle \frac{\kappa}{\rho}}\, M^2|^2\,
\ln\Bigg(\frac{\rho^2 (|S|^2 + {\textstyle \frac{\kappa}{\rho}}
M^2)}{Q^2}\Bigg)\, +\,
|S^2 - {\textstyle \frac{\kappa}{\rho}} M^2|^2\,
\ln\Bigg(\frac{\rho^2 (|S|^2 - {\textstyle \frac{\kappa}{\rho}}
M^2)}{Q^2}\Bigg)\Bigg]\nonumber\\
&&\hspace{-0.17cm} -\, |S|^4\, \Bigg[\, 2{\cal N}\kappa^4\,
\ln\Bigg(\frac{\kappa^2\,|S|^2}{Q^2}\Bigg)\: +\: 4\lambda^4\,
\ln\Bigg(\frac{\lambda^2\,|S|^2}{Q^2}\Bigg)
\: +\:
3\rho^4\,\ln\Bigg(\frac{\rho^2\, |S|^2}{Q^2}\Bigg)\,\Bigg]\,\Bigg\}\; .
\end{eqnarray}
Given~(\ref{dZs})  and~(\ref{V1loop}),  it  can  be checked  that  the
expression $V^{(0)}_{\rm  inf} + V^{(1)}_{\rm inf}$  is independent of
$\ln Q^2$, as it should be.

Finally,  the  SUGRA contribution  $V_{\rm  SUGRA}$  to $V_{\rm  inf}$
in~(\ref{Vinf}) is highly model-dependent.  In general, one expects an
infinite series  of non-renormalizable  operators to occur  in $V_{\rm
SUGRA}$, i.e.~\cite{CLLSW,CP,LR}
\begin{eqnarray}
  \label{VSUGRA}
V_{\rm SUGRA}\ =\ -\, c^2_H\, H^2\, |S|^2\: +\:
\kappa^2 M^4\, \frac{|S|^4}{2\,m^4_{\rm Pl}}\: +\: {\cal O}(|S|^6)\ .
\end{eqnarray}
where $H^2 = \kappa^2 M^4/(3 m^2_{\rm Pl})$ is the squared Hubble rate
during  inflation.   The  first  term in~(\ref{VSUGRA})  represents  a
Hubble-induced mass  term, which is preferably defined  to be negative
for observational reasons  to be discussed in Section~\ref{inflation}.
In  a model  with a  minimal K\"ahler  potential, the  parameter $c_H$
vanishes  identically.\footnote{Strictly  speaking, curvature  effects
related to  an expanding de  Sitter background will contribute  to the
potential a term given by $-\frac{3}{16\pi^2}\, (2{\cal N} \kappa^2 + 4
\lambda^2 + 3 \rho^2)\, H^2 |S|^2 \ln(|S|^2/Q^2)$, even in the minimal
K\"ahler potential case~\cite{BG}.  Such a term, however, turns out to
be  negligible to  affect  the inflation  dynamics  in the  $F_D$-term
hybrid model.  Finally,  this term may be partially  absorbed into the
RG running of $c_H^2 (Q^2)$.} In fact, if $|c_H| \stackrel{<}{{}_\sim}
10^{-2}$, its  influence on the CMB  data~\cite{JP} gets marginalized.
In our analysis in Section~\ref{inflation}, we present results for two
representative  models:  (i)  the  scenario with  a  minimal  K\"ahler
potential  ($c_H  = 0$);  (ii)  a  next-to-minimal K\"ahler  potential
scenario with  $c_H \stackrel{<}{{}_\sim} 0.2$, where  only the effect
of  the  term  $(\widehat{S}^\dagger \widehat{S})^2/m^2_{\rm  Pl}$  is
considered  and  all  higher  order non-renormalizable  operators  are
ignored  in  the K\"ahler  manifold.   Moreover,  we neglect  possible
1-loop       contributions      to       $V_{\rm       inf}$      from
$A_{\kappa,\lambda,\rho}$-terms, which are insignificant for values $M
\stackrel{>}{{}_\sim} 10^{15}$~GeV.  We  only include the tadpole term
$\kappa a_S M^2\, S$, which  may become relevant for values of $\kappa
\stackrel{<}{{}_\sim}   10^{-4}$,   but    ignore   all   other   soft
SUSY-breaking    terms,    since    they   are    negligible    during
inflation~\cite{SS}.

The stability  of the inflationary  trajectory in the presence  of the
Higgs  doublets  $H_{u,d}$   and  the  right-handed  scalar  neutrinos
$\widetilde{N}_{1,2,3}$ provides further restrictions on the couplings
$\lambda$  and  $\rho$.   In  order  to  successfully  trigger  hybrid
inflation,  the fields  at  the  start of  inflation  should obey  the
following conditions:
\begin{equation}
  \label{initial}
{\rm Re}\, S^{\rm in}\ =\ |S^{\rm in}|\ \stackrel{>}{{}_\sim}\ M\,,\qquad
X^{\rm in}_{1,2}\ =\ 0\,,\qquad
H^{\rm in}_{u,d}\ =\ 0\,,\qquad
\widetilde{N}^{\rm in}_{1,2,3}\ =\ 0\; .
\end{equation}
The precise start  values of the inflaton ${\rm  Re}\, S^{\rm in}$ are
determined by the number of $e$-folds ${\cal N}_e$, which is a measure
of Universe's  expansion during inflation (see also  our discussion in
Section~\ref{inflation}).    After   inflation   and   the   waterfall
transition mechanism  have been completed,  it is important  to ensure
that  the   waterfall  fields  acquire  a  high   VEV,  i.e.   $X^{\rm
end}_{1,2}\ =\ M$, while all other fields have small electroweak-scale
VEVs.  This  can be achieved  by requiring that the  Higgs-doublet and
the  sneutrino mass  matrices  stay positive  definite throughout  the
inflationary  trajectory up  to a  critical value  $|S_c|  \approx M$.
Instead, the corresponding mass matrix  of $X_{1,2}$ will be the first
to develop  a negative eigenvalue  and tachyonic instability  close to
$|S_c|$.  As a consequence, the  fields $X_{1,2}$ will be the first to
start  moving away  from 0  and set  in to  the `good'  vacuum $X^{\rm
end}_1\  =\  X^{\rm end}_2\  =\  M$,  well  before the  other  fields,
e.g.~$H^{\rm in}_{1,2}$ and  $\widetilde{N}^{\rm in}_{1,2,3}$, go to a
`bad' vacuum  where $X^{\rm end}_{1,2}\ =\ 0$,  $H^{\rm end}_{1,2}\ =\
\frac{\kappa}{\lambda}\,  M$  and  $\widetilde{N}^{\rm  in}_{1,2,3}  =
\frac{\kappa}{\rho}\,  M$.  To  better understand  this point,  let us
write down the mass matrix in the weak field basis $(H_d\,,\ H_u^* )$:
\begin{equation}
  \label{Mdoublet}
M^2_{\rm Higgs}\ =\ \left(\! \begin{array}{cc}
\lambda^2 |S|^2 & -\,\kappa \lambda (M^2 - X_1 X_2 ) \\
-\,\kappa\lambda (M^2- X^*_1 X^*_2) & \lambda^2 |S|^2 \end{array}\!\right)\ .
\end{equation}
Then,  positive definiteness  of $M^2_{\rm  Higgs}$ implies that
\begin{equation}
  \label{Scondition}
\lambda\, |S|^2\ \ge\ \kappa \, |M^2 - X_1 X_2 |\ .
\end{equation}
From~(\ref{Scondition}),  it is  evident that  the  condition $\lambda
\stackrel{>}{{}_\sim}   \kappa$  is   sufficient  for   ending  hybrid
inflation  to the  `good'  vacuum. Finally,  one  obtains a  condition
analogous to~(\ref{Scondition}) from  the sneutrino mass matrix, which
is  equivalent  to having  $\rho  \stackrel{>}{{}_\sim} \kappa$.   The
above two constraints on $\lambda$ and $\rho$, i.e.~$\lambda,\ \rho\ >
\kappa$,   will   be   imposed    in   the   analysis   presented   in
Section~\ref{inflation}.

\subsection{Radiative Lifting of MSSM Flat Directions}\label{RadLift}

Flat    directions   in    supersymmetric   theories,    e.g.~in   the
MSSM~\cite{GKM}, play an important role in cosmology~\cite{AD,DK}.  As
we  will demonstrate  in  this section,  however,  their influence  on
$F_D$-term  hybrid   inflation  is  minimal   under  rather  realistic
assumptions.

One possible consequence of flat directions could be the generation of
a   primordial  baryon   asymmetry  $\eta_B^{\rm   in}$   through  the
Affleck--Dine  mechanism~\cite{AD}.  However,  if this  initial baryon
asymmetry $\eta_B^{\rm in}$ is generated at temperatures $T > m_N$, it
will rapidly  be erased  by the strong  $(B-L)$-violating interactions
mediated  by electroweak-scale  heavy  Majorana neutrinos  at $T  \sim
m_N$. The BAU  will then reach the present observed  value by means of
the thermal  resonant leptogenesis mechanism  and will only  depend on
the   basic   theoretical   parameters   of  the   $F_D$-term   hybrid
model~\cite{APRL,PU2}.  More details are given in Section~\ref{BAU}.

In  addition,  one  might   argue  that  large  VEVs  associated  with
quasi-flat directions  in the  MSSM would make  all MSSM  particles so
heavy  after  inflation, such  that  all  perturbative  decays of  the
inflaton would  be kinematically blocked and hence  the Universe would
never  thermalize~\cite{AM}.  The  system  may fall  into a  false
vacuum with  a large VEV at  the start of inflation,  which could, for
example, be  triggered by a negative Hubble-induced  squared mass term
of order $H^2$~\cite{DRT}, along the flat direction. In the $F_D$-term
hybrid  model, however, spontaneous  SUSY breaking  due to  a non-zero
$\langle S  \rangle$ is communicated  radiatively to the  MSSM sector,
via  the renormalizable  operators $\lambda  \widehat{S} \widehat{H}_u
\widehat{H}_d$  and  $\rho  \widehat{S} \widehat{N}_i  \widehat{N}_i$.
Consequently, their effects  on the MSSM flat directions  can be large
and  so  affect  the  inflaton  decays  which  proceed  via  the  same
renormalizable operators. In the  following, we will present a careful
treatment  of this  radiative  lifting of  MSSM  flat directions,  and
examine  the  conditions,  under  which the  directions  would  remain
sufficiently  flat  so  as  to  prohibit  the  Universe  from  thermal
equilibration, shortly after inflation.

To  obtain a  flat direction  in supersymmetric  theories, one  has to
impose the conditions of $D$- and $F$-flatness on the scalar potential
$V$, namely  the vanishing  of all $F$-  and $D$-terms for  a specific
field configuration $\sigma$.  $D$-flatness  is automatic for any flat
direction associated with a  gauge-invariant operator, which is absent
in the MSSM,  e.g.~$\widehat{D}_i \widehat{D}_j \widehat{U}_k$.  Based
on   this   observation,   let   us  therefore   consider   here   the
gauge-covariant field configuration
\begin{equation}
\sigma\ =\ \frac{1}{\sqrt{3}}\,
\left(\,\
\frac{\tilde{u}^*_{R\,k}}{|\tilde{u}_{R\,k}|}\;\tilde{d}^*_{R\,i}
\: +\:
\frac{\tilde{u}^*_{R\,k}}{|\tilde{u}_{R\,k}|}\;\tilde{d}^*_{R\,j}
\: +\:
\tilde{u}_{R\,k}\, \right)\; ,
\end{equation}
where $i\not=j$.   It can be straightforwardly checked  that the field
configuration $\sigma$, with the constraint
\begin{equation}
\frac{\tilde{u}^*_{R\,k}}{|\tilde{u}_{R\,k}|}\;\tilde{d}^*_{R\,i}  \ =
\  \frac{\tilde{u}^*_{R\,k}}{|\tilde{u}_{R\,k}|}\;\tilde{d}^*_{R\,j} \
= \ \tilde{u}_{R\,k}\ \neq \ 0
\end{equation}
and all remaining fields being set  to zero, is a flat direction, with
vanishing  $F$- and $D$-terms.   It is  then easy  to verify  that the
scalar potential $V (\sigma )$ is truly flat, i.e.~$d V/d\sigma = 0$.
Although we  will consider  here the case  of $\sigma =  \tilde u_{R\,
k}$, the  discussion of  other squark and  slepton flat  directions is
completely  analogous.   For   notational  convenience,  we  drop  all
generation  indices from  the fields,  and denote  the  flat direction
simply by $\tilde u_R$.

Because of the  spontaneous SUSY breaking induced by  the non-zero VEV
of $S$, the flatness of the potential along the $\tilde u_R$-direction
gets lifted,  once radiative corrections  are taken into  account. The
non-renormalization  theorem  related to  theories  of  SUSY is  still
applicable and entails that this radiative lifting should be UV finite
and therefore calculable.  We start our calculation by considering the
pertinent  mass spectrum  in  the  background of  a  non-zero $S$  and
$\tilde{u}_R$.   The fermionic  sector  consists of  2 Dirac  higgsino
doublets, with  squared masses $m^2_{\tilde h}= \lambda^2  |S|^2 + h^2
|\tilde{u}_R|^2$, while the mass spectrum of the bosonic sector may be
deduced by the mass matrix
\begin{equation}
  \label{Mbosonic}
{\cal M}^2_H\ =\ \left(
\begin{array}{ccc}
\lambda^2 |S|^2 & - \kappa\lambda M^2 & h \lambda S \tilde u_R^* \\
- \kappa \lambda M^2 & \lambda^2 |S|^2 + h^2 |\tilde u_R|^2 & 0\\
h\lambda S \tilde u_R & 0 & h^2 |\tilde u_R|^2
\end{array} \right)\; ,
\end{equation}
which   is   defined   in   the   weak   basis   $(H_d\,,\   H_u^*\,,\
\widetilde{Q})$.   The coupling  $h$ in~(\ref{Mbosonic})  represents a
generic up-type quark Yukawa coupling.

In the  renormalization scheme  of dimensional reduction  with minimal
subtraction~$\overline{\rm  DR}$~\cite{Jones},  the  1-loop  effective
potential $V^{(1)}$ related to $\tilde{u}_R$ is given by
\begin{equation}
  \label{VuR}
V^{(1)} (\tilde{u}_R)\ =\ \frac{2\,Q^2}{16\pi^2}\; {\rm STr}\,{\cal M}^2\
+\ \frac{2}{32\pi^2}\; {\rm STr}\, \Bigg\{ {\cal M}^4\, \Bigg[\,\ln\Bigg(
\frac{ {\cal M}^2}{Q^2}\Bigg)\ -\ \frac{3}{2}\, \Bigg]\,\Bigg\}\ ,
\end{equation}
where ${\rm STr}$ denotes  the usual supertrace, e.g.~${\rm STr} {\cal
M}^2 = {\rm  Tr}\, {\cal M}^2_H - 2 m^2_{\tilde  h}$, ${\rm STr} {\cal
M}^4  = {\rm  Tr}\, {\cal  M}^4_H  - 2  m^4_{\tilde h}$  etc.  In  the
absence of soft  SUSY-breaking terms, one finds that  ${\rm STr} {\cal
M}^2 =  0$ and ${\rm STr}  {\cal M}^4 =  2\kappa^2\lambda^2 M^4$.  The
first  condition implies the  absence of  quadratic UV  divergences in
SUSY theories, whereas  the first together with the  second one ensure
the UV  finiteness along the $\tilde{u}_R$ direction,  namely the fact
that $d V^{(1)} (\tilde{u}_R)/d\tilde{u}_R$ is $Q^2$ independent.

It  would  be  more  illuminating  to  compute  the  1-loop  effective
potential in~(\ref{VuR}) in a  Taylor series expansion with respect to
$h^2|\tilde{u}_R|^2$.   To  order $h^4  |\tilde{u}_R|^4$,  the 3  mass
eigenvalues of ${\cal M}^2_H$ are approximately given by
\begin{eqnarray}
M^2_\pm & = & \lambda^2 |S|^2\: \pm\: \kappa\lambda M^2\: +\:
\frac{\kappa M^2 \pm 2 \lambda |S|^2}{2(\kappa M^2 \pm
\lambda |S|^2)}\; h^2|\tilde{u}_R|^2\: \pm\:
\frac{\kappa M^2(\kappa M^2 \pm 3 \lambda |S|^2)}{8\lambda(\kappa
  M^2 \pm \lambda |S|^2)^3}\;
h^4 |\tilde{u}_R|^4\; ,\nonumber\\
M_0^2 &=&
\frac{\kappa^2 M^4}{\kappa^2M^4 -\lambda^2 |S|^4}\; h^2 |\tilde{u}_R|^2\:
-\:
\frac{2\kappa^2\lambda^2 M^4 |S|^6}{(\kappa^2 M^4-\lambda^2 |S|^4)^3}\;
h^4 |\tilde{u}_R|^4\; .
\end{eqnarray}
Notice  that   in  the  limit   $\tilde{u}_R  \to  0$,   one  obtains:
$m^2_{\tilde{h}}  = \lambda^2  |S|^2$,  $M^2_\pm =  \lambda |S|^2  \pm
\kappa\lambda M^2$ and $M^2_0 =  0$, as expected.  Moreover, it is not
difficult  to  check  that ${\rm  STr}  {\cal  M}^2  = {\cal  O}  (h^6
|\tilde{u}_R|^6)$ and ${\rm STr} {\cal M}^4 = 2\kappa^2\lambda^2 M^4 +
{\cal  O} (h^6  |\tilde{u}_R|^6)$, in  accordance with  our discussion
given above.

Employing the fact that $|S|^2\gg \frac{\kappa}{\lambda}\, M^2$ at the
start   of  inflation,   the  1-loop   effective   potential  $V^{(1)}
(\tilde{u}_R)$ may further be approximated as follows:
\begin{eqnarray}
\label{VuRappr}
V^{(1)} (\tilde{u}_R) \!&=&\!
\frac{\kappa^2\lambda^2 M^4}{8\pi^2}\, \Bigg[
\ln\Bigg(\frac{\lambda^2 |S|^2}{Q^2}\Bigg)\: -\:  \frac{3}{2}\, \Bigg]\
-\ \frac{1}{48\pi^2}\frac{h^2 \kappa^4 M^8}{\lambda^2 |S|^6}\;
|\tilde{u}_R|^2\
+\
\frac{1}{16\pi^2}\frac{h^4\kappa^2 M^4}{\lambda^2 |S|^4}\; |\tilde{u}_R|^4
\nonumber\\
&&+\; \frac{1}{16\pi^2}
\left(\frac{h^2\kappa^2 M^4}{\lambda^2 |S|^4}\; |\tilde{u}_R|^2\, \right)^2
\ln\left(\frac{h^2\kappa^2 M^4}{\lambda^4 |S|^6}\; |\tilde{u}_R|^2
\right)\quad
+\quad {\cal O}(h^6 |\tilde{u}_R|^6)\,.
\end{eqnarray}
The   first  term   in~(\ref{VuRappr})  contributes   to   the  1-loop
inflationary    potential~(\ref{V1loop}),    while    the    remaining
$Q^2$-independent     terms    lift     the     flatness    of     the
$\tilde{u}_R$-direction.  Assuming  that $\kappa^2 \ll  \lambda^2$ and
$M\simeq |S|$ towards  the end of inflation, we  find the well-defined
minimum
\begin{equation}
  \label{VEVuR}
\langle\, \tilde{u}_R\, \rangle\ =\ \frac{\kappa}{\sqrt{6}\,h}\ M\; .
\end{equation}
We should remark  here that the above minimum  would remain unaltered,
even if the  flat direction were a squark or  slepton doublet. In this
case, only the overall  normalization of the $Q^2$-independent part of
$V^{(1)}  (\tilde{u}_R)$ would  have  changed by  a  factor 1/2.   The
loop-induced   VEV   of  $\tilde{u}_R$   generates   a  squared   mass
$M^2_{\tilde{u}_R}$ via the Higgs mechanism, which is given by
\begin{equation}
  \label{MuR}
M^2_{\tilde{u}_R}\ =\ \frac{1}{24\pi^2}\,
\frac{h^2\,\kappa^4}{\lambda^2}\; M^2\ .
\end{equation}
This squared mass $M^2_{\tilde{u}_R}$ should be compared with the size
of possible negative Hubble-induced squared mass terms of order $H^2 =
\kappa^2   M^4/    (3m_{\rm   Pl}^2)$,   e.g.~terms    of   the   form
$-c^2_{\tilde{u}}\,  H^2 |\tilde{u}_R|^2$ that  may occur  in $V^{(1)}
(\tilde{u}_R)$ and originate from  SUGRA effects. These terms may play
some   role   in   our   model,  unless   $c^2_{\tilde{u}}\,   H^2   <
M^2_{\tilde{u}_R}$.  The  latter condition may be  translated into the
inequality
\begin{equation}
  \label{cuR}
c_{\tilde{u}}\ <\ \frac{1}{2\sqrt{2}\pi}\, \frac{h\,\kappa}{\lambda}\;
\frac{m_{\rm Pl}}{M}\ .
\end{equation}
As a typical  example, let us consider an  inflationary scenario, with
$\lambda = 2\kappa$, $\kappa = 10^{-3}$ and $M = 10^{16}$~GeV. In this
case,  (\ref{cuR}) implies  that $c_{\tilde{u}}  < 0.87\,  h$.  Hence,
although  the required  tuning of  the coefficient  $c_{\tilde{u}}$ to
fulfill  this last  inequality may  not  be significant  for the  third
generation squarks  and sleptons, it  becomes excessive for  the first
generation, unless a minimal K\"ahler potential is assumed.  It should
be  stressed   here,  however,  that   the  deepest  and   hence  most
energetically favoured  minimum for all squark  and slepton directions
is the  one related  to $\tilde{t}_R$. In  other words,  given chaotic
initial conditions, the fields are  most likely to settle to minima of
quasi-flat directions involving large Yukawa couplings.  In this case,
radiative  effects play  an important  role  in the  dynamics of  flat
directions\footnote{We  should   note  that  the   evolution  of  flat
directions during the waterfall  and coherent oscillation periods is a
non-equilibrium  dynamics problem.   Moreover, no  theoretical methods
yet exist  that would  lead to a  practical solution to  this problem,
even though effective potential  corrections to the flat directions as
the ones considered here are  expected to be relevant during the above
cosmological periods.}.

Let  us  finally  assume  that   we  are  in  a  situation  where  the
Hubble-induced mass  terms can be neglected,  i.e.~$c_{\tilde{u}} = 0$
as is the case for  a minimal K\"ahler potential, for example. Suppose
that  the  loop-induced  VEV  of  the  quasi-flat  direction  persists
throughout  the  coherent  oscillatory  regime.   In  this  case,  the
VEV~(\ref{VEVuR}) gives rise to masses $h\langle \tilde{u}_R \rangle =
\kappa M/\sqrt{6}$ in the $\widehat{Q} \widehat{H}_u$-sector, which do
not   depend  on   the   Yukawa  coupling   $h$.   Consequently,   the
inflaton-related   fields   of  mass   $\sqrt{2}   \kappa\,  M$   (see
Table~\ref{spectrum}) will  have a large  decay rate to  those massive
particles, thus creating a non-thermal distribution.  This non-thermal
distribution will in turn induce $T$-dependent mass terms which can be
larger than  the expansion  rate $H(T)$ at  some temperature  $T$ soon
after inflation, such that  $\langle \tilde{u}_R \rangle$ will rapidly
relax  to   zero.   Of  course,  one  might   think  of  contemplating
configurations  where   multiple  flat  directions   have  VEVs  which
contribute constructively to the masses  of both $H_u$ and $H_d$, such
that all inflaton and waterfall particle decays would be kinematically
forbidden.   However,  we  consider   such  a  possibility  as  a  bit
contrived.    It    is   therefore   reasonable    to   assume   that,
provided~(\ref{cuR}) is fulfilled,  reheating and equilibration of all
MSSM degrees of freedom will take place in the $F_D$-term hybrid model
and  in  all  supersymmetric  models  of  inflation  that  include  an
unsuppressed   renormalizable  operator   of  the   form  $\widehat{S}
\widehat{H}_u \widehat{H}_d$.

\subsection{Topological Defects and GUT Embeddings}\label{TDGUT}

As  we mentioned  in the  Introduction, topological  defects,  such as
domain walls, cosmic  strings or monopoles, may be  created at the end
of inflation,  when a symmetry  group $G$, local, global  or discrete,
breaks  down into  a  subgroup $H$,  in  a way  such  that the  vacuum
manifold  $M =  G/H$ is  not trivial.   Specifically,  the topological
properties  of the  vacuum  manifold $M$  under  its homotopy  groups,
$\pi_n    (M)$,   determine    the   nature    of    the   topological
defects~\cite{Vilenkin,HK}.  Thus, one  generally has the formation of
domain walls for  $\pi_0 (M) \neq {\bf I}$,  cosmic strings for $\pi_1
(M) \neq {\bf I}$, monopoles if  $\pi_2 (M) \neq {\bf I}$, or textures
if $\pi_{n  > 2} (M)  \neq {\bf I}$~\cite{Vilenkin}. For  example, for
the SSB breaking pattern U(1)$_X \to {\bf I}$ in the waterfall sector,
the  first homotopy  group  of  the vacuum  manifold  is not  trivial,
i.e.~$\pi_1({\rm  U}(1)/{\bf I})  = {\bf  Z}$.  In  this  case, cosmic
strings will  be produced  at the end  of inflation.  In  general, the
non-observation  of  any  cosmic  string  contribution  to  the  power
spectrum $P_{\cal R}$ at the 10\% level introduces serious constraints
on the theoretical parameters of hybrid inflation models.

A  potentially   interesting  inflationary  scenario   arises  if  the
waterfall sector possesses an SU(2)$_X$ gauge symmetry.  In this case,
the SSB  breaking pattern  is: SU(2)$_X \to  {\bf I}$,  i.e.~the group
SU(2)$_X$ breaks completely. It is worth stressing here that this is a
unique  property of  the SU(2)  group,  since the  breaking of  higher
SU($N$)  groups,  with $N  >  2$, into  the  identity~{\bf  I} is  not
possible.    Moreover,   an  homotopy   group   analysis  gives   that
$\pi_{0,1,2} ({\rm SU}(2)_X/{\bf I}) = {\bf I}$, implying the complete
absence  of domain  walls,  cosmic strings  and  monopoles.  The  only
non-trivial homotopy  group is $\pi_3  ({\rm SU}(2)_X/{\bf I})  = {\bf
Z}$, thus signifying the formation  of textures, in case the SU$(2)_X$
group is global.  If the SU(2)$_X$ group is local, however, observable
textures do not occur.  Since their corresponding field configurations
never leave the  vacuum manifold, the would-be textures  can always be
compensated by  local SU(2)$_X$ gauge transformations~\cite{Vilenkin}.
It  is therefore  essential  that the  $X$-symmetry  of the  waterfall
sector is local in the $F_D$-term hybrid model.

It is now interesting to  explore whether generic scenarios exist, for
which the waterfall gauge groups  ${\rm U}(1)_X$ or ${\rm SU}(2)_X$ of
the $F_D$-term hybrid model  may, partially or completely, be embedded
into a  GUT. As a key  element for such a  model-building, we identify
the maintenance of $D$-parity  conservation in the $X$-gauged waterfall
sector, which  is discussed  in detail in  Section~\ref{postinfl}.  In
order to  preserve $D$-parity, the waterfall sector  should be somehow
`hidden'  from the  perspective of  the SM  gauge group  $G_{\rm SM}$.
This  means that  the SM  fields must  be neutral  under $X$  and vice
versa, the $X$-gauge and waterfall sector fields should not be charged
under  $G_{\rm SM}$.   Consequently,  we  have to  require,  as a  GUT
breaking   route,  that   the  waterfall   $X$-gauge  group   and  the
GUT-subgroup that contains  $G_{\rm SM}$ factor out into  a product of
two independent groups without overlapping charges.

It is reasonable to assume  that the GUT-subgroup is broken to $G_{\rm
SM}$ before  or while inflation takes place.   Then, possible unwanted
topological  defects due to  the various  stages of  symmetry breaking
from the  GUT-subgroup down to the  SM will be  inflated away.  Notice
that we do not have to require that the GUT-subgroup breaking scale is
higher than the respective  $X$-symmetry breaking scale, but only that
the  reheat temperature  $T_{\rm  reh}$  is low  enough  such that  no
symmetries  of  the GUT-subgroup  are  restored  during reheating.   A
related  discussion  within  the   context  of  SO(10)  may  be  found
in~\cite{SO10inflation}.

Let us first investigate whether a `hidden' gauge group ${\rm U}(1)_X$
related to the waterfall sector  can be embedded into a GUT.  Although
`hidden'  ${\rm   U}(1)$'s  naturally   arise  in  models   of  string
compactification~\cite{ExtraU1fromStrings},  our interest  here  is to
identify possible ${\rm  U}(1)_X$ factors that can be  embedded into a
simple GUT.   Given the above criterion, the  frequently discussed GUT
based on ${\rm SO}(10)$ should  be excluded, since it does not contain
`hidden'  U(1)$_X$ groups~\cite{Slansky:1981}.   As  a next  candidate
theory,  we may  consider the  exceptional  group E(6),  with the  SSB
breaking  path ${\rm E}(6)\to  {\rm U}(1)  \times {\rm  SO}(10)$.  The
fundamental representation of ${\rm  E}(6)$ is the chiral ${\bf 27}_F$
representation, which branches under  ${\rm U}(1) \times {\rm SO}(10)$
as follows:
\begin{equation}
{\bf 27}_F\  =\ (4,{\bf 1})\: +\: (-2, {\bf 10})\:  +\: (1,{\bf 16})\; .
\end{equation}
Although  the SM  particles  may fit  into  ${\bf 16}$,  they are  not
neutral  under the extra  U(1). Higher  representations, such  as $(0,
{\bf 45})$ stemming from ${\bf 78}$ of ${\rm E}(6)$, are neutral under
the  ${\rm  U}(1)$ factor,  but  they  are  not suitable  to  properly
accommodate all the SM particles.

We  therefore turn  our  attention to  possible  breaking patterns  of
maximal  groups that  contain a  `hidden' ${\rm  SU}(2)_X$  factor.  A
promising  example is  ${\rm  E}(6)\supset {\rm  SU}(2)_X \times  {\rm
SU}(6)$, where the fundamental representation ${\bf 27}_F$ follows the
branching:
\begin{equation}
  \label{27F}
{\bf 27}_F\ =\ ({\bf 2} ,{\bf \overline 6})\: +\: ({\bf 1},{\bf 15})\; .
\end{equation}
Under ${\rm SU}(6) \supset {\rm  SU}(5) \times {\rm U}(1)$, ${\bf 15}$
is an antisymmetric  representation of SU(6) and one  of its branching
rules is
\begin{equation}
  \label{15SM}
{\bf 15}\ =\ ({\bf 5},-4)\: +\: ({\bf 10},2)\; .
\end{equation}
However, we need a ${\bf \overline 5}$ of ${\rm SU}(5)$, together with
${\bf 10}$ in~(\ref{15SM}), in  order to appropriately describe all SM
fermions.   This shortcoming may  be circumvented  by adding  an extra
${\bf  \overline{27}}_F$ of ${\rm  E}(6)$ to  the spectrum,  where the
missing ${\bf \overline 5}$ may be obtained from the complex conjugate
branching of~(\ref{15SM}).  Such an extension of the particle spectrum
may even be welcome to resolve the proton stability problem, through a
kind of split multiplet mechanism~\cite{proton}.  Within the framework
of SUSY, the quark and lepton Yukawa interactions may be generated via
the  introduction of  a pair  of  the multiplets  ${\bf 27}_H$,  ${\bf
\overline{27}}_H$.  Finally,  in such an E(6) unified  scenario, the 3
right-handed neutrinos can only appear as singlets.

Another possible  GUT scenario that  complies with our criterion  of a
hidden  SU(2)$_X$  is ${\rm  E}(7)\supset  {\rm  SU}(2)_X \times  {\rm
SO}(12)$. The fundamental representation  is ${\bf 56}_F$ and branches
under ${\rm SU}(2)_X \times {\rm SO}(12)$ as follows:
\begin{equation}
  \label{56E7}
{\bf 56}_F\ =\ ({\bf 2} , {\bf \overline{12}})\: +\:
                                                ({\bf 1},{\bf 32})\; .
\end{equation}
Subsequently,  SO(12) breaks  spontaneously  into ${\rm  SO}(10)\times
{\rm  U}(1)$,  where  ${\bf  32}  = ({\bf  16},1)  +  ({\bf  \overline
{16}},-1)$  is a  vector-like representation.   However, one  may well
envisage   a    string-theoretic   framework,   in    which   orbifold
compactification  projects  out  the undesirable  anti-chiral  states.
Then,  all SM  particles,  including right-handed  neutrinos, will  be
contained in one of the ${\bf 16}$'s of ${\bf 32}$. Related discussion
of missing  or incomplete multiplets due  to orbifold compactification
may be found in~\cite{orbifold}.

Building  a realistic  GUT model  from  the blocks  stated above  lies
beyond the scope  of this paper.  We have  demonstrated here, however,
that the  embedding of an SU(2)$_X$  gauge group into a  GUT, which is
hidden but nevertheless takes  part in the gauge coupling unification,
appears feasible within E(6) and E(7) unified theories.

We conclude this section by observing that the presence of the singlet
inflaton  field~$S$ offers  alternative options,  for  suppressing the
heavy Majorana  neutrino masses within  SUSY GUTs.  As an  example, we
mention the  breaking scenario, where  ${\rm SO}(10) \to  {\rm SU}(5)$
via  the VEV of  a ${\bf  126}_H$ Higgs  representation and  the usual
superpotential  term  ${\bf  16}_F  \langle  {\bf  126}_H\rangle\,{\bf
16}_F$ induces heavy  Majorana masses of the GUT  scale $M_{\rm GUT}$.
Given  that the  above renormalizable  operator is  forbidden  by some
$R$-symmetry,  the presence of  an $R$-charged  inflaton $S$  may give
rise to a drastic suppression  of the GUT-scale Majorana mass, through
a superpotential  term of the  form $\widehat{S}\, {\bf  16}_F \langle
{\bf 126}_H\rangle\,{\bf 16}_F/m_{\rm Pl}$.   Since $S$ receives a VEV
of  order  $M_{\rm SUSY}/\kappa$  in  general  $F$-term hybrid  models
[cf.~(\ref{mu})], one naturally obtains heavy Majorana neutrino masses
of order $M_{\rm SUSY}$,  if $\kappa \sim \langle {\bf 126}_H\rangle\,
/m_{\rm  Pl} \sim  10^{-3}$. Such  values of  $\kappa$ do  satisfy the
current inflationary constraints which we discuss in the next section.

\setcounter{equation}{0}
\section{Inflation}\label{inflation}

Here,  we  first  briefly  review  in  Section~\ref{intro}  the  basic
formalism   of   inflation,  including   the   constraints  from   the
non-observation of  cosmic strings in the power  spectrum $P_{\cal R}$
of  the  CMB data.   Then,  in  Section~\ref{numinf},  we present  our
numerical results  for two scenarios:  (i) the minimal  SUGRA (mSUGRA)
scenario and  (ii) the  next-to-minimal SUGRA (nmSUGRA)  scenario.  In
particular, we  exhibit numerical  predictions for the  spectral index
$n_{\rm  s}$  and  discuss  its  possible  reduction  in  the  nmSUGRA
scenario.  Finally,  we  analyze   the  combined  constraints  on  the
fundamental  theoretical parameters  $\kappa$, $\lambda$,  $\rho$, and
$M$, which result from the recent CMB observations and inflation.

\subsection{Basic Formalism}\label{intro}

According to the  inflationary paradigm~\cite{review}, the horizon and
flatness  problems   of  the   standard  Big-Bang  Cosmology   can  be
technically  addressed, if  our observable  Universe has  undergone an
accelerated  expansion   of  a  number  50--60   of  $e$-folds  during
inflation.  In  the slow-roll approximation, the  number of $e$-folds,
${\cal N}_e$, is related to the inflationary potential through:
\begin{equation}
  \label{Nefold}
{\cal N}_e\ =\ \frac{1}{m^2_{\rm Pl}}\; \int_{\phi_{\rm
    end}}^{\phi_{\rm exit}}\, d\phi\: \frac{V_{\rm inf}}{V'_{\rm inf}}\
    \simeq\ 55\; .
\end{equation}
Hereafter, a  prime on $V_{\rm inf}$ will  denote differentiation with
respect  to  the inflaton  field  $\phi=\sqrt{2}\,  {\rm Re}\,S$.   In
addition, $\phi_{\rm exit}$  is the value of $\phi$,  when our present
horizon scale crossed outside inflation's horizon and $\phi_{\rm end}$
is the  value of  $\phi$ at  the end of  inflation.  In  the slow-roll
approximation, the field value $\phi_{\rm end}$ is determined from the
condition:
\begin{equation}
  \label{slow}
{\sf max}\{\epsilon(\phi_{\rm end}),|\eta(\phi_{\rm
end})|\}\ =\ 1\, ,
\end{equation}
where
\begin{equation}
  \label{epseta}
\epsilon\ =\ \frac{m_{\rm Pl}^2}{2}\ \left(
\frac{V'_{\rm inf}}{V_{\rm inf}}\right)^2\,,\qquad
\eta\ =\  m_{\rm Pl}^2\ \frac{V''_{\rm inf}}{V_{\rm inf}}\ .
\end{equation}
We  have checked  that  the slow-roll  condition~(\ref{slow}) is  well
satisfied up to the critical point $\phi_{\rm end}=\sqrt{2} M$, beyond
which  the waterfall  mechanism takes  place.  We  also find  that the
slow-roll condition  remains valid,  even within the  nmSUGRA scenario
with $c_H\neq0$ and with appreciable non-renormalizable SUGRA effects.
Finally,  we note that  the assumed  value of  ${\cal N}_e  \simeq 55$
in~(\ref{Nefold})   is   slightly  higher   than   the  one   computed
consistently  from~(\ref{Ng}), which  is about  50 for  our low-reheat
cosmological  scenario.   However,  our numerical  results  concerning
$P_{\cal R}$ and $n_{\rm s}$ do not depend on such a 10\% variation of
${\cal N}_e$ in any essential way.

The power  spectrum $P_{\cal R}$  is a cosmological observable  of the
curvature perturbations, which  sensitively depends on the theoretical
parameters of the inflationary potential. The square root of the power
spectrum, $P^{1/2}_{\cal R}$, may be conveniently written down as
\begin{equation}
    \label{PR}
P^{1/2}_{\cal R}\ =\ \frac{1}{2\sqrt{3}\, \pi m^3_{\rm Pl}}\;
\frac{V_{\rm inf}^{3/2}(\phi_{\rm exit})}{|V'_{\rm inf}(\phi_{\rm
exit})|}\ .
\end{equation}
The recent  WMAP~\cite{WMAP,WMAP3} results, which  are compatible with
the  ones suggested  for the  COBE  normalization~\cite{COBE}, require
that
\begin{equation}
  \label{Pr}
P^{1/2}_{\cal R}\ \simeq\ 4.86\times 10^{-5}\, .
\end{equation}

In addition  to scalar  curvature perturbations, tensor  gravity waves
and cosmic string effects may  also contribute to $P_{\cal R}$. In the
$F_D$-term  hybrid model  with an  Abelian $U(1)_X$  waterfall sector,
cosmic strings arise after the SSB of the gauge symmetry (see also our
discussion   in  Section~\ref{TDGUT}).    In  this   case,  additional
constraints  are obtained  from the  non-observation of  cosmic string
effects on~$P_{\cal  R}$~\cite{cstrings1,cstrings}.  The evaluation of
such effects involves a certain degree of uncertainty in the numerical
simulations  of string  networks \cite{cstrings2}.   Nevertheless, the
common approach  taken to cosmic string  effects~\cite{mairi,JP} is to
require that  their contribution $(P_{\cal R})_{\rm cs}$  to the power
spectrum $P_{\cal  R}$ does not exceed the  10\% level, i.e.~$(P_{\cal
R})_{\rm  cs}/P_{\cal R}  \stackrel{<}{{}_\sim}  0.1$.  In~detail,  we
require that
\begin{equation}
  \label{Prcs}
(P^{1/2}_{\cal R})_{\rm cs}\ \leq\ 1.54 \times 10^{-5}\; .
\end{equation}
The cosmic string contribution $(P_{\cal  R})_{\rm cs}$ to the power
spectrum may be computed by
\begin{equation}
(P^{1/2}_{\cal R})_{\rm cs}\ = \
{\sqrt{15}\over 4\pi}\: {\mu_{\rm cs}\over m^2_{\rm Pl}}\ y_{\rm cs}\; ,
\end{equation}
where the tension of the cosmic strings, $\mu_{\rm cs}$, is calculated
using the formulae:
\begin{equation}
  \label{mucs}
\mu_{\rm cs}\ =\ 2\pi M^2\epsilon_{\rm cs}(\beta)\,,\qquad
\epsilon_{\rm cs}(\beta)\ \simeq\ \left\{\matrix{
1.04\ \beta^{0.195}\hfill ,  & \mbox{for}~~\beta>10^{-2},
\hfill \cr
2.4\,/ \ln(2/\beta) \hfill ,  &\mbox{for}~~\beta\leq10^{-2}\; .
\hfill \cr}
\right.
\end{equation}
In~(\ref{mucs})   the  argument   $\beta$   is  given   by  $\beta   =
\kappa^2/(2g^2)$, while  the U(1)$_X$  gauge coupling constant  $g$ is
considered to  assume the value $g \simeq  0.7$ as is the  case in GUT
models.  The  central value of the  parameter $y_{\rm cs}$  is 8.9 and
its error  margin lies  in the interval  [6.7,11.6], according  to the
analysis in~\cite{cstrings}.

The  recently  announced   three-years  results  of  WMAP~\cite{WMAP3}
improved  upon  the  precision  of  a  number  of  other  cosmological
observables.  The  merits of  an inflationary model  can be  judged by
comparing its predictions for  the scalar spectral index, $n_{\rm s}$,
the  tensor to  scalar ratio,  $r$, and  the running  of  $n_{\rm s}$,
$dn_{\rm s}/d\ln\kappa$, with the  CMB data.  In the $F_D$-term hybrid
model,  $r=16\epsilon(\phi_{\rm exit})$  is much  lower than  the WMAP
bound,  i.e.~well  below~$10^{-2}$,  and  $dn_{\rm  s}/d\ln\kappa$  is
always smaller  than $10^{-3}$ and  so unobservable. In  addition, the
spectral index  $n_{\rm s}$ in our  model may well  be approximated as
follows:~\cite{review}
\begin{equation}
  \label{nS}
n_{\rm s}\ =\ 1-6\epsilon(\phi_{\rm exit})\ +\ 2\eta(\phi_{\rm exit})\
\simeq\ 1\ +\ 2\eta(\phi_{\rm exit}),
\end{equation}
since  $\epsilon$  is negligible.  The  predicted  value  needs to  be
compared with the recent WMAP results~\cite{WMAP3}:
\begin{equation}
  \label{nswmap}
n_{\rm  s}\ =\ 0.951_{-0.019}^{+0.015}\ .
\end{equation}
The latter is translated into the double inequality,
\begin{equation}
  \label{ns95}
0.913\ \lesssim\  n_{\rm s}\ \lesssim\ 0.981\; ,
\end{equation}
at the 95\% confidence level (CL).

The  result~(\ref{ns95})  brings  under  considerable  stress  minimal
$F$-term hybrid  inflation models~\cite{DSS}. This is due  to the fact
that these models predict $n_{\rm s}$ extremely close to unity without
much running.  To be  precise, when the radiative corrections dominate
the slope of the potential, we obtain
\begin{equation}
  \label{nSrc}
n_{\rm s}\ \simeq\ 1\ -\ {1/{\cal N}_e}\ \simeq\ 0.98\; ,
\end{equation}
for ${\cal  N}_e = 55$. On  the other hand,  if the non-renormalizable
operator  $|S|^4$ in $V_{\rm  SUGRA}$ of~(\ref{VSUGRA})  dominates the
slope of  the potential of  a mSUGRA model with  $c_H=0$~\cite{SS}, we
obtain a blue-tilted spectrum, with
\begin{equation}
  \label{nSrc2}
n_{\rm s}\ \simeq\ 1\ +\ {6M^2\over m_{\rm Pl}^2-2M^2{\cal N}_e}\
\stackrel{>}{{}_\sim}\ 1\; .
\end{equation}
A possible Hubble-induced positive  term $+c_H^2 H^2 |S|^2$ in $V_{\rm
SUGRA}$~\cite{CP, JP}  implies an  even more pronounced  blue spectrum
and is therefore excluded by the current WMAP data.

As  noticed  earlier   in~\cite{hilltop}  and  elaborated  further  in
Ref.~\cite{king},  agreement of  theory's  prediction for~$n_{\rm  s}$
with  observation  strongly  suggests   the  presence  of  a  negative
Hubble-induced  mass  term  $-c^2_H  H^2 |S|^2$  in  $V_{\rm  SUGRA}$,
thereby  clearly disfavouring the  minimal K\"ahler  potential. In~our
analysis, we therefore consider the following next-to-minimal form for
the K\"{a}hler manifold~\cite{CP}:
\begin{equation}
  \label{qK}
K_S\ =\ |S|^2\ +\ k_S\;{|S|^4\over 4\,m^2_{\rm Pl}}\ ,
\end{equation}
where  the  constant  $k_S$   can  be  either  positive  or  negative.
Substituting (\ref{qK}) into the general formula for the $F$-term type
contributions to the SUGRA potential~(see, e.g.~\cite{CLLSW}),
\begin{equation}
  \label{VF}
V_F\ =\ e^{K_S/m^2_{\rm Pl}}\, \Bigg[\, F^i (K^{-1}_S)_i^j F_j\ -\ 3\,
\frac{|W|^2}{m^2_{\rm Pl}}\,\Bigg]\ ,
\end{equation}
we  arrive at  the result~(\ref{VSUGRA})  with $c^2_H  =  3k_S$, after
neglecting   higher-order   terms    that   are   small   for   $|c_H|
\stackrel{<}{{}_\sim}   0.2$.    In    (\ref{VF}),   $F^i$   are   the
SUGRA-generalized  $F$-terms  and  $(K^{-1}_S)_i^j$ is  the  so-called
inverse  metric   of  the  K\"ahler  manifold,   where  a  superscript
(subscript)  index $i$ or  $j$ on  $K_S$ denotes  differentiation with
respect to $S$ ($S^*$).

The  aforementioned nmSUGRA  inflationary potential,  with  a negative
Hubble-induced mass term,  reaches a local minimum and  maximum at the
points  $\phi_{\rm min}$  and $\phi_{\rm  max}$,  respectively.  These
points can be estimated by
\begin{equation}
 \label{phimax}
\phi_{\rm max}\ \simeq\  {m_{\rm Pl}\over 4\pi c_H}\
\Big(\, 6\kappa^2{\cal N}\: +\: 12\lambda^2\: +\:
9\rho^2\,\Big)^{1/2}\,, \qquad
\phi_{\rm min}\ \simeq\ \sqrt{2\over3}\; c_H m_{\rm Pl}\ .
\end{equation}
For relevant  parameter values,  for which $\phi_{\rm  max} <\phi_{\rm
min}$, and under convenient  initial conditions, the so-called hilltop
inflation~\cite{hilltop}  can  take  place,  where $\phi$  rolls  from
$\phi_{\rm max}$ down to smaller  values, such that $\phi_{\rm exit} <
\phi_{\rm max}$.  In  this nmSUGRA scenario, the value  of $n_{\rm s}$
can be significantly lowered and can be approximately given by
\begin{equation}
  \label{nShilltop}
n_{\rm s}\ \simeq\ 1\: -\ \frac{1}{{\cal N}_e}\ -\ c_H^2\; .
\end{equation}
As  we will show  more explicitly  in the  next section,  the spectral
index $n_{\rm s}$ can be easily driven into the range of~(\ref{ns95}),
for  values   of  $c_H  \sim   0.1$.   The  presence  of   the  second
next-to-minimal term proportional  to $k_S$ in~(\ref{qK}) modifies the
analytic    expressions    of~(\ref{Nefold}),    (\ref{epseta})    and
(\ref{Pr})~\cite{king}.  However,  these modifications turn  out to be
numerically insignificant for the predicted values of ${\cal N}_e$ and
$P_{\cal   R}$,    if   $c_H$    is   not   very    large,   e.g.~$c_H
\stackrel{<}{{}_\sim} 0.2$.

\begin{figure}[t]
\centerline{\epsfig{file=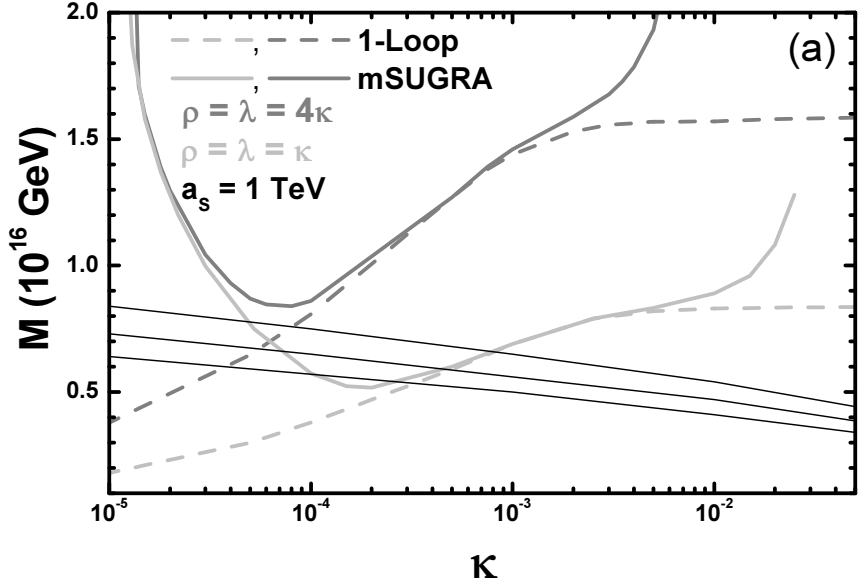,angle=-90,width=13.cm}} \hfill
\vspace*{-.15in} \hfill
\centerline{\epsfig{file=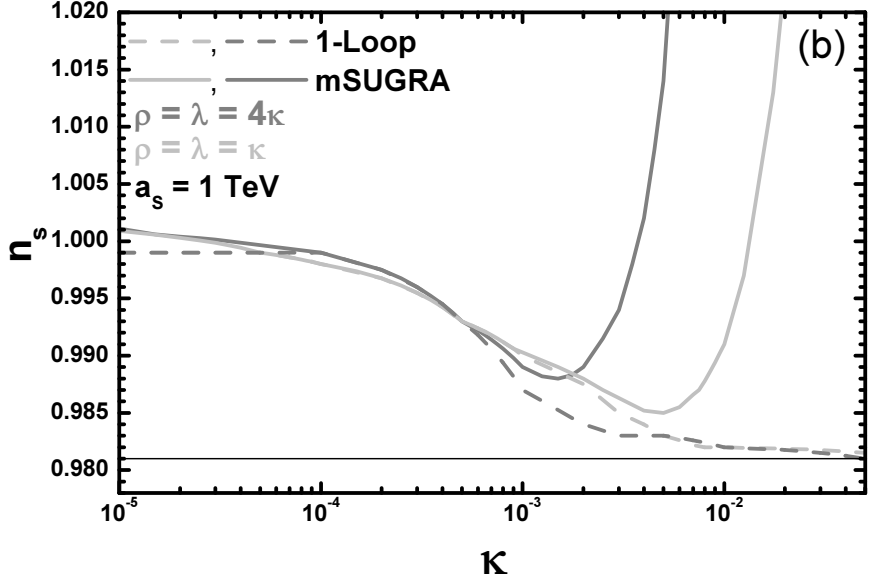,angle=-90,width=13.cm}}\hfill
\caption{\sl\small The values of the inflationary scale $M$ allowed
by~(\ref{Nefold}) and (\ref{Pr}) {\sf (a)} and the predicted values of
the spectral index $n_{\rm s}$ {\sf (b)} as a function of $\kappa$ for
${\cal N}=1$ and $\rho=\lambda=\kappa$ (light grey lines) or
$\rho=\lambda=4\kappa$ (grey lines), including the one-loop radiative
corrections (dashed lines) or the mSUGRA ($c_H=0$) contributions with
a$_S=1~{\rm TeV}$ (solid lines). The upper bound of~(\ref{Prcs}) for
$y_{\rm cs} =6.7,~8.9,~11.6$ (from top to bottom) [cf.~(\ref{nswmap})]
is also shown by thin lines {\sf (a)} [{\sf (b)}].}
\label{fig:mSUGRAPR}
\end{figure}

\subsection{Numerical results}\label{numinf}

\begin{figure}[t]
\centerline{\epsfig{file=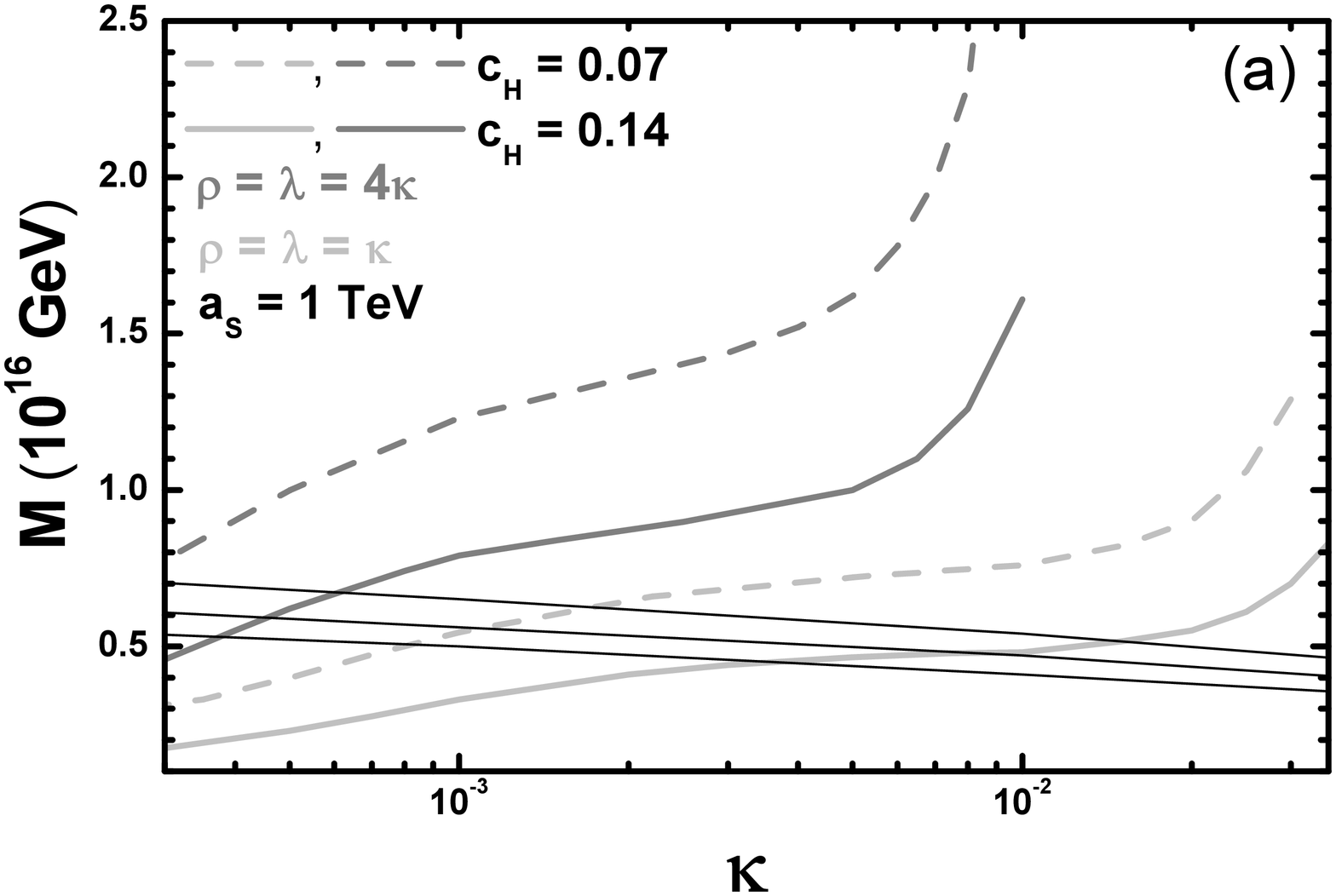,angle=-90,width=13.cm}} \hfill
\vspace*{-.15in} \hfill
\centerline{\epsfig{file=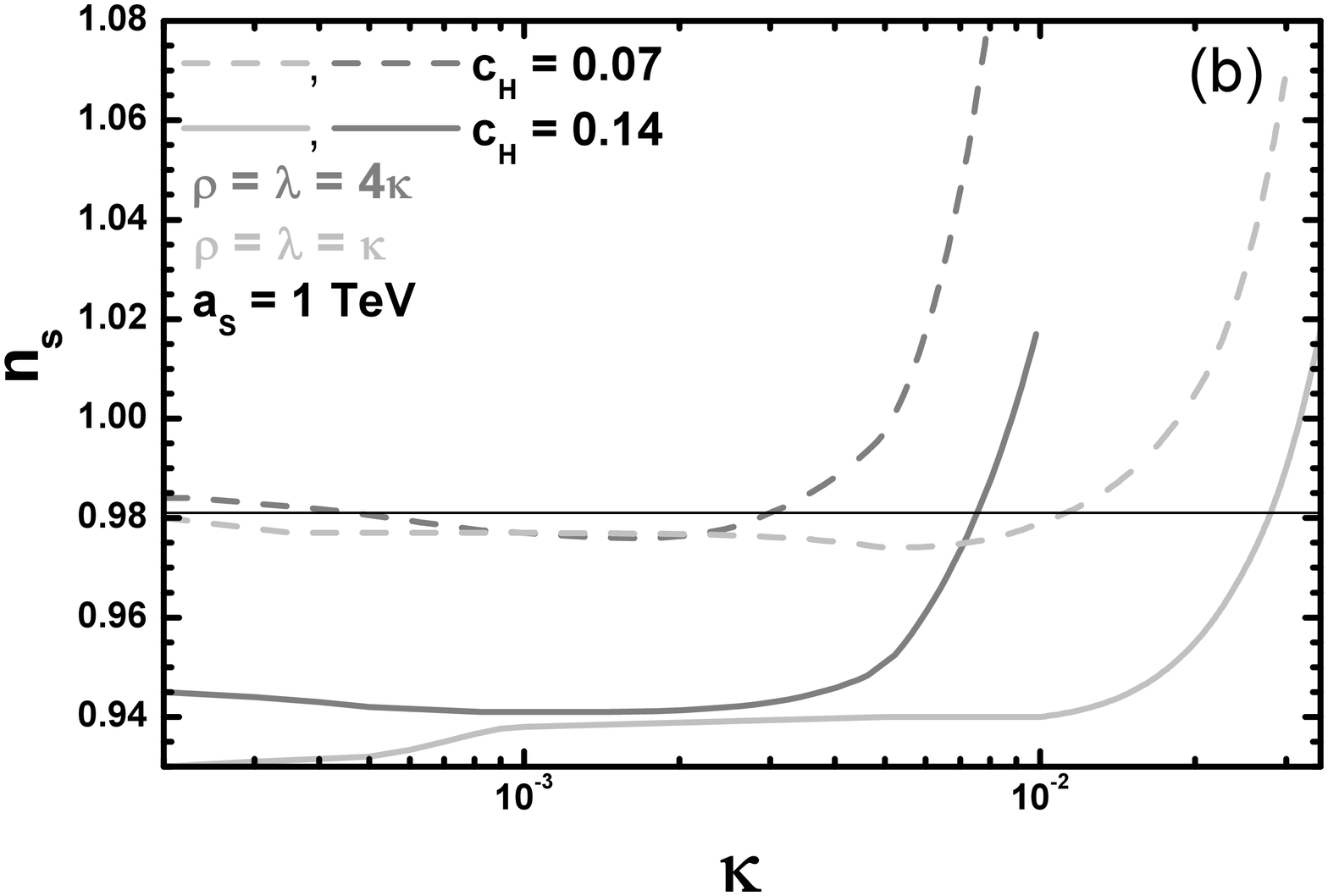,angle=-90,width=13.cm}}\hfill
\caption{\sl\small The values of the inflationary scale $M$ allowed
by~(\ref{Nefold}) and (\ref{Pr}) {\sf (a)} and the predicted values of
the spectral index $n_{\rm s}$ {\sf (b)} as a function of $\kappa$ for
${\cal N}=1$ and $\rho=\lambda=\kappa$ (light grey lines) or
$\rho=\lambda=4\kappa$ (grey lines), for the nmSUGRA scenario with
$c_H=0.07$ (dashed lines) or $c_H=0.14$ (solid lines). In both cases
we take a$_S=1~{\rm TeV}$.  The upper bound given in~(\ref{Prcs}) (for
$y_{\rm cs}=6.7,~8.9,~11.6$ from top to bottom) [cf.~(\ref{nswmap})]
is also depicted by thin lines {\sf (a)} [{\sf (b)}].}
\label{fig:nmSUGRAPR}
\end{figure}

In  our  numerical estimates,  we  use  the  full expression  for  the
inflationary  potential  $V_{\rm  inf}$ given  in~(\ref{Vinf}),  which
consists  of the  tree-level,  1-loop and  SUGRA contributions,  given
in~(\ref{V0inf}), (\ref{V1loop}) and (\ref{VSUGRA}), respectively.  We
will ignore all soft SUSY-breaking  terms, but the tadpole term $a_S$.
To facilitate  our numerical analysis,  we introduce the  real tadpole
parameter $\mbox{a}_S$,  which is defined, in terms  of the Lagrangian
parameter $a_S$, by the relation:
\begin{equation}
  \label{aS}
\mbox{a}_S\  =\ -\, 2  |a_S|\, \cos{(\arg{a_S}  + \arg{S})}\; .
\end{equation}
For any  given value of  $\kappa,~\lambda,~\rho$, a$_S$ and  $c_H$, we
determine    $\phi_{\rm   exit}$    and   $M$,    by    imposing   the
conditions~(\ref{Nefold}) and  (\ref{Pr}) for the  number ${\cal N}_e$
of $e$-folds and the  power spectrum $P^{1/2}_{\cal R}$, respectively.
In  addition, we  compute  $n_{\rm s}$  by  means of~(\ref{nS}).   Our
results  are  presented  in  Fig.~\ref{fig:mSUGRAPR}  for  the  mSUGRA
scenario  and in  Fig.~\ref{fig:nmSUGRAPR} for  the  nmSUGRA scenario.
They will be analyzed in more detail in the following two subsections.

\begin{figure}[!th]
\centerline{\epsfig{file=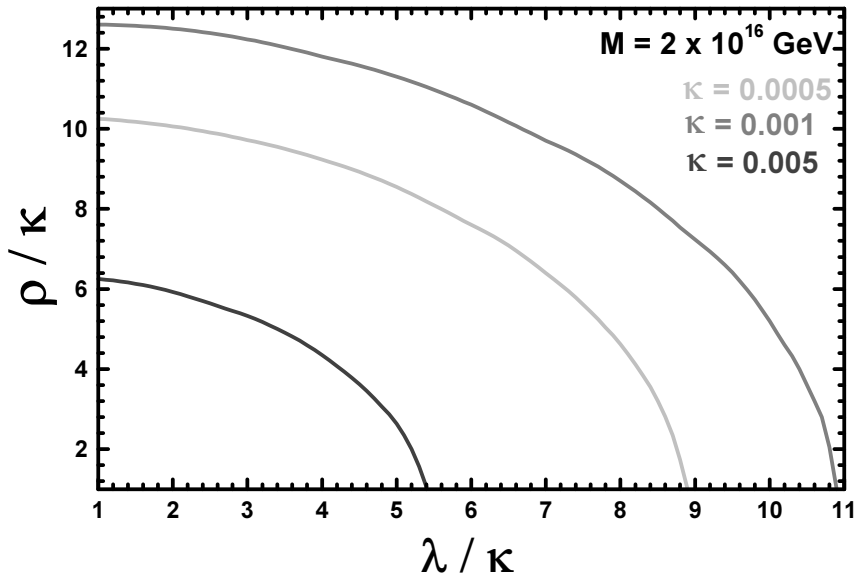,angle=-90,width=13.cm}} \hfill
\caption{\sl\small The allowed values of $\lambda/\kappa$ versus
$\rho/\kappa$ for the mSUGRA scenario with $M=2\times10^{16}~{\rm
GeV}$ and $\kappa=0.05$ (dark grey line), $\kappa=0.01$ (grey line) or
$\kappa=0.005$ (light grey line).}\label{fig:mSUGRAlr}
\end{figure}

\subsubsection{The minimal SUGRA scenario}\label{msugra}

Here,  we present  numerical  results for  the  mSUGRA scenario.   The
values  of the  inflationary scale  $M$ allowed  by~(\ref{Nefold}) and
(\ref{Pr}) and  the predicted values  of $n_{\rm s}$, as  functions of
$\kappa$,   for   $\rho=\lambda=\kappa$   (light   grey   lines)   and
$\rho=\lambda=4\kappa$     (grey    lines),    are     displayed    in
Fig.~\ref{fig:mSUGRAPR}(a)   and~\ref{fig:mSUGRAPR}(b),  respectively.
Dashed  lines   indicate  results  obtained,  when   only  the  1-loop
contribution to $V_{\rm inf}$ is considered and $\mbox{a}_S$ is set to
zero, whilst  solid lines represent numerical values  obtained, if the
remaining   contributions   are    included,   namely   those   coming
from~(\ref{V0inf})  with  a$_S=1~{\rm  TeV}$  and~(\ref{VSUGRA})  with
$c_H=0$.  In  Fig.~\ref{fig:mSUGRAPR}, we  observe that as  the common
value for  $\rho$, $\lambda$ and  $\kappa$ increases, $M$  and $n_{\rm
s}$ increase as well. In  particular, $M$ gets closer to the GUT-scale
value $2  \times 10^{16}~{\rm GeV}$ for $\kappa  \sim 10^{-3}$, unlike
the case $\lambda=\rho=0$,  where $M$ takes on much  smaller values at
this  point~\cite{DSS,SS,JP}.

It  is now  not difficult  to identify  in~Fig.~\ref{fig:mSUGRAPR} the
regimes,  in  which  the  different  contributions  to  $V_{\rm  inf}$
dominate.   More explicitly, for  $\kappa \gtrsim  4\times10^{-3}$ and
$\rho=\lambda=\kappa$     or    $\kappa    \gtrsim     10^{-3}$    and
$\rho=\lambda=4\kappa$,    the     non-renormalizable    SUGRA    term
of~(\ref{VSUGRA}) dominates and drives  $n_{\rm s}$ to values close to
or larger than~1 [cf.~Fig.~\ref{fig:mSUGRAPR}(b)].  On the other hand,
for  $4\times10^{-4}  \lesssim \kappa  \lesssim  4\times 10^{-3}$  and
$\rho = \lambda =  \kappa$ or $4\times10^{-4} \lesssim \kappa \lesssim
10^{-3}$   and    $\rho   =    \lambda   =   4\kappa$,    the   1-loop
corrections~(\ref{V1loop}) dominate, in  which case the spectral index
$n_{\rm  s}$   takes  on  the   predicted  value  $\sim   0.98$  given
in~(\ref{nSrc}).  Finally,  for $\kappa \lesssim  4\times 10^{-3}$ and
$\rho=\lambda=\kappa$  or  $\rho=\lambda=4\kappa$,  the  tadpole  term
in~(\ref{V0inf}) starts playing an  important role.  As $M$ increases,
the  non-renormalizable  SUGRA  term of~(\ref{VSUGRA})  becomes  again
important~\cite{SS,JP}.    In   this    case,   the   prediction   for
$P^{1/2}_{\cal R}$ and $n_{\rm s}$ is almost independent of $\rho$ and
$\lambda$, as expected.  For lower values of a$_S$, the solid lines in
the   latter    regime   would   eventually    approach   the   dashed
lines~\cite{JP}.  In Fig.~\ref{fig:mSUGRAPR}(b), we also indicate with
a  thin  line  the  $95\%$  CL  upper  limit  on  $n_{\rm  s}$  stated
in~(\ref{ns95}). Clearly,  a mSUGRA  version of the  $F_D$-term hybrid
model appears to be disfavoured by the most recent WMAP results.

In addition, we show in Fig.~\ref{fig:mSUGRAPR}(a) upper limits due to
cosmic   string  effects  based   on~(\ref{Prcs}),  for   $y_{\rm  cs}
=6.7,~8.9,~11.6$  (from top  to  bottom).  Such  constraints are  only
relevant for an Abelian realization of the waterfall-gauge sector.  We
observe  that the  presence of  cosmic strings  severely  restrict the
available parameter space of the U(1)$_X$ $F_D$-term hybrid model.  As
we discussed  in Section~\ref{TDGUT}, however, these  constraints do no
longer  apply, if  the  waterfall-gauge sector  realizes an  SU(2)$_X$
gauge  symmetry.  Since  the dimensionality  of the  representation is
${\cal N}=2$ in  this case, the allowed range of $M$  as a function of
$\kappa$ slightly changes.  In fact,  the allowed values of $M$ become
marginally     larger     than      the     ones     already     shown
in~Fig.~\ref{fig:mSUGRAPR}(a)      by     up     to      12\%,     for
$\rho=\lambda=\kappa$,  while   they  stay  at  the   1\%  level,  for
$\rho=\lambda=4\kappa$. Likewise, the  predicted values of $n_{\rm s}$
remain  almost  unaffected at  the  2\%  level,  from those  presented
in~Fig.~\ref{fig:mSUGRAPR}(b). Obviously, as $\rho$ and $\lambda$ gets
larger  than $\kappa$,  the difference  between the  ${\cal  N}=1$ and
${\cal N}=2$ cases becomes practically unobservable.

Finally,  in Fig.~\ref{fig:mSUGRAlr}  we  plot the  allowed values  of
$\lambda/\kappa$     versus    $\rho/\kappa$,    subject     to    the
constraints~(\ref{Nefold})   and  (\ref{Pr}),  for   $M  =   2  \times
10^{16}$~GeV  (close to  the GUT  scale) and  for different  values of
$\kappa$: $\kappa=0.005$ (dark  grey line), $\kappa=0.001$ (grey line)
or  $\kappa=0.0005$ (light  grey  line).  Along  these contour  lines,
$\phi_{\rm exit}$  and $n_{\rm  s}$ remain constant  and equal  to the
values  presented in  Table~\ref{tab1}.  We  observe that  as $\kappa$
increases, $\phi_{\rm exit}$ and $n_{\rm s}$ increase as well.

\begin{table}[!t]
\begin{center}
\begin{tabular}{|l|c|c||c|c|c|c|c|}\hline
\multicolumn{3}{|c||}{Fig.~\ref{fig:mSUGRAlr}}&\multicolumn{5}{|c|}{Fig.~\ref{fig:nmSUGRAlr}}\\\hline
\multicolumn{3}{|c||}{$M=2\times10^{16}~{\rm GeV}$}&\multicolumn{5}{|c|}{
$\kappa=0.005,~M=10^{16}~{\rm GeV}$}\\\hline
{$~~~\kappa$}&{$\phi_{\rm
exit}$}&{$n_{\rm s}$}&{$c_H$}
&{$\phi_{\rm min}$}&{$\phi_{\rm max}$}&{$\phi_{\rm exit}$}&{$n_{\rm s}$}\\
\hline\hline
$0.005$&{$6.28$}&{$1.017$}&{$0.07$}&{$8.4$}&{$-$}&{$5.1$}&{$0.978$}\\
$0.001$&{$2.14$}&{$0.99$}&{$0.14$}&{$16.5$}&{$10.5$}&{$8.1$}&{$0.955$}\\
$0.0005$&{$1.51$}&{$0.99$}&{$0.18$}&{$21.7$}&{$12.8$}&{$11.2$}&{$0.941$}\\ \hline
\end{tabular}
\end{center}
\caption{\sl\small The values of $\phi_{\rm exit}$ (in units
$\sqrt{2}M$) and $n_{\rm s}$ for several $\kappa$'s along the curves
in Fig.~\ref{fig:mSUGRAlr} and the values of $\phi_{\rm min}$, $\phi_{\rm
max}$, $\phi_{\rm exit}$ (in units $\sqrt{2}M$) and $n_{\rm s}$ for
several $c_H$'s along the curves in Fig.~\ref{fig:nmSUGRAlr}.} \label{tab1}
\end{table}

\subsubsection{The next-to-minimal SUGRA scenario}\label{qsugra}

We now turn  our attention to the nmSUGRA  scenario.  Although we
take the tadpole  term to be a$_S=1~{\rm  TeV}$, its impact  on
our results turns out  to be  insignificant for the  whole range
of  parameters we have  scanned.   The values  of  the
inflationary  scale $M$  allowed by~(\ref{Nefold}) and  (\ref{Pr})
and the  predicted $n_{\rm s}$  as a function  of  $\kappa$  are
presented  in  Fig.~\ref{fig:nmSUGRAPR}(a)
and~\ref{fig:nmSUGRAPR}(b),  respectively,  for
$\rho=\lambda=\kappa$ (light  grey  lines)  and
$\rho=\lambda=4\kappa$  (grey  lines).   We consider  the  two
cases:  $c_H=0.07$  (dashed  lines) and  $c_H=0.14$ (solid lines).
As in the case of mSUGRA, $M$ and $n_{\rm s}$ increase, with
increasing  $\rho, \lambda$  and $\kappa$. Moreover,  as $\kappa$
decreases, the non-renormalizable SUGRA contribution in
(\ref{VSUGRA}) becomes  subdominant  and $n_{\rm  s}$  decreases.
Such a  reduction becomes  even more  drastic with  increasing
$c_H$,  as can  be easily inferred  from
Fig.~\ref{fig:nmSUGRAPR}(a),  where  the 95\%~CL  upper bound  on
$n_{\rm  s}$  [cf.~(\ref{ns95})] is  indicated  with a  thin
horizontal line  on the  same plot.  In  stark contrast to  the
mSUGRA scenario, we  observe that our  model can become  perfectly
consistent with  the recent  WMAP result  for $0.04  \lesssim c_H
\lesssim 0.22$. Note that the various lines terminate at large
values of $\kappa$, for which  the two restrictions~(\ref{Nefold})
and~(\ref{Pr}) cannot  be simultaneously met.

\begin{sidewaystable}
\begin{center}
\begin{tabular}{|l||c|c|c|c|c|c||c|c|c|c|c|c||c|c|c|c|c|c|}\hline
{$~~\kappa$}&{$c_H$}&{$M$}&{$\phi_{\rm min}$}&{$\phi_{\rm max}$}&{$\phi_{\rm
exit}$}&{$\Delta_{\rm exit}$}
&{$c_H$}&{$M$}&{$\phi_{\rm min}$}&{$\phi_{\rm max}$}&{$\phi_{\rm
exit}$}&{$\Delta_{\rm exit}$}&{$c_H$}&
{$M$}&{$\phi_{\rm min}$}&{$\phi_{\rm max}$}&{$\phi_{\rm exit}$}&{$\Delta_{\rm
exit}$}\\
\cline{2-19}
&\multicolumn{6}{|c||}{$n_{\rm s}=0.913$}&\multicolumn{6}{|c||}{$n_{\rm
s}=0.951$}
&\multicolumn{6}{|c|}{$n_{\rm s}=0.981$}\\ \hline\hline
\multicolumn{19}{|c|}{$\lambda=\rho=\kappa$}\\ \hline
$0.01$&{$0.179$}&{$0.34$}&{$73.6$}&{$11.9$}&{$11.3$}&{$0.050$}
      &{$0.130$}&{$0.53$}&{$32.0$}&{$10.8$}&{$8.75$}&{$0.19$}&
      {$0.065$}&{$0.78$}&{$16.7$}&{$-$}&{$7.50$}&{$-$}\\
$0.005$&{$0.176$}&{$0.34$}&{$73.1$}&{$6.0$}&{$5.7$}&{$0.053$}
       &{$0.120$}&{$0.53$}&{$32.2$}&{$6.2$}&{$4.48$}&{$0.18$}
       &{$0.040$}&{$0.78$}&{$8.20$}&{$-$}&{$3.90$}&{$-$}\\
$0.001$&{$0.173$}&{$0.25$}&{$95.6$}&{$1.64$}&{$1.6$}&{$0.028$}
        &{$0.120$}&{$0.38$}&{$45.0$}&{$1.55$}&{$1.42$}&{$0.09$}
        &{$0.060$}&{$0.58$}&{$19.0$}&{$2.10$}&{$1.36$}&{$0.34$}\\
$0.0005$&{$0.165$}&{$0.19$}&{$121$}&{$1.23$}&{$1.21$}&{$0.014$}
        &{$0.116$}&{$0.28$}&{$58.8$}&{$1.19$}&{$1.15$}&{$0.04$}
        &{$0.060$}&{$0.43$}&{$20.0$}&{$1.37$}&{$1.13$}&{$0.17$}\\\hline
\multicolumn{19}{|c|}{$\lambda=\rho=4\kappa$}\\ \hline
$0.01$&{$0.216$}&{$0.56$}&{$49$}&{$23.0$}&{$22.0$}&{$0.046$}
      &{$0.190$}&{$0.83$}&{$23.0$}&{$21.9$}&{$17.0$}&{$0.22$}
      &{$0.169$}&{$1.12$}&{$26.0$}&{$-$}&{$14.3$}&{$-$}\\
$0.005$&{$0.188$}&{$0.61$}&{$41$}&{$11.4$}&{$10.8$}&{$0.050$}
       &{$0.146$}&{$0.96$}&{$26.0$}&{$9.1$}&{$8.30$}&{$0.19$}
       &{$0.103$}&{$1.36$}&{$8.6$}&{$-$}&{$7.05$}&{$-$}\\
$0.001$&{$0.177$}&{$0.57$}&{$43$}&{$2.48$}&{$2.38$}&{$0.043$}
       &{$0.125$}&{$0.89$}&{$24.6$}&{$2.28$}&{$1.96$}&{$0.14$}
        &{$0.058$}&{$1.30$}&{$4.7$}&{$-$}&{$1.82$}&{$-$}\\
$0.0005$&{$0.178$}&{$0.46$}&{$54$}&{$1.53$}&{$1.49$}&{$0.028$}
        &{$0.129$}&{$0.68$}&{$26$}&{$1.45$}&{$1.33$}&{$0.08$}
        &{$0.070$}&{$1.00$}&{$9.6$}&{$1.83$}&{$1.30$}&{$0.29$}\\ \hline
\end{tabular}
\end{center}
\caption{\sl\small The  values of  $c_H$, $M$ (in  units $10^{16}~{\rm
GeV}$) $\phi_{\rm min}$, $\phi_{\rm max}$, $\phi_{\rm exit}$ (in units
$\sqrt{2}M$)  and $\Delta_{\rm  exit}  = (\phi_{\rm  max} -  \phi_{\rm
exit})/\phi_{\rm max}$, for selected values of $\kappa$, $\lambda$ and
$\rho$, and for fixed values of the spectral index $n_{\rm s}$.}\label{tab2}
\end{sidewaystable}

It  is interesting  to further  investigate the  inflationary dynamics
described   by  $V_{\rm   inf}$  in   the  presence   of   a  negative
Hubble-induced mass term.  To this end, we exhibit in Table~\ref{tab2}
the values of $c_{H}$,  $\phi_{\rm min}$, $\phi_{\rm max}$, $\phi_{\rm
exit}$ (in units $\sqrt{2}M$) and the inflationary scale $M$ (in units
of $10^{16}$~GeV) which are obtained for different values of $\kappa$,
assuming that $\lambda = \rho = \kappa$ or $\lambda = \rho = 4\kappa$,
and for fixed values of $n_{\rm s}$, i.e.~$n_{\rm s} = 0.913,\ 0.951,\
0.981$, compatible  with the 95\% CL limits  given in~(\ref{ns95}). In
addition,  we present values  for the  parameter $\Delta_{\rm  exit} =
(\phi_{\max}  -  \phi_{\rm   exit})/\phi_{\rm  exit}$,  which  somehow
quantifies the degree of tuning  required in the initial conditions of
inflation.  The entries without a value assigned (in Tables~\ref{tab1}
and~\ref{tab2})  mean  that   the  respective  inflationary  potential
$V_{\rm inf}$  has no distinguishable nearby  local maximum $\phi_{\rm
max}$. We notice from  Table~\ref{tab2}, that as $n_{\rm s}$ decreases
with  fixed  values  of  $\kappa$,  $c_{H}$ increases  while  $M$  and
$\Delta_{\rm exit}$  decrease. Moreover,  for fixed values  of $n_{\rm
s}$  and decreasing $\kappa$,  $c_H$ and  $M$ decrease  and $\phi_{\rm
exit}$  approaches $\phi_{\max}$.   On the  contrary,  with increasing
$\kappa$, $\lambda$  and $\rho$, the inflationary  scale $M$ increases
and the parameter $\Delta_{\rm  exit}$ becomes larger. We have checked
that the  inequality $\phi_{\max}>\phi_{\rm exit}$  is fulfilled along
the lines presented in  Fig.~\ref{fig:nmSUGRAPR}.  In this respect, we
also  note  that $\phi_{\rm  min}$  is  in  general much  larger  than
$\phi_{\max}$ especially for low values of $n_{\rm s}$.

It  is important  to observe  from  Table~\ref{tab2} that  there is  a
degree  of  tuning required  for  the  values  $\phi_{\rm exit}$  with
respect to $\phi_{\max}$.  For values of $\kappa \stackrel{>}{{}_\sim}
10^{-3}$,  we find  that the  degree of  tuning required  is  not very
serious,   i.e.~$\Delta_{\rm    exit}   \stackrel{>}{{}_\sim}   10\%$.
However,  the  situation  becomes  rather delicate  as  $\kappa$  gets
smaller  than $10^{-3}$, for  $n_{\rm s}  \stackrel{<}{{}_\sim} 0.97$.
In  this case,  we find  that $\phi_{\max}  \approx  \phi_{\rm exit}$,
leading  to a  substantial tuning  at the  few per  cent level  in the
initial conditions of inflation.

As in the mSUGRA case, we also show in Fig.~\ref{fig:nmSUGRAPR}(a) the
upper bounds resulting  from cosmic-string effects [cf.~(\ref{Prcs})],
for $y_{\rm cs}  = 6.7,~8.9,~11.6$ (from top to  bottom). As mentioned
above,   these  constraints   are   only  relevant   for  an   Abelian
waterfall-gauge sector  with dimensionality  ${\cal N} =  1$. However,
unlike  in the mSUGRA  case, these  restrictions appear  less harmful,
since  the  inflationary scale  $M$  assumes  smaller  values
(see also Table~\ref{tab2}) and  the tadpole term becomes
unimportant. Thus, larger values of $\kappa$ up to order $10^{-2}$ can be
tolerated in this case.  For the non-Abelian SU(2)$_X$   $F_D$-term    hybrid
model,   the    restrictions   from considerations  of cosmic-string
effects are  totally lifted  and the
lines depicted  in Fig.~\ref{fig:nmSUGRAPR}  only vary within  the few
per  cent level.  Such  a variation  becomes even  smaller if  $\rho >
\kappa$ and/or $\lambda > \kappa$.

\begin{figure}[!t]
\centerline{\epsfig{file=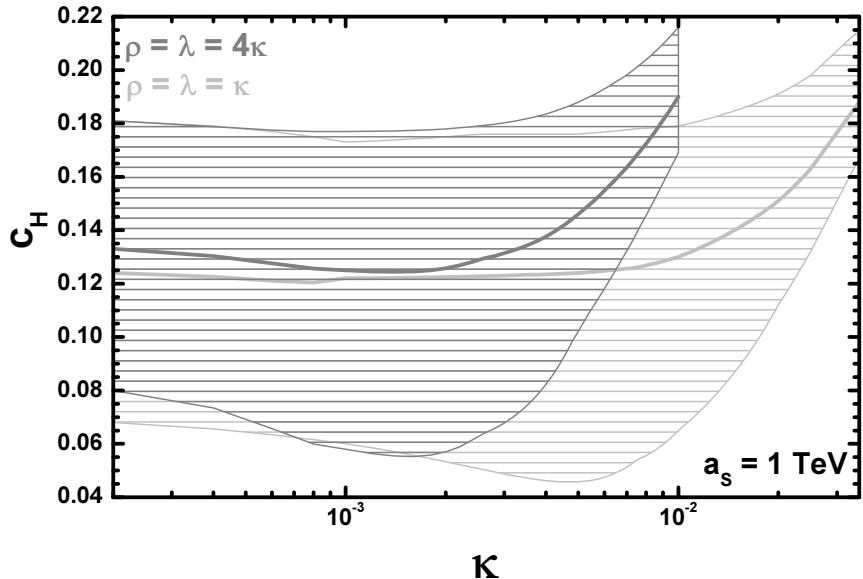,angle=-90,width=13.cm}} \hfill
\caption{\sl\small The parameter values $(\kappa,c_H)$ allowed by
(\ref{Nefold}), (\ref{Pr}) and (\ref{nswmap}) in the nmSUGRA scenario,
for $\rho=\lambda=\kappa$ (light grey hatched area) and
$\rho=\lambda=4\kappa$ (grey hatched area). The grey (light grey) line
has been obtained by fixing $n_{\rm s}$ to its central value given in
(\ref{nswmap}), for $\rho=\lambda=4\kappa$
($\rho=\lambda=\kappa$).}\label{fig:cHk}
\end{figure}

In Fig.~\ref{fig:cHk},  we present the  parameter space
$(\kappa,c_H)$ which  is  allowed by  the
conditions~(\ref{Nefold}), (\ref{Pr})  and (\ref{nswmap}) in the
nmSUGRA  scenario. The light grey (grey) hatched area   indicates
the   allowed   region  for   $\rho=\lambda=4\kappa$
($\rho=\lambda=\kappa$). The lower (upper) boundaries of the
allowed regions correspond to the upper (lower) bound on $n_{\rm
s}$, cf.~(\ref{ns95}), while the solid lines correspond to the
central value of $n_{\rm s}$, cf.~(\ref{nswmap}). We find that
values of $c_H \sim  0.2$ and $\kappa \sim  0.05$ are still
possible  in a nmSUGRA  extension of the $F_D$-term hybrid model.

\begin{figure}[!t]
\centerline{\epsfig{file=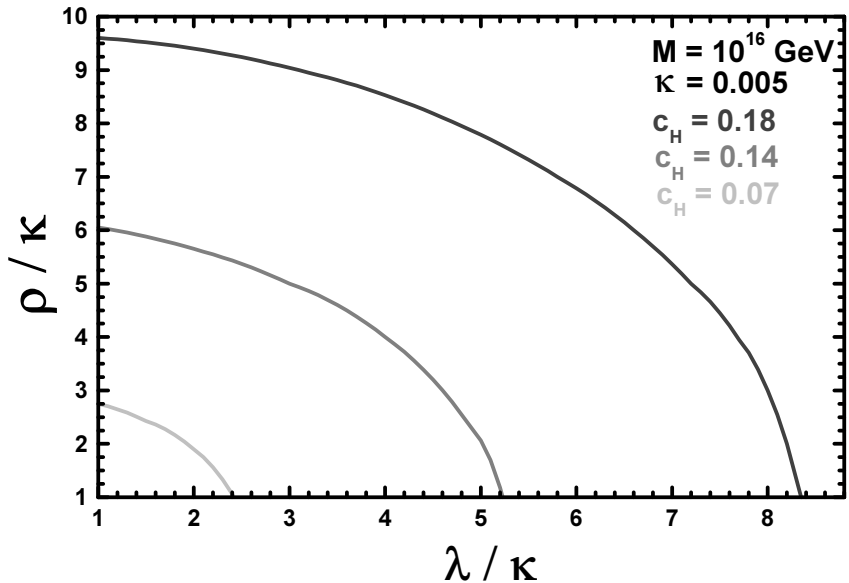,angle=-90,width=13.cm}} \hfill
\caption{\sl\small The allowed values of $\rho/\kappa$ versus $\lambda/\kappa$
for the nmSUGRA scenario with $\kappa=0.005,~M=10^{16}~{\rm GeV}$ and
$c_H=0.18$ (dark grey line), $c_H=0.14$ (grey line) or
$c_H=0.07$ (light grey line).}\label{fig:nmSUGRAlr}
\end{figure}

Finally,  we plot  in Fig.~\ref{fig:nmSUGRAlr}  the allowed  values of
$\lambda/\kappa$ versus $\rho/\kappa$,  on account of the inflationary
constraints~(\ref{Nefold})  and (\ref{Pr}),  for $\kappa=0.005$,  $M =
10^{16}$~GeV, and  for $c_H=0.18$  (dark grey line),  $c_H=0.14$ (grey
line) and  $c_H=0.07$ (light grey  line). We have selected  a slightly
lower value for $M$, because no viable nmSUGRA scenarios seem to exist
with acceptable values  for $n_{\rm s}$, if $M  = 2\times 10^{16}$~GeV
and    $c_H    \ge    0.07$.     Along   the    contour    lines    in
Fig.~\ref{fig:nmSUGRAlr},   $\phi_{\rm    min}$,   $\phi_{\rm   max}$,
$\phi_{\rm exit}$  and $n_{\rm  s}$ remain constant  and equal  to the
values  presented  in  Table~\ref{tab1}.    We  observe  that  as  $c_H$
increases, $\phi_{\rm  exit}$ approaches $\phi_{\rm  max}$, $\phi_{\rm
min}$  increases,   while  $n_{\rm  s}$   decreases.   This  kinematic
behaviour    is   in   agreement    with   our    discussion   related
to~(\ref{phimax}) and~(\ref{nShilltop}).

\setcounter{equation}{0}
\section{Preheating}\label{Preheat}

As  stated  in the  Introduction,  gravitinos,  if thermally  produced
during the early  stages of the evolution of  the Universe, will spoil
the  successful predictions  of  BBN~\cite{Sarkar}.  Their  disastrous
consequences may  be avoided, if the reheat  temperature $T_{\rm reh}$
of the  Universe is  not very  high. In fact,  depending on  the decay
properties   of   the   gravitino,   it   should   be   $T_{\rm   reh}
\stackrel{<}{{}_\sim}  10^7$--$10^{10}$~GeV~\cite{kohri,oliveg}.  This
fact leads to a tension between the allowed range of $T_{\rm reh}$ and
the natural  scale of  hybrid inflation $M$,  which is of  order $\sim
10^{16}$~GeV. The traditional way taken  to get around this problem is
to  consider scenarios  where the  decay rate  of the  inflaton  to SM
particles  is extremely suppressed,  e.g.~by suppressing  all possible
couplings of the inflaton to the SM fields.

\begin{figure}[t]
\begin{center}
\epsfig{file=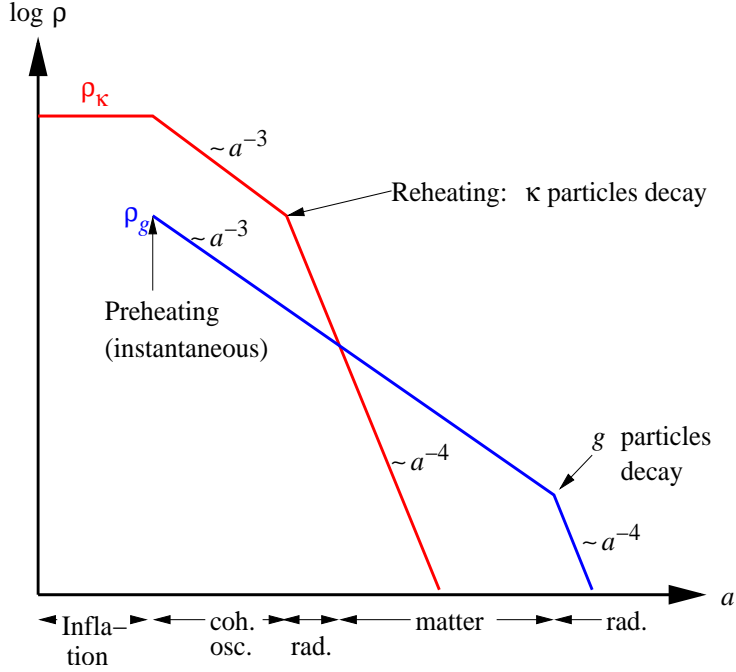, height=3.5in,width=4in}
\end{center}
\caption{\sl\small Schematic representation of the thermal history of
the Universe in the $F_D$-term hybrid model.}\label{figure:cartoon}
\end{figure}

In  this  and next  sections,  we  present  in detail  an  alternative
solution to the  above gravitino overabundance problem~\cite{GP}.  Our
solution  relies on  the huge  entropy  release caused  from the  late
out-of-equilibrium decays of the supermassive waterfall particles. The
entropy produced  through this mechanism  is sufficient to  reduce the
gravitino abundance $Y_{\widetilde{G}}$  to levels compatible with BBN
limits     discussed     in     detail    in     Section~\ref{reheat}.
Figure~\ref{figure:cartoon}  gives a  schematic representation  of the
post-inflationary dynamics  of the early Universe, as  is predicted by
the $F_D$-term hybrid model.  Shortly after inflation ends, the energy
density  $\rho_\kappa$  of the  Universe  is  predominantly stored  to
coherently oscillating  inflaton condensates which  scale as $a^{-3}$,
where  $a$  is the  usual  cosmological  scale  factor describing  the
expansion  of   the  Universe.   The  coherent   oscillations  of  the
inflaton-related    condensates   also    give    rise   to    another
non-perturbative  mechanism  called  preheating.   During  preheating,
waterfall  gauge particles  of  energy density  $\rho_g$ are  produced
almost instantaneously,  which are absolutely stable  if a $D$-parity,
an analogue of the usual  $R$-parity in the MSSM, is conserved.  Then,
the   following  scenario   visualized   in  Fig.~\ref{figure:cartoon}
emerges.   First,  $\rho_g/\rho_\kappa$  remains constant  during  the
epoch of coherent oscillations,  since both $\rho_g$ and $\rho_\kappa$
behave as  matter energy densities  and scale as $a^{-3}$  during this
period. The constancy of $\rho_g/\rho_\kappa$ ceases to hold, when the
coherently  oscillating inflaton  condensates decay  and  their energy
density  $\rho_\kappa$  gets   distributed  among  light  relativistic
degrees  of freedom.  As  a consequence  of the  latter, $\rho_\kappa$
will  be $\propto  a^{-4}$,  whilst $\rho_g$  will  still be  $\propto
a^{-3}$.   If the initial  value of  $\rho_g/\rho_\kappa$ is  not very
suppressed,  e.g.~it is of  order $10^{-4}$--$10^{-5}$,  the waterfall
gauge  particles will eventually  dominate the  energy density  of the
Universe, leading to  a second matter dominated epoch  which will last
until these particles decay  via $D$-parity violating couplings.  This
is expected to  produce an enormous entropy release  and so reduce the
gravitino-to-entropy  ratio $Y_{\widetilde{G}}$  to  values compatible
with BBN constraints.

The  discussion   in  this  section   is  organized  as   follows:  in
Section~\ref{postinfl},  we pay  special attention  to  $D$-parity and
derive the particle spectrum of the combined inflaton-waterfall sector
in the  supersymmetric limit of  the theory.  In addition,  we compute
the decay rates of all inflaton-related and waterfall gauge particles.
Finally, in Section~\ref{preheat}, we  discuss how the waterfall gauge
particles   are  instantaneously   produced  through   preheating  and
calculate  the  resulting energy  density  $\rho_g$  carried by  these
particles.

\subsection{{\boldmath $D$}-Parities and the
Inflaton-Waterfall Sector}\label{postinfl}

Let  us  first  consider  a  model  with  a  U(1)$_X$  gauge-symmetric
waterfall  sector.   The  case  of  a  waterfall  sector  realizing  a
non-Abelian  SU(2)$_X$  gauge  symmetry   is  analogous  and  will  be
discussed later.  In terms  of superfields, the  minimal gauge-kinetic
Lagrangian of the U(1)$_X$ model reads:
\begin{equation}
  \label{U1kin}
{\cal L}_{\rm kin}\ =\
\int d^4 \theta\, \left(
\frac{1}{2}\, W^\alpha W_\alpha\,\delta^{(2)}(\bar{\theta} )\ +\
\frac{1}{2}\, \overline{W}_{\dot{\alpha}}
\overline{W}^{\dot{\alpha}}\, \delta^{(2)}(\theta )\  +\
\widehat X_1^\dagger {\rm e}^{2 g \widehat V _X} \widehat X_1\
+\ \widehat X_2^\dagger
{\rm e}^{-2 g \widehat{V}_X} \widehat X_2 \right)\; ,
\end{equation}
where $\widehat{V}_X$ is the U(1)$_X$ vector superfield and $W_\alpha$
($\overline{W}_{\dot{\alpha}}$)    are    their   respective    chiral
(anti-chiral) field strengths. The latter are given by
\begin{equation}
W_\alpha\ =\ -\, \frac{1}{8g}\, \bar{D}^2\, ( e^{-2g\widehat{V}_X} D_\alpha\,
e^{2g\widehat{V}_X} )\,,\qquad
\overline{W}_{\dot{\alpha}}\ =\
\frac{1}{8g}\, D^2\, ( e^{2g\widehat{V}_X}
\bar{D}_{\dot{\alpha}}\, e^{-2g\widehat{V}_X} )\,,
\end{equation}
where   $D_\alpha$   and   $\bar{D}_{\dot{\alpha}}$  are   the   usual
SUSY-covariant  derivatives which  are irrelevant  for  our discussion
here.   The minimal  gauge-kinetic  Lagrangian~(\ref{U1kin}) possesses
the discrete symmetry
\begin{equation}
  \label{DsuperU1}
D:\qquad \widehat{X}_1\ \leftrightarrow\ \widehat{X}_2\,,\qquad
\widehat{V}_X\ \to\ -\,\widehat{V}_X\; ,
\end{equation}
whereas all other superfields do not transform. It is not difficult to
verify  that  the  complete  $F_D$-term hybrid  model,  including  the
superpotential~(\ref{Wmodel})  and its  associated  soft SUSY-breaking
sector, is  invariant under the  discrete symmetry~(\ref{DsuperU1}) in
the  unbroken phase  of the  theory. After  the SSB  of  U(1)$_X$, the
waterfall fields receive the VEVs:  $\langle X_1 \rangle = \langle X_2
\rangle = M$.  Thus, the  above discrete symmetry survives even in the
spontaneously broken phase of the theory.  Since the discrete symmetry
acts on  a gauged waterfall sector,  it manifests itself as  a kind of
parity, which we call $D$-parity.

It therefore proves convenient to choose a weak basis where the fields
are  eigenstates of  $D$-parity.  To  this end,  we define  the linear
combinations in terms of the waterfall superfields
\begin{equation}
  \label{Xparity}
\widehat{X}_\pm \ =\ \frac{1}{\sqrt{2}}\, \bigg(\, \widehat{X}_1\
\pm\ \widehat{X}_2\,\bigg)\; .
\end{equation}
Evidently, the  superfield $\widehat{X}_+$ ($\widehat{X}_-$)  has even
(odd) $D$-parity;  its $D$-parity quantum number is  $+1$ ($-1$).  The
vector  superfield  $\widehat{V}_X$,  which  is already  a  $D$-parity
eigenstate, has  odd $D$-parity.  All remaining  fields, including the
inflaton superfield $\widehat{S}$ and the other MSSM superfields, have
positive $D$-parity.

As  a consequence  of $D$-parity  conservation, all  $D$-odd particles
will  be stable,  in as  much  the same  way as  the usual  $R$-parity
guarantees  that the  LSP of  the MSSM  is stable.   As  we explicitly
mentioned in Section~\ref{FD}, the simplest way to break $D$-parity is
to  add  a FI  $D$-term  to  the  model, e.g.
\begin{equation}
  \label{LFI}
{\cal L}_{\rm FI}\ =\ -\, \frac{g}{2}\, m^2_{\rm FI}\,
                       \int d^4 \theta\; \widehat{V}_X\ =\ -\,
                       \frac{g}{2}\, m^2_{\rm FI}\, D\; ,
\end{equation}
where  $D$  is  the   auxiliary  component  of  the  vector  superfield
$\widehat{V}_X$.  It  is obvious that  ${\cal L}_{\rm FI}$  flips sign
under  the  discrete  symmetry~(\ref{DsuperU1}). Other  mechanisms  of
explicitly      breaking     $D$-parity      are      discussed     in
Appendix~\ref{Dappendix}.

We now calculate the particle spectrum of the inflaton-waterfall sector
in the presence  of a subdominant FI $D$-term $m_{\rm  FI}$ and in the
supersymmetric  limit of  the theory.   With this  aim, we  expand the
scalar $D$-parity eigenstates $X_\pm$ about their VEVs:
\begin{equation}
  \label{Xpm}
X_\pm \ =\ \langle X_\pm \rangle\: +\:  \frac{1}{\sqrt{2}}\,
\Big(\, R_\pm\: +\: {\rm i}I_\pm\,\Big)\; .
\end{equation}
The VEVs $\langle X_\pm \rangle$ are determined from the minimization
conditions of the combined $F$- and $D$-term scalar potential
\begin{eqnarray}
  \label{VFD}
V_{FD}\ =\  F_S^* F_S\: +\: \frac 12\, D^2\,,
\end{eqnarray}
where
\begin{equation}
F_S \ =\ \frac{\kappa}{2}\, \Big(\, X^2_+\: -\: X^2_-\:
-\: 2\,M^2\,\Big)\,,\qquad
D \ =\ \frac{g}{2}\, \Big(\,
X^*_+ X_-\: +\: X^*_- X_+\: -\: m^2_{\rm FI}\, \Big)\; .
\end{equation}
Since  SUSY  is  preserved  after  the SSB  of  U(1)$_X$,  the  scalar
potential  $V_{FD}$ will  vanish  at its  ground state,  i.e.~$\langle
V_{\rm FD}  \rangle =  0$. Consequently, to  leading order  in $m_{\rm
FI}/M$, the VEVs of the scalar inflaton-waterfall fields are
\begin{equation}
  \label{Xdec}
\langle S\rangle\ =\ 0\,,\qquad \langle X_+\rangle\ =\ \sqrt{2}\,M \,,\qquad
\langle X_-\rangle \ =\ \frac{v}{\sqrt{2}} \ ,
\end{equation}
where $v  = m^2_{\rm FI}/(2M)$. Notice  that the VEVs of  the $F$- and
$D$-terms   vanish   through    order   $m_{\rm   FI}/M$   considered,
i.e.~$\langle  D \rangle  = 0$  and $\langle  F_S \rangle  =  {\cal O}
(m^4_{\rm FI}/M^2)$.

To derive the mass spectrum,  we expand the potential about its ground
state  up to terms  quadratic in  all the  fields involved.   We first
consider the $F$-terms.  To order $v/M\ (= m^2_{\rm FI}/M^2)$, we find
the approximate mass eigenstates:
\begin{equation}
S\ =\ \frac{1}{\sqrt{2}}\; \Big(\, \phi\: +\: {\rm i} a\, \Big)\,,\qquad
R_+\: -\: \frac{v}{2M}\,R_-\,,\qquad I_+\: -\: \frac{v}{2M}\, I_-\ .
\end{equation}
All the  above fields, consisting of  4 bosonic degrees  of freedom in
total, share the common mass
\begin{equation}
  \label{Mkappa}
m_\kappa\ =\ \sqrt 2\, \kappa M\; .
\end{equation}
As a  consequence of  SUSY, the corresponding  4 fermionic  degrees of
freedom form  a Dirac  spinor $\psi_\kappa$, which  also has  the same
mass~(\ref{Mkappa}).  We refer  to these particles as inflaton-related
or $\kappa$-sector particles.

\begin{table}[t]
\begin{center}
\begin{tabular}{|c|c|c|c|}
\hline
Sector & Boson & Fermion & Mass\\
\hline\hline
\begin{tabular}{l}
Inflaton\\
($\kappa$-sector)\\
$D$-parity: $+1$
\end{tabular}
&
\begin{tabular}{l}
$S\,$,\\
$R_+-\frac{v}{2M}R_-\,$,\\
$I_+-\frac{v}{2M}I_-$
\end{tabular}
&
$\psi_\kappa=
\left(
\begin{array}{c}
\psi_{X_+} - \frac{v}{2M}\,\psi_{X_-}\\
\psi_S^\dagger
\end{array}
\right)^{\phantom{X}}
$
&
$\sqrt 2 \kappa M$
\\
\hline
\begin{tabular}{l}
U(1)$_X$\\ Waterfall Gauge\\
($g$-sector)\\
$D$-parity: $-1$
\end{tabular}
&
\begin{tabular}{l}
$V_\mu\, [I_-+\frac v{2M}I_+]\,$,\\
$R_-+\frac v{2M}R_+$
\end{tabular}
&
$\psi_g=
\left(
\begin{array}{c}
\psi_{X_-} + \frac{v}{2M}\,\psi_{X_+}\\
-{\rm i}\lambda^\dagger
\end{array}
\right)^{\phantom{X}}
$
&
$g M$
\\
\hline
\end{tabular}
\end{center}
\caption{\sl\small Particle spectrum of the inflaton and the ${\rm
U(1)}_X$ waterfall-gauge sectors after inflation, where the
approximate $D$-parity for each sector is displayed.  The field
$V_\mu$ denotes the ${\rm U(1)}_X$ gauge boson and $\lambda$~its
associate gaugino.  The would-be Goldstone boson related to the
longitudinal degree of $V_\mu$ appears in the square
brackets.}\label{spectrum}
\end{table}

The remaining scalar fields receive  their masses from the $D$-term of
the scalar potential  $V_{FD}$ in~(\ref{VFD}). Performing an analogous
calculation as  outlined above,  we obtain to  order $v/M$  the scalar
mass eigenstates:
\begin{equation}
I_-\: +\: \frac{v}{2M}\, I_+\,,\qquad R_-\: +\: \frac{v}{2M}\, R_+\ .
\end{equation}
The  first field  is absorbed  by  the longitudinal  component of  the
U(1)$_X$  gauge  field  $V_\mu$,  via  the Higgs  mechanism.   In  the
supersymmetric  limit,  all  these  fields, which  mediate  4  bosonic
degrees  of freedom, are  degenerate and  characterized by  the common
mass
\begin{equation}
  \label{Mg}
m_g\ =\ g\, M\ .
\end{equation}
Like in  the $\kappa$-sector case, the respective  4 fermionic degrees
of freedom will  make up a 4-component Dirac spinor  of mass $m_g$. We
refer  to this  group of  particles as  waterfall gauge  or $g$-sector
particles.  In  Table~\ref{spectrum}, we present a summary  of all the
inflaton-related  ($\kappa$-sector)  and waterfall-gauge  ($g$-sector)
particles. As  can also been seen from  the same Table~\ref{spectrum},
$\kappa$-sector  particles  are  predominantly $D$-even,  whereas  the
$g$-sector ones have approximately $D$-odd parity.

It is  now interesting to calculate  the decay rates  of the $\kappa$-
and $g$-sector particles and analyze their implications for the reheat
temperature of the Universe.  Starting  with the singlet field $S$, it
decays  predominantly into  pairs  of charged  and neutral  higgsinos,
$\tilde{h}^\pm_{u,d}$, $\tilde{h}^0_{u,d}$, $\tilde{\bar{h}}^0_{u,d}$,
and  into pairs of  right-handed Majorana  neutrinos $\nu_{1,2,3\,R}$.
On  the  other  hand, the  scalars  $R_+$  and  $I_+$ decay  into  the
SUSY-conjugate partners of the aforementioned fields at the same rate.
In fact, we  find a common decay rate for  each of the $\kappa$-sector
particles:
\begin{equation}
  \label{infldecay}
\Gamma_\kappa\ =\ \frac{1}{32\pi}\:  \Big(\, 4\lambda^2\: +\: 3 \rho^2\,
\Big)\: m_\kappa\; .
\end{equation}
The  reheat temperature $T_\kappa$  resulting from  these perturbative
decays  of the $\kappa$-sector  particles may  be estimated  using the
relation $\Gamma_\kappa  = H (T_\kappa )$, where  the Hubble parameter
$H(T)$ is  given in the radiation  dominated era of  the Universe.  In
this way, we obtain
\begin{equation}
  \label{Tkappa}
T_\kappa\ =\ \left( \frac{90}{\pi^2\, g_*}\right)^{1/4}\,
\sqrt{\Gamma_\kappa\: m_{\rm Pl} }\ ,
\end{equation}
where $g_* = 240$ is the number of the relativistic degrees of freedom
in   the  $F_D$-term  hybrid   model.   Substituting~(\ref{infldecay})
and~(\ref{Mkappa}) into~(\ref{Tkappa}), we arrive at the expression:
\begin{equation}
T_\kappa\ =\ 8.1 \cdot
10^{15}~{\rm GeV} \times
\Big[\kappa (4\lambda^2+3 \rho^2)\Big]^{1/2}
\left(\frac M{10^{16}{\rm GeV}}\right)^{1/2} \,.
\end{equation}
Assuming that  no relevant  amount of entropy  is released  during the
subsequent thermal  history of the Universe,  the gravitino constraint
on  the reheat  temperature $T_\kappa  \stackrel{<}{{}_\sim} 10^9$~GeV
requires that each individual  coupling $\kappa$, $\lambda$ and $\rho$
must be smaller than about $10^{-5}$, if $M \sim 10^{16}$~GeV. Further
details are given in Section~\ref{reheat}.

The above unnatural tuning of  all inflaton couplings to SM fields may
be avoided, if  the large entropy release from the  late decays of the
$g$-sector particles is  properly considered.  An extensive discussion
of  this  issue is  given  in  Section~\ref{reheat}.  Here, we  simply
compute the decay rates of  the $g$-sector particles which are induced
by  a  non-vanishing FI  term  $m_{\rm  FI}$.  In~fact,  the  relevant
interaction Lagrangian is given by
\begin{equation}
  \label{Lint}
{\cal L}_{\rm int}\  =\ \frac{g^2 m^2_{\rm FI}}{8 M}\;  R_-\, (R_+^2 +
I_+^2)\; .
\end{equation}
As  mentioned  above, this  induces  a  decay  width for  the  $D$-odd
particle $R_-$, which is easily calculated to be
\begin{equation}
  \label{GammaR}
\Gamma_g\ =\ \frac{g^3}{128 \pi}\, \frac{m^4_{\rm FI}}{M^3}\,.
\end{equation}
In close  analogy with  the $\kappa$-sector, each  $g$-sector particle
decay rate is equal to $\Gamma_g$.

Let  us now  consider a  model with  a waterfall  sector based  on the
SU(2)$_X$  gauge  group. As  was  mentioned  in Section~\ref{FD},  the
waterfall superfields  $\widehat{X}_1$ and $\widehat{X}_2$  are chosen
to   belong  in  this   case  to   the  2-component   fundamental  and
anti-fundamental representations of SU(2)$_X$, respectively.  Although
the  two  representations  are   equivalent  for  the  SU(2)  case,  we
nevertheless  use this  convention,  such that  its generalization  to
${\rm   SU}(N)$  theories,  with   $N>2$,  is   straightforward.   The
superpotential is almost identical to the one given in~(\ref{Wmodel}),
with  the obvious  substitution: $\widehat{X}_{1}  \widehat{X}_{2} \to
\widehat{X}_{1}^T  \widehat{X}_{2}$.   Extending~(\ref{U1kin}) to  the
SU(2)$_X$ case, the minimal gauge-kinetic Lagrangian is written down
\begin{eqnarray}
  \label{SU2kin}
{\cal L}_{\rm kin} &=&
\int d^4 \theta\ \Bigg[\,
\frac{1}{2}\, {\rm Tr}\,(W^\alpha W_\alpha)\,\delta^{(2)}(\bar{\theta} )\ +\
\frac{1}{2}\, {\rm Tr}\,(\overline{W}_{\dot{\alpha}}
\overline{W}^{\dot{\alpha}})\, \delta^{(2)}(\theta )\nonumber\\
&&  +\
\widehat X_1^\dagger {\rm e}^{2 g \widehat V _X} \widehat X_1\
+\ \widehat X_2^\dagger
{\rm e}^{-2 g \widehat{V}^T_X} \widehat X_2\, \Bigg]\; .
\end{eqnarray}
In the above, $\widehat{V}_X = \widehat{V}^a_X\, T^a$ is the SU(2)$_X$
vector    superfield    and     $W_\alpha    =    W^a_\alpha\,    T^a$
($\overline{W}_{\dot{\alpha}}  = \overline{W}^a_{\dot{\alpha}}\, T^a$)
are the corresponding non-Abelian chiral (anti-chiral) field strengths
in the  so-called Wess--Zumino~(WZ)  gauge.  The superscript  `$T$' on
$\widehat{V}_X$, i.e.~$\widehat{V}^T_X$,  indicates transposition that
acts on  the generators $T^a  = \frac 12  \tau^a$ of the  SU(2) group,
where $\tau^{1,2,3}$ are the usual Pauli matrices.  Finally, the trace
in~(\ref{SU2kin}) is understood to be taken over the group space.

The minimal SU(2)$_X$ gauge-kinetic Lagrangian is invariant under the
discrete transformations
\begin{equation}
  \label{D1}
D_1:\qquad    \widehat{X}_1\    \leftrightarrow\    \widehat{X}_2\,,\qquad
\widehat{V}_X\ \to\ -\widehat{V}^T_X\; .
\end{equation}
Notice  that  under  the  action  of $D_1$  in~(\ref{D1}),  the  field
strengths   transform   as:   $W_\alpha   \to  -   (W_\alpha)^T$   and
$\overline{W}_{\dot{\alpha}} \to - (\overline{W}_{\dot{\alpha}})^T$ in
{\em  any}  SUSY  gauge,  including   the  WZ  gauge.   If  all  other
superfields  do   not  transform,  the  complete   Lagrangian  of  the
non-Abelian  $F_D$-term  hybrid  model  will be  invariant  under  the
discrete  transformation~(\ref{D1})  in  the  unbroken  phase  of  the
theory.

Our discussion so far has made no reference to the specific properties
of  SU(2)$_X$ and  so applies  equally well  to any  SU($N>2$) theory.
However, in  the SU(2)$_X$ case, the $F_D$-term  hybrid model exhibits
an  additional Abelian  or diagonal  discrete symmetry.   This  may be
defined by
\begin{equation}
  \label{D2}
D_2:\qquad \widehat{X}_1\ \to\ \tau^3 \widehat{X}_1\,,\qquad
\widehat{X}_2\ \to\ \tau^3 \widehat{X}_2\,,\qquad
\widehat{V}_X\ \to\ \tau^3\,\widehat{V}_X\,\tau^3\; ,
\end{equation}
whereas all other superfields do not transform. It is then easy to see
that  (\ref{D2}) implies:  $W_\alpha \to  \tau^3\,W_\alpha\,\tau^3$ in
any SUSY gauge  and likewise for $\overline{W}_{\dot{\alpha}}$.  Since
$\tau^3 = (\tau^3)^T$ and $(\tau^3)^2  = {\bf 1}_2$, the invariance of
the Lagrangian~(\ref{SU2kin}) and of  the whole model under the action
of~$D_2$ is evident.\footnote{In general, for a waterfall-gauge sector
based  on  an  SU($N>2$)   group,  there  are  $N$  distinct  discrete
symmetries.   The first  is given  by~(\ref{D1}), while  the remaining
$N-1$  symmetries result  from  replacing $\tau^3$  with  $D_n =  {\rm
diag}\, (1,1,\dots,1,  -1,1,\dots, 1)$. The  entry $-1$ occurs  at the
$n$ position  of the $N$-dimensional  diagonal matrix $D_n$,  with the
restriction $1< n \leq N$.  Obviously,  it is $D_n = D^T_n$ and $D^2_n
=  {\bf  1}_N$.  These  discrete  symmetries  are  non-Abelian in  the
adjoint group space, in the sense that the eigenvalue matrix~$c^{ab}$,
determined by means of the relation $D_n T^a D_n = c^{ab} T^b$, is not
diagonal.}

We now  proceed to  compute the particle  spectrum of  the non-Abelian
$F_D$-term hybrid model after the  SSB of SU(2)$_X$. For this purpose,
it is useful to introduce the notation
\begin{equation}
Z\ =\
\left(\begin{array}{c}
^+\!Z\\
^-\!Z
\end{array}\right)
\,,
\end{equation}
where  $Z$  is  a  generic  ${\rm  SU}(2)_X$-doublet  or  anti-doublet
(conjugate)  field.  The left  superscripts~$\pm$  on  $Z$ denote  the
eigenvalues of  the discrete  symmetry transformation operator  $D_2 =
\tau^3$ defined  in~(\ref{D2}), and they  should not be  confused with
the  corresponding eigenvalues of  the isospin  operator $T^3$  of the
SU(2)$_X$  group.  In  the unitary  gauge, the  minimum of  the scalar
potential occurs for the field values
\begin{equation}
\label{SU2:VEV}
^+\!X_1\ =\ ^+\!X_2\ =\ M\,,\qquad ^-\!X_1\ =\ ^-\!X_2\ =\ 0\;.
\end{equation}
Consequently,   the  discrete   symmetries  $D_1$   and   $D_2$  given
in~(\ref{D1})  and~(\ref{D2})  remain  intact  after the  SSB  of  the
SU(2)$_X$ gauge  group. Since they  act on a gauged  waterfall sector,
they  are  actually  parities.   We   refer  to  them  as  $D_1$-  and
$D_2$-parities, or collectively as $D$-parities.

Analogously to  the U(1)$_X$ case,  we express the  SU(2)$_X$ doublets
$X_{1,2}$   in  terms   of  eigenstates   of   the  $D_{1,2}$-parities
[cf.~(\ref{Xparity})]. In terms of  their components, these fields may
be conveniently expressed as follows:
\begin{equation}
^\pm\!X_\pm \ =\
\langle X_\pm \rangle\: +\: \frac{1}{\sqrt{2}}\,
\Big(\, ^\pm\!R_\pm\ +\ {\rm i}\,^\pm\!I_\pm\, \Big)\; ,
\end{equation}
with $\langle X_+  \rangle = \sqrt{2}\, M$ and  $\langle X_- \rangle =
0$ in the absence of  any $D$-parity violating coupling in the theory.
Moreover, the SU(2)$_X$ $D$-terms are given by
\begin{equation}
D^a\ =\ \frac g2 \left(
X_1^\dagger \tau^a X_1\: -\: X_2^T \tau^a X_2^* \right)\,.
\end{equation}
In the $D$-parity eigenbasis~(\ref{Xparity}), they take on the form
\begin{equation}
D^a\ =\ \frac g2 \times
\left\{
\begin{array}{l}
X_+^\dagger \tau^a  X_- + X_-^\dagger \tau^a X_+\,, \quad
                \textnormal{for }\tau^a \textnormal{ symmetric } (a=1,3)\\
X_+^\dagger \tau^a X_+ + X_-^\dagger \tau^a X_-\,, \quad
                \textnormal{for }\tau^a \textnormal{ antisymmetric } (a=2)
\end{array}
\right.
\;.
\end{equation}
Exactly as in the U(1)$_X$ case,  we find that there are two groups of
mass-degenerate   fields,  $\kappa$-   and  $g$-sector,   with  masses
$m_\kappa$   and  $m_g$  given   in  (\ref{Mkappa})   and  (\ref{Mg}),
respectively.   The complete  inflaton-waterfall spectrum,  along with
their    $D_1$     and    $D_2$    parities,     is    exhibited    in
Table~\ref{spectrum:SU2}.

\begin{table}
\begin{center}
\begin{tabular}{|c|c|c|c|c|c|}
\hline
Sector & Boson & Fermion & Mass & $\,D_1\mbox{-parity}\!\!$
& $\,D_2 \textnormal{-parity}\!\!$
\\
\hline\hline
\begin{tabular}{l}
Inflaton\\
($\kappa$-sector)
\end{tabular}
&
$S\,$,
$\!^+\!R_+$,
$\!^+I_+$
&
$\psi_\kappa=
\left(
\begin{array}{c}
\psi_{^+\!X_+}
\\
\psi_S^\dagger
\end{array}
\!\!\!\right)^{\phantom{X}}_{\phantom{X}}
$
&
$\,\sqrt 2 \kappa M$
&
$+$
&
$+$
\\
\hline

&
\begin{tabular}{l}
$V_\mu^1 [^-\!I_-]\,,$\\
${^-\!R_-}\,$;
\end{tabular}
&
$\psi_g^1=
\left(
\begin{array}{c}
\psi_{^-\!X_-}
\\
-{\rm i} {\lambda^1}^\dagger
\end{array}
\!\!\!\right)_{\phantom{X}}^{\phantom{X}}
$
&
$g M$
&
$-$
&
$-$
\\
\begin{tabular}{c}
SU(2)$_X$\\ Waterfall Gauge\\
($g$-sector)
\end{tabular}
&
\begin{tabular}{l}
$V_\mu^2 [^-\!R_+]\,,$\\
${^-\!I_+}\,$;
\end{tabular}
&
$\psi_g^2=
\left(
\begin{array}{c}
{\rm i}\,\psi_{^-\!X_+}
\\
-{\rm i} {\lambda^2}^\dagger
\end{array}
\!\!\!\right)_{\phantom{X}}^{\phantom{X}}
$
&
$g M$
&
$+$
&
$-$
\\
&
\begin{tabular}{l}
$V_\mu^3 [^+\!I_-]\,,$\\
${^+\!R_-}$
\end{tabular}
&
$\psi_g^3=
\left(
\begin{array}{c}
\psi_{^+\!X_-}
\\
-{\rm i} {\lambda^3}^\dagger
\end{array}
\!\!\!\right)^{\phantom{X}}
$
&
$g M$
&
$-$
&
$+$
\\
\hline
\end{tabular}
\end{center}
\caption{\sl\small Particle  spectrum of  the inflaton
and  an  SU(2)$_X$-gauged   waterfall  sectors  after  inflation.  The
would-be Goldstone bosons of the respective SU(2)$_X$ gauge fields are
given in the square brackets \label{spectrum:SU2}}
\end{table}

The conservation of both  $D_{1,2}$-parities enforces the stability of
all $g$-sector  particles. Instead, if only the  $D_1$-parity, but not
$D_2$,  is   conserved,  then   only  the  $D_1$-odd   particles  from
Table~\ref{spectrum:SU2}  will  be   stable,  and  {\it  vice  versa}.
Obviously, both $D_1$-  and $D_2$-parities need be broken  to make all
$g$-sector   particles  unstable.   In   Appendix~\ref{Dappendix},  we
discuss  possible mechanisms  of explicit  $D$-parity breaking  for an
SU(2)$_X$   waterfall-gauge  sector.   In   general,  there   are  two
mechanisms  for  breaking  $D$-parity.   The  first  one  consists  of
including  higher-order non-renormalizable  operators in  the K\"ahler
potential  whose  presence explicitly  breaks  $D$-parity, whilst  the
second one is  very analogous to the ${\rm  U}(1)_X$ case.  Although a
bare FI  $D$-term is not  possible in non-Abelian  theories, effective
$D^a$-tadpole  terms  may appear  after  the  SSB  of SU(2)$_X$.   The
effective $D^a$-tadpole  terms do not break SUSY.   They get generated
either  from a  non-renormalizable K\"ahler  potential or  are induced
radiatively,  after integrating out  Planck-scale degrees  of freedom.
Thus, without excessive tuning,  the effective $D^a$-tadpole terms can
in general  be small of  the size required  to obtain a  second reheat
phase in the evolution of the Universe.

Independently of  the mechanism which is invoked  to break $D$-parity,
we  may in general  parameterize the  $g$-sector particle  decay rates
through the  $D$-parity-violating mass $m_{\rm FI}$,  which enters the
relation~(\ref{GammaR}).   In the  next section,  we will  discuss how
these  relatively  long-lived $g$-sector  particles  are produced  via
preheating.

\subsection{Preheating and Thermalization}\label{preheat}

After the inflaton field $\phi$  passes below a certain critical value
$\phi_c \approx M$, the  so-called waterfall mechanism gets triggered.
In this case, the inflaton $\phi$ and all other $\kappa$-sector fields
(see  Tables~\ref{spectrum}  and~\ref{spectrum:SU2})  oscillate  about
their true supersymmetric minima:  $\langle S\rangle = 0$ and $\langle
X_+ \rangle  = \sqrt{2}\,  M$.  In this  waterfall epoch, most  of the
energy  density  of  the   Universe  is  stored  to  these  coherently
oscillating $\kappa$-sector  field condensates and  is given initially
by  $\rho_\kappa  =  \kappa^2  M^4$.   During  the  waterfall  regime,
however,  there is  an  additional mechanism  for particle  production
called {\em preheating}.

In  general, there  are two  phenomena associated  with the  notion of
preheating:
\begin{itemize}

\item  The  first  effect  of  preheating  arises  from  the  negative
  curvature  of  the potential  with  respect  to the  $\kappa$-sector
  fields.   Such  a  negative  curvature  corresponds  to  a  negative
  tachyonic  mass  term  in  the  potential.  As  a  consequence,  the
  particle  number within infrared  modes of  momentum less  than this
  tachyonic mass grows exponentially.  This phenomenon is known as the
  {\em    negative    coupling    instability}   or    \emph{tachyonic
  preheating}~\cite{TACHYPREH}.  Numerical simulations have shown that
  the  field  amplitudes  suffer   strong  damping  during  the  first
  oscillation, due to  the energy transfer to the  infrared modes.  In
  the  $F_D$-term  hybrid model,  only  $\kappa$-sector particles  are
  produced  by tachyonic preheating.   A full  study of  this process,
  including thermal equilibration of the $\kappa$-sector particles, is
  a highly  nontrivial matter  and has so  far only been  achieved for
  very  particular models of  preheating.  Since  the fraction  of the
  energy   density  transferred  instantaneously   to  $\kappa$-sector
  particles through tachyonic preheating  is rather small, compared to
  their  initial energy  density $\rho_\kappa$,  these model-dependent
  details fortunately have no dramatic impact on the expansion and the
  thermal history of the Universe.   Therefore, we do not consider the
  phenomenon of tachyonic preheating in the $F_D$-term hybrid model.

\item  Particle   production  may  also  occur   during  the  coherent
  oscillation  regime,  because  both  the  $\kappa$-  and  $g$-sector
  particles have masses  that can vary very strongly  with time.  This
  effect is called {\em preheating  via a time-varying mass} or simply
  {\em preheating}~\cite{PREHEATING,GBRM}.   As we will  show below, a
  small but  significant fraction of  the total energy density  of the
  Universe $\rho_\kappa$  can be transferred,  almost instantaneously,
  to the $g$-sector particles, e.g.~$\rho_g \sim 10^{-4} \rho_\kappa$,
  for $\kappa  \sim 10^{-2}$.  As we illustrated  in the  beginning of
  this section and will  show more explicitly in Section~\ref{reheat},
  this small  fraction of the $g$-sector energy  density is sufficient
  to alter dramatically the thermal history of the Universe.

\end{itemize}

Our interest lies therefore in computing the production energy density
$\rho_g$ of the $g$-sector particles  via preheating. A key element in
such  a computation is  the profile  of the  time-varying mass  of the
$g$-sector  particles, $m_g (t)  = g\,  X_+ (t)/\sqrt{2}$.   The exact
time  dependence of  $m_g (t)$  depends crucially  on the  dynamics of
tachyonic preheating.  Comparative  numerical studies strongly suggest
that a sufficiently accurate description  of the time evolution of the
$g$-sector  mass is  obtained by~\cite{GBRM}\footnote{To  be specific,
the mass-term  time-variation studied  in~\cite{GBRM} was for  a model
with  a single field  rolling from  the top  of a  local maximum  of a
quartic potential.  It was  found that a $\tanh$-functional dependence
accurately  captures the  evolution  of the  time-varying mass.   Even
though the  model considered~\cite{GBRM}  is still different  from our
hybrid inflationary potential,  the derived $\tanh$-functional profile
for  the  time-varying  mass  should  be  regarded  as  a  substantial
improvement over the one assumed in~\cite{GP}.}
\begin{equation}
\label{tanhmass}
m_g(t)\ =\ \frac{g M}{2}\; \Big[\, \tanh ( \kappa M t)\: +\: 1\,
  \Big]\; .
\end{equation}
Notice that the time-dependent function $m_g(t)$ properly interpolates
between the values $m_g (t \to -\infty) = 0$ and $m_g (t \to \infty) =
g M$ that  occur in the beginning and the end  of the waterfall epoch,
respectively.

Given  the time-dependent  mass~(\ref{tanhmass}), we  may  compute the
occupation  number of the  fermionic $g$-sector  modes by  solving the
Dirac equation
\begin{equation}
  \label{uheq}
\Big[\, {\rm i}\, \gamma^0\, \partial_t\: -\:
\mbox{\boldmath $\gamma$}\cdot\mathbf{k}\: -\: m_g(t)\,\Big]\; u_h(t)\ =\ 0\, .
\end{equation}
The  solution  to   the  above  equation  may  be   expressed  by  the
time-dependent Dirac spinor $u_h (t)$ in the chiral representation:
\begin{equation}
u_h (t)\ =\ \left( \begin{array}{c} L_h (t) \\ R_h (t) \end{array} \right)\,
\otimes\, \xi_h\; ,
\end{equation}
where $\xi_h$  is the helicity two-component  eigenspinor for helicity
$h=\pm$.   The  occupation  number  of  Dirac  fermions  produced  via
preheating in the true supersymmetric vacuum at $t\to \infty$ is given
by
\begin{equation}
  \label{nFh}
n^{\rm F}_h(k)\ =\ \frac 1{2\omega(k)}\, \Big[\,
hk (|R_h|^2\, -\, |L_h|^2)\: -\: m_g (L_hR_h^*\, +\, L_h^*R_h)\,\Big]
\ +\ \frac 12\ ,
\end{equation}
where $k  = |\mathbf{k}|$ is the  modulus of the  3-momentum. With the
help  of~(\ref{nFh}), the  $k$-mode  energy density  is calculated  by
$\rho(k)    =    \sum_h    \omega(k)\,    n^{\rm    F}_h(k)$,    where
$\omega(k)=\sqrt{k^2+m_g^2(t\to   \infty)}$.   To   obtain   a  unique
solution to  the linear differential  equation~(\ref{uheq}), we impose
initial  conditions  that  correspond  to a  zero  occupation  number,
i.e.~$n^{\rm F}_h(k)=0$.  These are given at $t\to -\infty$ by
\begin{equation}
  \label{initF}
L_h\ =\ \sqrt\frac{\omega(k) + hk}{2\omega}\ ,\qquad
R_h\ =\ \sqrt\frac{\omega(k) - hk}{2\omega}\ .
\end{equation}

By analogy, the occupation number  of the bosonic $g$-sector modes are
determined by solving the Klein--Gordon equation of motion
\begin{equation}
  \label{phieq}
\Big[\, \partial_t^2\: +\: \mathbf{k}^2\: +\: m_g^2(t)\,\Big]\,
\varphi(t) \ =\ 0\; ,
\end{equation}
and imposing the initial conditions at $t\to -\infty$,
\begin{equation}
\varphi\ =\ \frac{1}{2\sqrt{\omega(k)}}\ ,\qquad
\frac{\partial \varphi}{\partial t}\ =\ -\,{\rm i}\,\sqrt{\frac\omega{2}}\ .
\end{equation}
As in the Dirac case, these initial conditions correspond to vanishing
occupation  numbers.  The occupation  number of  the bosonic  modes at
$t\to \infty$ is given by
\begin{equation}
  \label{initB}
n^{\rm B}(k)\ =\ \frac 12\; \omega(k)\, |\varphi|^2\ +\
\frac 1{2\omega(k)}\; \left|\frac{d \varphi}{dt}\right|^2\ -\ \frac 12\ .
\end{equation}

Using  the time-dependent  mass-term~(\ref{tanhmass}), along  with the
initial  conditions~(\ref{initF}) and  (\ref{initB}),  one may  obtain
analytical     expressions      in     terms     of     hypergeometric
functions~\cite{GBRM},  for  the  particle  production  between  $t\to
-\infty$ and $t\to \infty$.  For $\kappa\ll  g$ and $k \ll g M$, these
analytical expressions reduce to
\begin{equation}
n(k)\ =\  \frac{2}{\exp{\left(\frac{\displaystyle \pi k}
{\displaystyle \kappa M}\right)}\: \pm\:  1}\ \ ,
\end{equation}
where the sign $+$ applies  for $n(k)=n^{\rm F}_h(k)$ and the sign $-$
for $n(k)=n^{\rm  B}(k)$.  Recalling that there are  2 helicity states
for  a  $g$-sector  fermion  and  4  real degrees  of  freedom  for  a
$g$-sector  boson,  we may  calculate  the  occupation  number of  all
$g$-sector modes as follows:
\begin{equation}
n_g(k)\ =\ N_{\rm b}\, \left(\, \sum\limits_{h=\pm}\, n_h^{\rm F}(k)\: +\:
4 n^{\rm B}(k)\,\right)\,,
\end{equation}
where $N_{\rm b}$ is the  number of broken generators of the waterfall
gauge symmetry.  In particular, it is $N_{\rm b}=1$ for ${\rm U}(1)_X$
and  $N_{\rm  b}=3$  for  ${\rm  SU}(2)_X$  [cf.~Tables~\ref{spectrum}
and~\ref{spectrum:SU2}].     Since   the   produced    particles   are
non-relativistic,   i.e.~$k   \ll   gM$,   their   occupation   number
distribution  $n_g (k)$  can easily  be integrated  to give  the total
energy density carried by the $g$-sector fields, i.e.
\begin{equation}
  \label{rhog}
\frac{\rho_g}{\rho_\kappa}\ \approx\
\frac{gM}{\rho_\kappa\,2\pi^2}\, \int\limits_0^\infty k^2 dk\, n_g(k)\
\approx\ 2.1 \times 10^{-2}\, N_{\rm b}\; \kappa g\; .
\end{equation}
Here  $\rho_{\kappa}=\kappa^2  M^4$  is  the  energy  density  of  the
$\kappa$-sector  particles shortly  before  the waterfall  transition.
Equation~(\ref{rhog}) will be a valuable input for the next section to
compute  the true reheat  temperature $T_{\rm  reh}$ of  the Universe,
which arises from the combined  effect of the $\kappa$- and $g$-sector
particle decays.

\setcounter{equation}{0}
\section{Coupled Reheating and Gravitino Abundance}\label{reheat}

In the previous  section, we have seen that  the $g$-sector particles,
e.g.~$\psi_g$, $R_-$  and $V_\mu$,  can be abundantly  produced during
the  preheating epoch.  Assuming  that they  dominate the  Universe at
some  later  time,  their  decays  induced  by  the  small  $D$-parity
violating  couplings will give  rise to  a second  reheat temperature,
which we  denote here by $T_g$. As  we will show in  this section, the
large entropy, which is released  by the late decays of the $g$-sector
particles,  will be  sufficient  to dilute  the  gravitinos to  levels
compatible  with BBN  limits.

More  explicitly, we  present  a detailed  numerical  analysis of  the
gravitino abundance~$Y_{\widetilde{G}}$, where  the combined effect of
the $\kappa$-  and $g$-sector particle decays is  carefully taken into
account.   As  we  mentioned  in  the Introduction,  we  call  such  a
two-states'   mechanism  of  reheating   the  Universe   {\em  coupled
reheating}.  In Section~\ref{Beqs}, we set the BEs relevant to coupled
reheating  and give  numerical  estimates of  the gravitino  abundance
$Y_{\widetilde{G}}$ and  the energy densities  $\rho_\kappa$, $\rho_g$
and $\rho_{\rm rad}$ related to the $\kappa$- and $g$-sector particles
and their radiation,  respectively.  In Section~\ref{Seqs}, we present
a semi-analytic approach to  BEs, where useful approximate expressions
for~$Y_{\widetilde{G}}$ are  obtained.  Finally, in Section~\ref{para}
we   derive  gravitino  abundance   constraints  on   the  theoretical
parameters.

\subsection{Boltzmann Equations}\label{Beqs}

The number density $n_{{\tilde G}}$ of gravitinos, the energy density
$\rho_\kappa$ ($\rho_g$) of the $\kappa$~($g$)-sector particles and
the energy density $\rho_{\rm rad}$ of the radiation produced by their
decays satisfy the following system of BEs~\cite{kolb}:
\begin{eqnarray}
  \label{ng}
\dot n_{{\widetilde G}}\: +\: 3Hn_{{\widetilde G}} &=&
                                         C_{\widetilde G}\, T^6\,,\nonumber\\
  \label{nf}
\dot \rho_\kappa\: +\: 3H\rho_\kappa &=& -\,\Gamma_\kappa\,
\rho_\kappa\,,\nonumber\\
  \label{nfb}
\dot\rho_g\: +\: 3H\rho_g &=& -\,\Gamma_g\,\rho_g\,,\nonumber\\
  \label{rR}
\dot\rho_{\rm rad}\: +\: 4H\rho_{\rm rad} &=&
\Gamma_\kappa\,\rho_\kappa\: +\: \Gamma_g\,\rho_g\; ,
\end{eqnarray}
where a dot on  $n_{\widetilde{G}}$, $\rho_{\kappa, g}$ and $\rho_{\rm
rad}$  indicates  differentiation  with  respect to  the  cosmic  time
$t$.  The quantity  $C_{{\widetilde G}}(T)$  is a  collision  term for
gravitino production  calculated in~\cite{brand,kohri} and  the Hubble
parameter $H$ is given by
\begin{equation}
  \label{Hini}
H\ =\ \frac{1}{\sqrt{3}\,m_{\rm Pl}}\;
\bigg(\, m_{{\widetilde G}}\,n_{{\widetilde G}}\:
+\: \rho_\kappa\: +\: \rho_g\: +\: \rho_{\rm rad}\, \bigg)^{1/2}\; ,
\end{equation}
where $m_{{\widetilde  G}}$ is the  mass of the  gravitino $\widetilde
G$. In addition,  the temperature $T$ and the  entropy density $s$ may
be determined through the relations:
\begin{equation}
  \label{rs}
\rho_{\rm rad}\ =\ \frac{\pi^2}{30}\; g_*\, T^4\,,\qquad
s\ =\ \frac{2\pi^2}{45}\; g_*\ T^3,
\end{equation}
where  $g_* (T)$  is the  effective number  of degrees  of  freedom at
temperature  $T$. Since  the initial  temperature is  $T_{\rm  in} \ll
\kappa M$, it is $g_* = 240$ for all $T > M_{\rm SUSY}$.

Here  we should  note that  in  BEs~(\ref{rR}) we  have neglected  the
collision  terms  related   to  the  self-annihilation  of  $g$-sector
particles.   Their thermally  averaged cross  section  times velocity,
$\langle \sigma_{\rm ann} v \rangle$, is estimated to be
\begin{equation}
\langle \sigma_{\rm ann}\, v \rangle\ \stackrel{<}{{}_\sim}\
10^{-35}~{\rm GeV}^{-2}\; ,
\end{equation}
which is numerically negligible.

The  numerical  analysis of  the  BEs  (\ref{rR})  gets simplified  by
absorbing the Hubble expansion terms into new variables.  To this end,
we define the following dimensionless quantities \cite{riotto}:
\begin{equation}
  \label{fdef}
f_{{\widetilde G}}\ =\ n_{{\widetilde G}} a^3\,,\quad f_\kappa\ =\ \rho_\kappa
a^3\,,\quad f_g\ =\ \rho_g a^3\,, \quad f_{\rm rad}\ =\ \rho_{\rm rad} a^4.
\end{equation}
where $a$  is the  usual expansion scale  factor of the  Universe.  We
also convert the  time derivatives to derivatives with  respect to the
logarithmic   time  $\ln\left(a/a_{\rm   I}\right)$~\cite{qui},  where
$a_{\rm I}$  is some initial or  reference value for  the scale factor
$a$.   With  the  above   substitutions,  the  BEs~(\ref{rR})  may  be
re-written as
\begin{eqnarray}
 \label{fg}
H f^\prime_{\widetilde G} &=& C_{{\widetilde G}}\, T^6 a^3\,,\nonumber\\
 \label{ff}
H f^\prime_\kappa &=& -\,\Gamma_\kappa f_\kappa\,,\nonumber\\
 \label{ffb}
H f^\prime_g &=& -\,\Gamma_g\, f_g\,,\nonumber\\
 \label{fR}
H f^\prime_{\rm rad} &=& \Gamma_\phi f_\phi a\: +\: \Gamma_g f_g a\,,
\end{eqnarray}
where   the   prime   now   denotes   differentiation   with   respect
to~$\ln\left(a/a_{\rm I}\right)$. Correspondingly, the Hubble parameter
$H$ and  temperature $T$ may  now be expressed  in terms of  the newly
introduced variables~(\ref{fdef}) as follows:
\begin{equation}
  \label{H2exp}
H\ =\ \frac{1}{\sqrt{3}\, a^{3/2}\, m_{\rm Pl}}\;
\bigg(\, m_{\widetilde{G}} f_{\widetilde{G}}\: +\:
f_\kappa\: +\: f_g\:  +\: a^{-1} f_{\rm rad}\,\bigg)^{1/2}\,,\quad
T\ =\ \left(\frac{30\, f_{\rm rad}}{\pi^2 g_{\ast} a^4}\right)^{1/4}\; .
\end{equation}
The transformed system of  BEs~(\ref{fR}) can be numerically solved by
imposing the following initial conditions:
\begin{equation}
  \label{init}
f_{\kappa,{\rm I}}\, a_{\rm I}^3\ =\ \kappa^2 M^4\,,\quad
f_{g,{\rm I}}\, a_{\rm I}^3\ =\ 2.1\times 10^{-2} g \kappa^3\, M^4\,,
\quad f_{{\rm rad},{\rm I}}\ =\ 0\; ,
\end{equation}
where    the   subscript~I   refers    to   quantities    defined   at
$\ln\left(a/a_{\rm  I}\right)=0$.   Notice   that  the  initial  value
$f_{g,{\rm  I}}\,  a_{\rm  I}^3$   is  equal  to  the  energy  density
$\rho_{g,{\rm  I}}$  of   the  $g$-sector  particles  produced  during
preheating and is given in~(\ref{rhog}) .

\begin{figure}[t]
\begin{center}
\epsfig{file=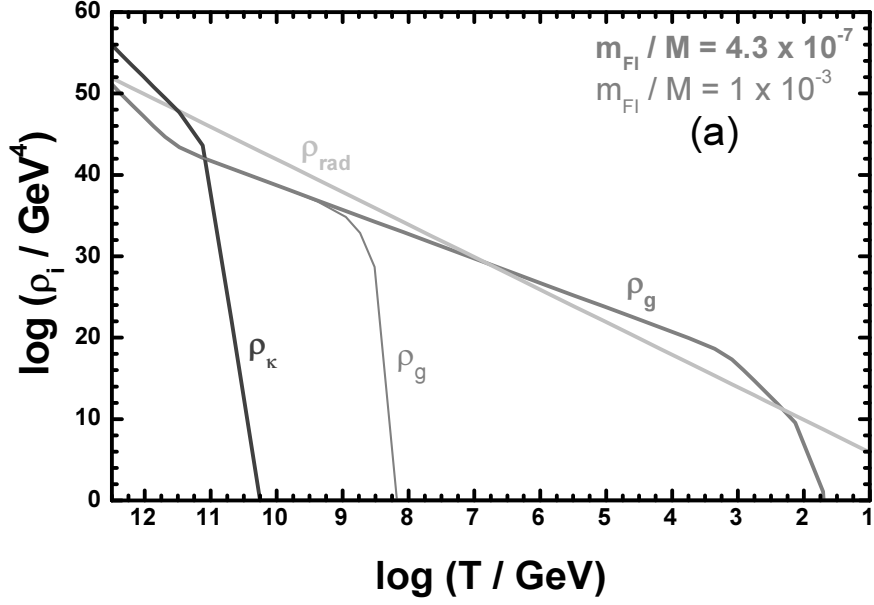,height=5.2in,angle=-90} \\[3mm]
\epsfig{file=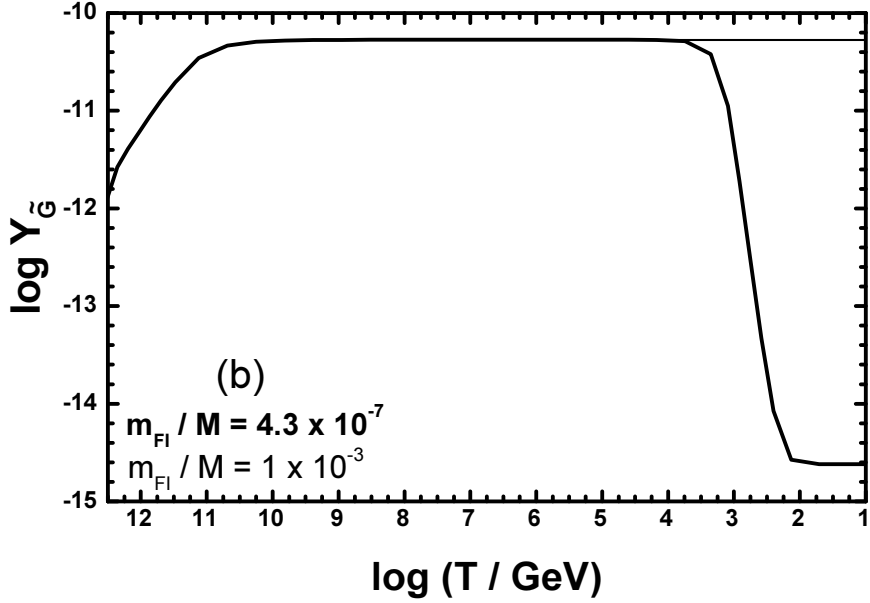,height=5.2in,angle=-90}
\end{center}
\caption{\sl\small The evolution as a function of $\log
T$ of the quantities: {\sf (a)} $\log\rho_i$ with $i=\kappa$ (dark
grey line), $i=g$ (grey line), $i={\rm rad}$ (light grey line)
{\sf (b)} $\widetilde G$ yield, $Y_{\widetilde G}$. In both cases, we take
$M=0.7\times10^{16}~{\rm GeV}$, $\rho=\lambda=\kappa=0.001$ and
$m_{\rm FI}/M=4.3\times10^{-7}~{\rm GeV}$ (bold lines) and $m_{\rm
FI}/M=1\times10^{-3}$ (thin lines).} \label{fig:rY}
\end{figure}

In  Fig.~\ref{fig:rY}(a),  we   present  numerical  estimates  of  the
cosmological evolution of energy densities $\rho_{\kappa,g,{\rm rad}}$
as functions  of the temperature  $T$ in a double  $x$-$y$ logarithmic
plot, where $\rho_\kappa$ is represented by a dark grey line, $\rho_g$
by  a grey line  and $\rho_{\rm  rad}$ by  a light  grey line.   As an
example,   we    use   $M    =   0.7   \times    10^{16}~{\rm   GeV}$,
$\rho=\lambda=\kappa=10^{-3}$ and  $m_{\rm FI}/M =  4.3\times 10^{-7}$
(bold  lines)  and  $m_{\rm  FI}/M  = 10^{-3}$  (thin  lines).   Since
$\rho_{\rm  rad}$ is  affected very  little  for the  larger value  of
$m_{\rm FI}/M$, it  has not been added to  the plot.  The intersection
point  of  the $T$-dependent  functions  $\rho_\kappa$ and  $\rho_{\rm
rad}$ signals  the completion of the  $\kappa$-sector particle decays.
For  the   specific  example,  this   point  occurs  at   $T_\kappa  =
3.2\times10^{11}~{\rm   GeV}$.   For  $m_{\rm   FI}/M  =   4.3  \times
10^{-7}~{\rm  GeV}$,  we  obtain   two  more  intersections:  one  for
$T=T_{\rm eq} \simeq 3.9  \times 10^6~{\rm GeV}$ where $\rho_g (T_{\rm
eq}) = \rho_{\rm rad} (T_{\rm  eq})$ and another one for $T=T_g \simeq
200~{\rm  GeV}$,  where  the  $g$-sector  particles  have  practically
decayed away and $\rho_g (T_g) = \rho_{\rm rad} (T_g)$.  Thanks to the
huge   entropy  release   in  this   case,  the   gravitino  abundance
$Y_{\widetilde  G}\ =\  n_{{\widetilde G}}/s$  gets  sharply decreased
from  about $2.2  \times  10^{-11}$ to  $2.4  \times 10^{-15}$.   This
dramatic    reduction   of    $Y_{\widetilde   G}$    is    shown   in
Fig.~\ref{fig:rY}(b).  On  the contrary, if $m_{\rm  FI}/M = 10^{-3}$,
no intersection of $\rho_g$ with  $\rho_{\rm rad}$ takes place and, in
consequence,  no  phase of  second  reheating  occurs.   This is  also
illustrated in Fig.~\ref{fig:rY}(a),  where the dependence of $\rho_g$
is   displayed   by   a   thin    line.    As   can   be   seen   from
Fig.~\ref{fig:rY}(b),  the   gravitino  abundance  $Y_{\widetilde  G}$
remains  unsuppressed  in   this  case,  i.e.~$Y_{\widetilde  G}  \sim
10^{-10}$.   As we will  see below  in Section~\ref{para},  such large
values  of   $Y_{\widetilde  G}$  are  in  gross   conflict  with  BBN
constraints.

\subsection{Semi-analytic Approach} \label{Seqs}

We now  present a more intuitive  and rather accurate  approach to the
dynamics  of  coupled   reheating,  and  find  approximate  analytical
expressions    that   describe   the    evolution   of    the   energy
densities~$\rho_{\kappa,g,{\rm  rad}}$.  In  addition,  we derive  the
conditions that ensure  the existence of a second  reheat phase in the
evolution  of  the  Universe.   Finally,  we  estimate  the  gravitino
abundance $Y_{\widetilde G}$ due to coupled reheating.

Shortly after inflation ends, the energy of our observable Universe is
dominated by the inflaton $S$ and the other $\kappa$-sector particles,
with an initial energy density  $\rho_{\kappa, \rm I} = \kappa^2 M^4$.
As   we  schematically   illustrated  in   Section~\ref{preheat},  the
$\kappa$-sector particles decay  into relativistic degrees of freedom,
producing  an energy  density  $\rho_{\rm rad}$.   The energy  density
$\rho_g$  of the $g$-sector  particles is  subdominant at  these early
stages after  the first reheating due to  the $\kappa$-sector particle
decays.  In  fact, for  temperatures $T >  T_{\rm eq}$,  where $T_{\rm
eq}$ is  the first intersection  point of the  $T$-dependent functions
$\rho_{\rm rad}$ and $\rho_g$  [see~(\ref{trg})], the evolution of all
relevant energy densities may be approximately described as follows:
\begin{equation}
  \label{rRfq}
\rho_\kappa\ =\ \rho_{\kappa,{\rm I}}\, \left(a/a_{\rm I}\right)^{-3}\,,\quad
\rho_g\ =\ \rho_{g,{\rm I}}\, \left(a/a_{\rm I}\right)^{-3}\,,\quad
\rho_{\rm rad}\ =\ \rho_{\rm rad}(T_\kappa)\,
\left(T/T_\kappa \right)^4\; .
\end{equation}
As mentioned  above, the  $T$-dependent function $\rho_{\rm  rad}$ may
first cross the corresponding $\rho_g$ at $T=T_{\rm eq}$, where
\begin{equation}
  \label{trg}
\rho_{\rm rad} (T_{\rm eq })\ =\ \rho_g (T_{\rm eq })\; .
\end{equation}
Using  the   fact  that  $\rho_{\rm  rad}   (T_\kappa)  =  \rho_\kappa
(T_\kappa)$ and assuming that the Universe expands isentropically with
$a  \propto T^{-1}$ when  $T_{\rm eq}\leq  T\leq T_\kappa$,  we obtain
from~(\ref{trg}) the approximate relation
\begin{equation}
  \label{Teq}
T_{\rm eq}\ \simeq\ T_\kappa\
\frac{\rho_{g, {\rm I}}}{\rho_{\kappa,{\rm I }}}\ .
\end{equation}
In deriving~(\ref{Teq}), we have also made use of~(\ref{rRfq}).

A second  reheat phase in the  evolution of the  Universe takes place,
only  if  $T_g  <  T_{\rm  eq}$,  where  $T_g$  is  the  naive  reheat
temperature  due to  the $g$-sector  particles  decays [see~(\ref{Tg})
below].  To better explore the consequences of this last condition, we
use  the  abbreviation  $g$-DAD  ($g$-DBD)  to  indicate  whether  the
$g$-sector  particles Decay  After  (Before) the  Domination of  their
energy  density.  With  the  aid  of~(\ref{Teq}),  the  following  two
conditions for $g$-DAD and $g$-DBD may be deduced:
\begin{equation}
  \label{cond}
\frac{T_g}{T_\kappa}\ <\
  \frac{\rho_{g,\rm I}}{\rho_{\kappa, \rm I}}\quad (g\mbox{-DAD}),\qquad
\frac{T_g}{T_\kappa}\ \geq\ \frac{\rho_{g,\rm I}}{\rho_{\kappa,\rm
  I}}\quad (g\mbox{-DBD})\; .
\end{equation}
These two possible scenarios  are illustrated in Fig.~\ref{fig:rY} for
$m_{\rm FI}/M  = 4.3\times 10^{-7}$ ($m_{\rm FI}/M  = 10^{-3}$), where
the bold (thin) lines correspond to $g$-DAD ($g$-DBD).

The gravitino abundance $Y_{\widetilde  G}$ can be calculated by simply
integrating  $f^\prime_{\widetilde G}$  that  occurs in  the first  BE
of~(\ref{fg})   and  using   the   fact  that   $Y_{\widetilde  G}   =
f_{\widetilde G}/sa^3$.   It turns out  that the main  contribution to
$Y_{\widetilde G}$  comes from the integration  after the commencement
of  the  radiation  dominated  era, i.e.~for  $T\leq  T_\kappa$.   The
so-derived  formula  reproduces rather  accurately  the one  presented
in~\cite{kohri} in the massless gluino limit, where
\begin{equation}
  \label{Yk}
Y^\kappa_{\widetilde G}\ =\ 1.6 \times 10^{-12}
\left(\frac{T_\kappa}{10^{10}~{\rm GeV}}\right)\ .
\end{equation}
Note that (\ref{Yk}) is only valid for the $g$-DBD case.

The  situation is  different for  the  $g$-DAD case,  where a  drastic
reduction  of the  gravitino  abundance, caused  by  the huge  entropy
release  from the $g$-sector  particle decays,  takes place.   In this
case, the gravitino abundance $Y^g_{\widetilde G}$ may be estimated in
the following way. We first notice that
\begin{equation}
  \label{Yg1}
Y^g_{\widetilde G}\ =\ Y^\kappa_{\widetilde G}\
\frac{s(T_{\rm eq})\, a^3 (T_{\rm eq})}{s(T_g)\, a^3(T_g)}\ .
\end{equation}
Then, with the help of (\ref{rs}) and (\ref{rRfq}), we may obtain the
relation
\begin{equation}
  \label{dilution}
\frac{s(T_{\rm eq})\, a^3 (T_{\rm eq})}{s(T_g)\, a^3(T_g)}\ =\
\left(\frac{T_{\rm eq}}{T_g}\right)^3\;
\left(\frac{\rho_g(T_{\rm    eq})}{\rho_g(T_g)}\right)^{-1}\
=\ \frac{T_g}{T_{\rm eq}}\ .
\end{equation}
Substituting    the   respective    expressions   of~(\ref{dilution}),
(\ref{Yk}) and~(\ref{Teq}) into~(\ref{Yg1}) yields
\begin{equation}
  \label{Yg}
Y^g_{\widetilde G}\ =\ 1.6\times10^{-12}\, \left(
\frac{T_g}{10^{10}~{\rm GeV}}\right)\; \frac{\rho_{\kappa , \rm
I}}{\rho_{g, \rm I}}\ \simeq\ \frac{7.6\times 10^{-11}}{\kappa
g}\; \left(\frac{T_g}{10^{10}~{\rm GeV}}\right)\ ,
\end{equation}
where  we  have  used  (\ref{rhog})  to derive  the  last  approximate
equality. We have checked that the semi-analytic formula (\ref{Yg}) is
in  remarkable  agreement  with  numerical estimates  in  the  $g$-DAD
regime.

Finally, we should  comment on the fact that  the number of $e$-folds,
${\cal  N}_e$  gets modified  in  the  $g$-DAD  case, because  of  the
occurrence  of  a  $g$-sector-matter  dominated era.   Making  use  of
standard methods~\cite{CLLSW,review}, we  are able to determine ${\cal
N}_e$  at the  WMAP pivotal  point $k_0=0.002~{\rm  Mpc}^{-1}$  by the
following relation:
\begin{equation}
  \label{Ng}
{\cal N}_e\ =\ 22.6\ +\ {1\over 6}\ln (\kappa^2M^4)\
+\ {1\over3}\ln T_g\ +\ {1\over3}\ln{
\rho_{\kappa,\rm I}\over\rho_{g,\rm I}}\ .
\end{equation}
This result,  however, does  not crucially alter  the value  of ${\cal
N}_e$,  which  remains  close  to  $55-60$  in  the  $g$-DAD  case  as
well. Interestingly  enough, $Y^g_{\widetilde G}$ and  ${\cal N}_e$ do
not directly  depend on $T_\kappa$ given  in~(\ref{Tkappa}).  In fact,
in the $g$-DAD  case, $Y^g_{\widetilde G}$ and ${\cal  N}_e$ are fully
independent of the superpotential  couplings $\lambda$ and $\rho$, and
only  have  a mild  linear  and  logarithmic  dependence on  $\kappa$,
respectively.  As we will discuss below, it is this last property that
leads to a significant  relaxation of the strict gravitino constraints
on these couplings, when compared to the $g$-DBD case.

\subsection{Gravitino Abundance Constraints}\label{para}

In order to avoid destroying the apparent success between the standard
theory  for BBN  and observation,  gravitinos must  have  an abundance
$Y_{\widetilde G}$ below certain  upper limits, which crucially depend
on  their decay  properties~\cite{oliveg,kohri}.   Some representative
upper  bounds  on~$Y_{\widetilde  G}$,   obtained  in  a  very  recent
analysis~\cite{kohri}, are
\begin{equation}
  \label{bYg}
Y_{\widetilde G}\ \stackrel{<}{{}_\sim}\ \left\{\matrix{
10^{-15}\hfill ,  & \mbox{for}~~m_{\widetilde G}\simeq360~{\rm GeV},
\hfill \cr
10^{-14}\hfill ,  &\mbox{for}~~ m_{\widetilde G}\simeq600~{\rm GeV},
\hfill \cr
10^{-13}\hfill ,  &\mbox{for}~~ m_{\widetilde G}\simeq7.5~{\rm TeV},
\hfill \cr
10^{-12}\hfill ,  &\mbox{for}~~m_{\widetilde G}\simeq9.3~{\rm TeV}\; .
\hfill \cr}
\right.
\end{equation}
The above bounds  pertain to the less restrictive  case of a gravitino
that  decays with a  small branching  ratio $B_h=0.001$  into hadronic
modes.    For   the   $g$-DBD   case  discussed   above,   the   upper
limits~(\ref{bYg}) imply the corresponding stringent bounds on $T_{\rm
reh}$:
\begin{equation}
  \label{bTr}
T_{\rm reh}\ \stackrel{<}{{}_\sim}\ \left\{\matrix{
9\times10^{6}~{\rm GeV}\hfill ,  & \mbox{for}~~m_{\widetilde
G}\simeq360~{\rm GeV}\,, \hfill \cr
7\times10^{7}~{\rm GeV}\hfill ,  &\mbox{for}~~ m_{\widetilde
G}\simeq600~{\rm GeV}\,, \hfill \cr
7\times10^{8}~{\rm GeV}\hfill ,  &\mbox{for}~~ m_{\widetilde
G}\simeq7.5~{\rm TeV}\,, \hfill \cr
7\times10^{9}~{\rm GeV}\hfill ,  &\mbox{for}~~m_{\widetilde
G}\simeq9.3~{\rm TeV}\; . \hfill \cr}
\right.
\end{equation}

The  aforementioned upper limits  lead to  serious constraints  on the
basic couplings  $\kappa$, $\lambda$ and $\rho$,  usually forcing them
to  acquire  very  small  values,  i.e.~$\kappa.\,  \lambda,  \,  \rho
\stackrel{<}{{}_\sim}  10^{-5}$.   For  the standard  $F$-term  hybrid
model within  mSUGRA and with  a soft SUSY-breaking  tadpole parameter
$\mbox{a}_S =  1$~TeV, the requirement of accounting  for the observed
power spectrum  $P_{\cal R}$, with  a number of $e$-folds  ${\cal N}_e
=50$--60, implies  that $\kappa> 10^{-4}$ and $T_\kappa  = T_{\rm reh}
\stackrel{>}{{}_\sim} 9  \times 10^{9}$~GeV.  Such a  high lower bound
on $T_{\rm reh}$ invalidates  all the limits presented in~(\ref{bTr}),
thereby ruling out the above $F$-term hybrid model.

\begin{figure}[t]
\begin{center}
\epsfig{file=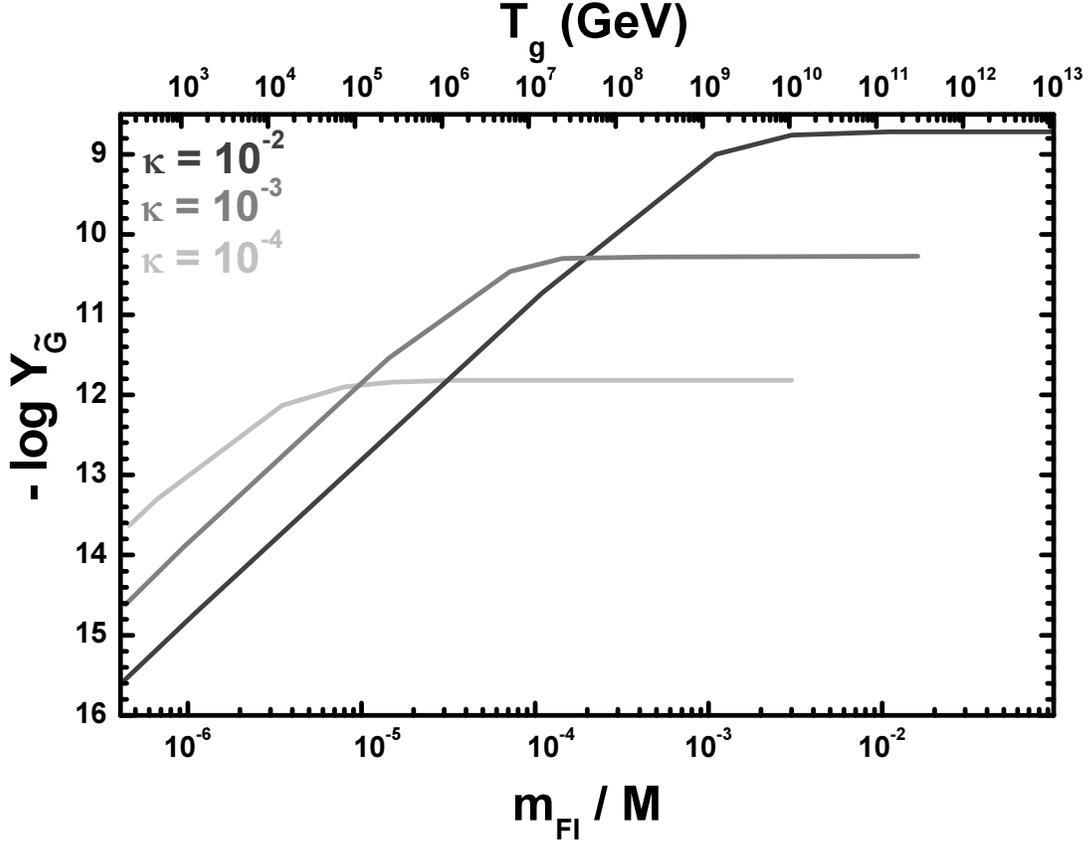,height=6.4in,angle=-90} \\[3mm]
\end{center}
\caption{\sl\small  The dependence  of $\log  Y_{\widetilde  G}$ on
$m_{\rm    FI}/M$,   for    $\kappa=10^{-2}$    (dark   grey    line),
$\kappa=10^{-3}$   (grey  line)   and  $\kappa=10^{-4}$   (light  grey
line). }\label{fig:Ygm}
\end{figure}

The above  situation, however,  changes drastically in  the $F_D$-term
hybrid model with small $D$-parity violation, e.g.~due to the presence
of a  subdominant FI $D$-term.   This corresponds to the  $g$-DAD case
described    in   the   previous    subsection,   where    the   upper
bounds~(\ref{bYg})  translate,  by  means  of~(\ref{Yg}),  into  upper
bounds on $m_{\rm FI}/M$ for $\kappa > 8\times 10^{-5}$.  The required
size of the $D$-parity violating parameter $m_{\rm FI}$ may naively be
estimated using a relation very analogous to (\ref{Tkappa}), viz.\
\begin{equation}
  \label{Tg}
T_g\ =\ \left( \frac{90}{\pi^2\, g_*}\right)^{1/4}\,
\sqrt{\Gamma_g\: m_{\rm Pl} }\ ,
\end{equation}
where $\Gamma_g$  is the decay width  of a $g$-sector  particle and is
given in~(\ref{GammaR}).  If we solve~(\ref{Tg}) for the ratio $m_{\rm
FI}/M$, we obtain
\begin{equation}
  \label{ratio:mFI:M}
\frac{m_{\rm FI}}{M} \approx \ 8.4 \cdot 10^{-4}\times \left(
\frac{0.5}{g}\right)^{3/4} \left(\frac{T_g}{10^9~{\rm
GeV}}\right)^{1/2}\, \left( \frac{10^{16}~{\rm
GeV}}{M}\right)^{1/4}\; .
\end{equation}
For  second  reheat   temperatures  $T_g$  of  cosmological  interest,
i.e.~$0.2~{\rm  TeV}  \stackrel{<}{{}_\sim} T_g  \stackrel{<}{{}_\sim}
10^9~{\rm GeV}$, the following double inequality for $M = 10^{16}$~GeV
may be derived:
\begin{equation}
  \label{FIcombined}
4\times 10^{-7}\ \stackrel{<}{{}_\sim}\ \frac{m_{\rm FI}}{M}\
\stackrel{<}{{}_\sim}\ 10^{-3}\; .
\end{equation}
The  lower  bounds  on  $T_g$  and  $m_{\rm  FI}/M$  result  from  the
requirement  that thermal  electroweak-scale resonant  leptogenesis be
successfully realized. More details are given in Section~\ref{BAU}.

\begin{figure}[t]
\begin{center}
\epsfig{file=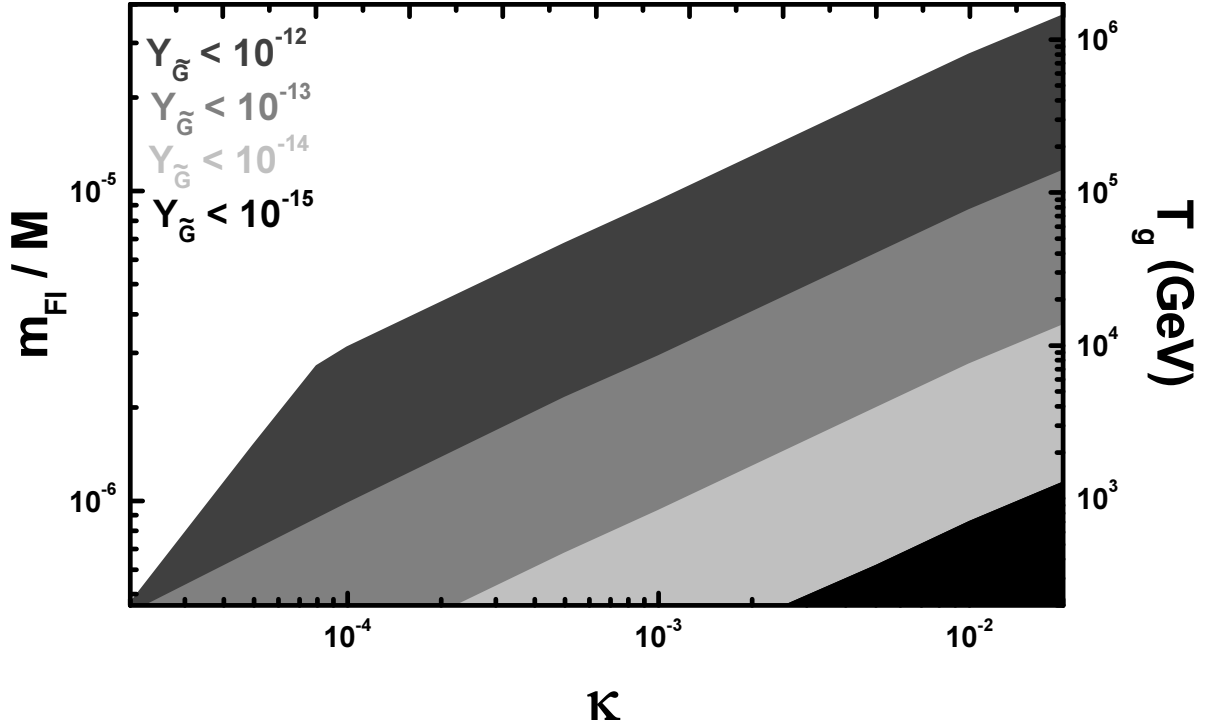,height=6.4in,angle=-90}
\end{center}
\caption{\sl\small The allowed region on the $(m_{\rm FI}/M,\kappa)$
plane for $Y_{\widetilde G} < 10^{-15}$ (black area), $Y_{\widetilde
G}<10^{-14}$ (light grey area), $Y_{\widetilde G}<10^{-13}$ (grey
area) and $Y_{\widetilde G} < 10^{-12}$ (dark grey area).}\label{fig:mFI}
\end{figure}

A  numerical  analysis  of  the gravitino  abundance  predictions  and
constraints  related to  the $g$-DAD  scenario has  been  performed in
Figs.~\ref{fig:Ygm}  and \ref{fig:mFI},  respectively.   Our numerical
results apply equally  well to both mSUGRA and  nmSUGRA scenarios.  In
detail, Fig.~\ref{fig:Ygm} shows $\log Y_{\widetilde G}$ as a function
of $m_{\rm  FI}/M$ and  $T_g$, for the  different values of  $\kappa =
10^{-4},\,  10^{-3}\,,  10^{-2}$, while  $M$  is  fixed  by the  usual
inflationary constraints on ${\cal N}_e$ and $P_{\cal R}$, for $\kappa
=  \lambda   =  \rho$  and  $c_H   =  0$.   The   different  lines  in
Fig.~\ref{fig:Ygm} terminate  at high values of  $m_{\rm FI}/M$, since
the inequality $T_g < T_\kappa$ does no longer hold.  The lowest value
of  $m_{\rm FI}/M$  is determined  by the  condition $T_g  > 200$~GeV,
which  results  from   the  aforementioned  requirement  that  thermal
electroweak-scale     resonant     leptogenesis    is     successfully
realized~\cite{APRD,APRL,PU2}.

In Fig.~\ref{fig:Ygm},  we also observe  the two regimes:  $g$-DBD and
$g$-DAD.   In  the $g$-DBD  regime,  $\log  Y_{\widetilde G}$  remains
constant  for given  $\kappa$ up  to some  value $m_{\rm  FI}/M$.  For
example, for  $\kappa = 10^{-3}$, $\log Y_{\widetilde  G}$ is constant
for  $m_{\rm  FI}/M  \stackrel{>}{{}_\sim}  10^{-4}$. This  result  is
consistent with~(\ref{Yk}). For smaller  values of $m_{\rm FI}/M$, one
enters  the $g$-DAD  regime.  In  this case,  $\log  Y_{\widetilde G}$
decreases   rapidly,  as  $m_{\rm   FI}/M$,  or   equivalently  $T_g$,
decreases.   This  behaviour  of  $Y_{\widetilde G}$  is  expected  on
account  of~(\ref{Yg}).   Also,   in  agreement  with~(\ref{Yg}),  the
reduction of~$Y_{\widetilde G}$ becomes more drastic for larger values
of $\kappa$.

In  Fig.~\ref{fig:mFI}  we  delineate   the  allowed  regions  on  the
$(\kappa,   m_{\rm  FI}/M)$   plane   for  the   discrete  values   of
$Y_{\widetilde  G}  =  10^{-15},  10^{-14}, 10^{-13},  10^{-12}$,  for
$\kappa \geq  8 \times10^{-5}$.  The  upper boundaries of  the various
areas are obtained using (\ref{Yg}).  For $\kappa < 8 \times 10^{-5}$,
we  are   in  the  $g$-DBD   region,  where  we  obtain   $10^{-13}  <
Y_{\widetilde  G}   <  10^{-12}$,  almost   independently  of  $m_{\rm
FI}/M$~[cf.~(\ref{Yk})]. Therefore, we only display values for $m_{\rm
FI}/M$, for which $g$-DAD becomes  relevant.  We observe that the most
stringent  limit on  $Y_{\widetilde  G}$ can  still  be fulfilled  for
$\kappa    \stackrel{>}{{}_\sim}    10^{-2}$    and   $m_{\rm    FI}/M
\stackrel{<}{{}_\sim}  10^{-6}$. Such large  values of  $\kappa$ would
have been excluded from  naive estimates of the $\kappa$-sector reheat
temperature  $T_\kappa$ due  to the  $\kappa$-sector  particle decays.
According to our analysis in this section, however, these large values
of $\kappa$,  $\lambda$ and  $\rho$ of order  $10^{-2}$--$10^{-1}$ are
allowed within  the $F_D$-term hybrid inflationary model.   As we will
see  in  the  next section,  this  is  a  distinctive feature  of  the
$F_D$-term hybrid  model that opens up novel  possibilities in solving
the CDM problem.

At  the  end  of this  section,  we  wish  to  comment on  a  possible
$F_D$-term  hybrid scenario, where  the $\kappa$-sector  particles can
decay  directly  into  the  $g$-sector  ones.  This  can  happen,  for
example, if $m_\kappa > 2 m_g$ or equivalently when $\kappa > \sqrt{2}
g$.   Since the gauge  coupling $g$  of the  waterfall sector  must be
smaller than 0.1  in this case, it would be difficult  to embed such a
$F_D$-term hybrid scenario into  a GUT. The energy density transferred
from  the $\kappa$-sector particles  into the  $g$-sector ones  may be
calculated by
\begin{equation}
\frac{\rho_g}{\rho_\kappa}\  =\ \frac{g}{\sqrt{2}\, \kappa}\ B_{\kappa
\to g}\ .
\end{equation}
Here $B_{\kappa \to  g}$ denotes the branching ratio  of the decays of
the  $\kappa$- to $g$-sector  particles. Assuming  conservatively that
$B_{\kappa  \to g}  \sim 10^{-2}$  and $\kappa  = 2  g$, we  obtain an
estimate for the gravitino abundance $Y_{\widetilde{G}} \sim 10^{-18}$
for  $m_{\rm FI}/M  \stackrel{<}{{}_\sim} 10^{-6}$,  thereby rendering
gravitinos quite harmless.

\setcounter{equation}{0}
\section{Baryon Asymmetry and Cold Dark Matter}\label{BAU}

In this  section we briefly discuss  further cosmological implications
of the $F_D$-term hybrid model for the BAU and the CDM.

\subsection{Resonant Flavour-Leptogenesis at the Electroweak Scale}

Earlier  studies  of  the  BAU  in  supersymmetric  models  of  hybrid
inflation  have  mainly  been  focused  on  scenarios  of  non-thermal
leptogenesis~\cite{LS}, with  an hierarchical heavy  Majorana neutrino
spectrum, e.g.~$m_{N_1} < m_{N_2}  \ll m_{N_3}$. The simplest model of
this type  is obtained by  identifying the waterfall gauge  group with
U(1)$_{B-L}$, which allows the presence of the operator $\gamma_{ij}\,
\widehat{X}_2 \widehat{X}_2 \widehat{N}_i \widehat{N_j}/m_{\rm Pl}$ in
the  superpotential. Notice  that  such  a term  is  forbidden in  the
$F_D$-term hybrid model by virtue of the $R$ symmetry~(\ref{RFD}).  In
the non-thermal leptogenesis  model, the reheat temperature consistent
with  the observed BAU  $\eta_B =  6.1\times 10^{-10}$  and low-energy
neutrino data is estimated to be~\cite{SS}
\begin{equation}
  \label{Tnonth}
T_{\rm reh}\ \stackrel{>}{{}_\sim}\  2.5\cdot 10^7~{\rm GeV} \times
\Bigg( \frac{10^{16}~{\rm GeV}}{M}\Bigg)^{1/2}\,
\Bigg(\frac{\kappa}{10^{-5}}\Bigg)^{3/4}\ ,
\end{equation}
where the superpotential couplings  $\lambda,\, \rho$ are set to zero.
If  $\lambda =  \kappa$,  the lower  bound~(\ref{Tnonth}) gets  larger
roughly  by  a  factor  20.   It  is  obvious  that  in  this  generic
non-thermal leptogenesis scenario, the gravitino constraint on $T_{\rm
reh}$  favours   rather  small  values  of   $\kappa$  and  $\lambda$,
e.g.~$\kappa,\,  \lambda  \stackrel{<}{{}_\sim}  10^{-5}$ for  $T_{\rm
reh}   \stackrel{<}{{}_\sim}   10^8$~GeV.    As   was   discussed   in
Section~\ref{inflation}.2,  however,  such  small values  of  $\kappa$
introduce strong tuning  at a less than 1\% level  to the horizon exit
values of the  inflaton field $\phi_{\rm exit}$ in  a nmSUGRA scenario
that accounts  for the recently  observed value of the  spectral index
$n_{\rm  s}$ given  by~(\ref{nswmap}).   Moreover, the  success of  this
scenario  relies heavily  on the  assumption  that there  is no  other
source of baryogenesis,  e.g.~through the Affleck--Dine mechanism, nor
of entropy release, e.g.~from possible late decays of moduli or flaton
fields~\cite{DLES},  between the energy  scales $m_{N_1}\  (\gg T_{\rm
reh})$ and the electroweak phase transition.

In  the  $F_D$-term  hybrid  model, non-thermal  leptogenesis  is  not
possible for one  of the reasons mentioned above.   The late decays of
the  $g$-sector  ($D$-odd) particles  generally  lead  to an  enormous
entropy release,  so that  not only gravitinos,  but also  any initial
lepton-number excess will be  diluted to unobservable values. However,
as has  already been discussed in~\cite{GP}, the  $F_D$-term model can
realize  electroweak-scale  resonant  leptogenesis~\cite{PU2}, if  the
coupling of  the inflaton  superfield $\widehat{S}$ to  the respective
right-handed   neutrinos   $\widehat{N}_i$  is   very   close  to   an
SO(3)-symmetric form, i.e.~$\rho_{ij} \approx \rho\, {\bf 1}_3$.  This
will give rise to 3 nearly  heavy Majorana neutrinos of mass $m_N$ and
so would enable a  successful realization of the resonant leptogenesis
mechanism at the electroweak  scale.  The required SO(3)-breaking may,
for  example, originate from  renormalization-group (RG)~\cite{Branco}
or possible GUT threshold effects~\cite{PU2,Lindner}.

\begin{figure}[t]
\begin{center}
\begin{picture}(600,320)
\put(50,0){\includegraphics[scale=0.55]{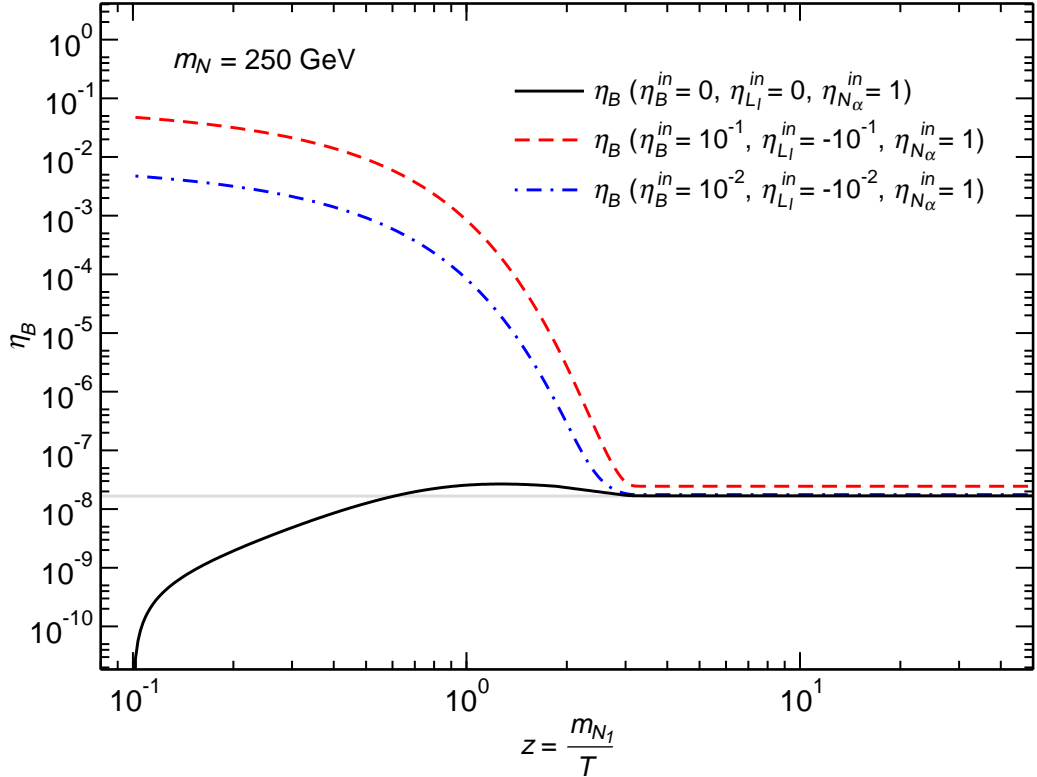}}
\end{picture}
\end{center}
\caption{\sl\small Numerical estimates of  the BAU for a scenario with
  $m_N = 250$~GeV and  for different initial lepton- and baryon-number
  abundances, $\eta^{\rm in}_{L_l}$ and $\eta^{\rm in}_B$, assuming an
  initial  thermal  distribution  for  the heavy  Majorana  neutrinos,
  i.e.~$\eta^{\rm  in}_{N_{1,2,3}}  = 1$.   The  horizontal grey  line
  shows   the   BAU   needed    to   agree   with   today's   observed
  value.}\label{mN250GeV}
\end{figure}

An order  of magnitude estimate  of the final BAU  $\eta_B$, including
single lepton flavour effects, may be obtained as~\cite{APRL,PU2}
\begin{equation}
  \label{Bestimate}
\eta_B\ \sim\ -\, 10^{-2}\,\times\, r(T_g/m_N)\
 \sum_{l=1}^3\, \sum_{N_i}\:
\delta^l_{N_i}\: \frac{K^l_{N_i}}{K_l\,K_{N_i}}\ ,
\end{equation}
where
\begin{equation}
  \label{KlN}
K^l_{N_i}\ =\ \frac{\Gamma (N_i \to L_l \Phi)\: +\:
\Gamma(N_i \to L^C_l \Phi^\dagger)}{H (T=m_N)}
\end{equation}
is a  lepton-flavour dependent wash-out factor, which  quantifies in a
way the degree of in- or  out-of-equilibrium of the decay rates of the
heavy Majorana neutrino mass eigenstates  $N_i$ ($i = 1,2,3$) into the
SM-like Higgs  doublet $\Phi$ and  the lepton doublet  $L_l$ ($l=e,\mu
,\tau$).  The  remaining $K$-factors in~(\ref{Bestimate})  are defined
with the help of $K^{l}_{N_i}$ as follows:
\begin{eqnarray}
  \label{Kfactors}
K_{N_i} \!&=&\!  \sum_{l = 1}^3\ K^{l}_{N_i}\;,\qquad
K_l \ =\ \sum_{N_i}\, K^{l}_{N_i}\; .
\end{eqnarray}
The  parameters $\delta^l_{N_i}$  denote the  different lepton-flavour
asymmetries related to  the decays $N_i \to L_l  \Phi$ and are defined
by
\begin{equation}
\delta^l_{N_i}\ =\ \frac{\Gamma (N_i \to L_l \Phi)\: -\:
\Gamma(N_i \to L^C_l \Phi^\dagger)}{\Gamma (N_i \to L_l \Phi)\: +\:
\Gamma(N_i \to L^C_l \Phi^\dagger)} \ .
\end{equation}
Finally, the prefactor $r(T_g/m_N)$ in~(\ref{Bestimate}) takes care of
a possible  dilution effect  on the  BAU that might  be caused  by the
entropy  release of  late  $g$-sector particle  decays. This  dilution
effect is  only relevant,  if the second  reheat temperature  $T_g$ is
smaller  than  the   leptogenesis  scale  $m_N$.   Employing  standard
arguments of thermodynamics, one may estimate that
\begin{equation}
r (T_g / m_N)\ \sim \ \Bigg(\frac{T_g}{m_N}\Bigg)^5\; .
\end{equation}
Instead,  if   $T_g  \gg   m_N$,  the  dilution   factor  $r(T_g/m_N)$
approaches~1 and can therefore be omitted.

In Fig.~\ref{mN250GeV}, we display  numerical estimates of the BAU for
a resonant leptogenesis scenario with  $m_N = 250$~GeV and an inverted
hierarchical  light-neutrino spectrum.  For  a detailed  discussion of
the  heavy and light  neutrino spectra  of this  model, the  reader is
referred to~\cite{PU2}.  As can  be seen from Fig.~\ref{mN250GeV}, one
advantageous feature of resonant leptogenesis is that the final baryon
asymmetry~$\eta_B$  does not  sensitively depend  on  any pre-existing
lepton-   or  baryon-number   abundance,   $\eta^{\rm  in}_{L_l}$   or
$\eta^{\rm  in}_B$.    For  instance,  assuming   an  initial  thermal
distribution  for   the  heavy  Majorana   neutrinos,  i.e.~$\eta^{\rm
in}_{N_{1,2,3}}  = 1$,  and primordial  baryon  asymmetries $\eta^{\rm
in}_B  \stackrel{<}{{}_\sim}  10^{-2}$,  we  observe  that  the  final
$\eta_B$ is  practically independent  of the initial  conditions, once
the  relevant particle-physics  model  parameters, such  as the  heavy
Majorana masses and their respective Yukawa couplings, are fixed.

It is important to comment here on the fact that the above property of
the  independence  of the  BAU  on  the  initial conditions  does  not
necessarily  get  spoiled,  if  the second  reheat  temperature  $T_g$
happens to be smaller than  the resonant leptogenesis scale $m_N$.  In
this  case, one  only needs  to make  sure that  the  entropy dilution
suppression factor  $\sim (T_g/m_N)^5$ does not lead  to a significant
reduction  of  the  BAU.   Therefore, we  have  rather  conservatively
assumed throughout our numerical analysis in Section~\ref{reheat} that
$T_g \stackrel{>}{{}_\sim} m_N \sim  250$~GeV, even though $T_g$ could
still be somewhat smaller than the resonant leptogenesis scale $m_N$.

Another  point that  deserves to  be  clarified here  is the  physical
significance of lepton-flavour effects  on the BAU.  In general, there
are  two sources  of lepton  flavour:  (i) the  charged lepton  Yukawa
couplings $h_l$  and (ii) the neutrino  Yukawa couplings $h^\nu_{ij}$.
The    former    has     been    extensively    discussed    in    the
literature~\cite{BAULFV} and may affect the predictions for the BAU by
up to one order of  magnitude, depending on the scale of leptogenesis.
For our electroweak-scale leptogenesis scenario, these effects are not
significant,  since  all   charged  lepton  Yukawa  couplings  mediate
interactions that  are in thermal  equilibrium.  The second  source of
flavour effects  is due to neutrino-Yukawa  couplings $h^\nu_{ij}$ and
has  been  studied only  very  recently in~\cite{APRL,PU2,Bari}.   The
effect on the  BAU is most relevant when  the heavy Majorana neutrinos
get   closer   in  mass.    In   models   of  resonant   leptogenesis,
neutrino-Yukawa  coupling effects can  have a  dramatic impact  on the
predictions  for  the BAU,  enhancing  its  value  by many  orders  of
magnitude~\cite{APRL,PU2}.

This last fact opens up  new vistas in the model-building of scenarios
that   can  be  phenomenologically   more  accessible   to  laboratory
experiments.   For   instance,  if  a  certain   hierarchy  among  the
Yukawa-neutrino couplings  $h^\nu_{ij}$ is assumed, e.g.~$h^{\nu}_{i2}
=  i  h^{\nu}_{i3}  \sim  10^{-2}  \sim  h_\tau$  and~$h^{\nu}_{i1}  =
10^{-6}$--$10^{-7} \sim h_e$,  resulting from the approximate breaking
of  some global  U(1)$_l$  symmetry,  the required  BAU  can still  be
generated  successfully from  an individual  lepton  number asymmetry,
namely $L_\tau$ in  this case.  For this particular  model of resonant
$\tau$-leptogenesis,  the   values  of  the   $K$-factors  defined  in
(\ref{KlN}) are:
\begin{equation}
K^\tau_{N_{1,2,3}}\ \sim\ 10\,,\
\qquad K^{e,\mu}_{N_3}\ \sim\ 30\,,
\qquad K^{e,\mu}_{N_{1,2}}\ \sim\ 10^{10}\; .
\end{equation}
Given that the leptonic asymmetry is $\delta^\tau_{N_3} \sim 10^{-6}$,
one  can  estimate from~(\ref{Bestimate})  that  the  right amount  of
baryon asymmetry is produced, with~$\eta_B \sim 10^{-9}$. This is also
shown in  Fig.~\ref{mN250GeV}.  Instead, older approaches  to BEs that
do  not appropriately treat  lepton flavour  effects via  the neutrino
Yukawa couplings $h^\nu_{ij}$ would  have predicted a value that would
have been short of a huge factor $\sim 10^{-6}$~\cite{APRL,PU2}.

As can be  seen from the above example,  the lepton-flavour directions
$L_{e,\mu}$ orthogonal  to $L_\tau$ can involve  large neutrino Yukawa
couplings  of  order  $10^{-2}$.   Such  couplings can  give  rise  to
distinctive    signatures   in   the    production   and    decay   of
electroweak-scale heavy  Majorana neutrinos at  high-energy colliders,
such  as the  LHC~\cite{NprodLHC}, the  International  Linear $e^+e^-$
Collider~(ILC)~\cite{NprodILC}         and         other        future
colliders~\cite{NprodEG}.  Moreover,  electroweak-scale heavy Majorana
neutrinos can give  rise to phenomena of lepton  flavour and/or number
violation,    such    as    the   neutrinoless    double-beta    decay
($0\nu\beta\beta$), the  decays $\mu \to  e\gamma$~\cite{CL}, $\mu \to
eee$,      $\mu       \to      e$      conversion       in      nuclei
etc~\cite{IP,LFVN,LFVrev,LLfit}.    A  detailed   discussion   of  the
low-energy phenomenology of resonant  leptogenesis models may be found
in~\cite{PU2}.

\subsection{Thermal Right-Handed Sneutrinos as CDM}

An  interesting  feature  of  the  $F_D$-term  hybrid  model  is  that
$R$-parity is conserved, even though the lepton number $L$, as well as
$B-L$,    are   explicitly   broken    by   the    Majorana   operator
$\frac{1}{2}\,\rho\,  \widehat{S}  \widehat{N}_i  \widehat{N}_i$.   In
fact, in  our model, all superpotential couplings  either conserve the
$B-L$ number  or break it by  even number of units.   For example, the
coupling $\rho$  breaks explicitly $L$, along with  $B-L$, by 2~units.
Since the $R$-parity of  each superpotential operator is determined to
be $R  = (-1)^{3(B-L)}  = +1$, the  $F_D$-term hybrid  model conserves
$R$-parity.  As a  consequence, the LSP of the  spectrum is stable and
so  becomes a  viable  candidate to  address  the CDM  problem of  the
Universe.

In addition to the standard  CDM candidates of the MSSM, e.g.~a stable
neutralino,  it  would  be  interesting  to  explore  whether  thermal
right-handed sneutrinos as LSPs could solve the CDM problem. Before we
estimate  their   relic  abundance,   we  first  observe   that  light
right-handed sneutrinos  may easily appear in  the spectrum.  Ignoring
the small  neutrino-Yukawa coupling terms,  the right-handed sneutrino
mass matrix  ${\cal M}^2_{\widetilde N}$  is written down in  the weak
basis $(\widetilde{N}_{1,2,3}, \widetilde{N}^*_{1,2,3})$:
\begin{equation}
{\cal   M}^2_{\widetilde    N}\ =\ \frac{1}{2}\, \left(\! \begin{array}{cc}
\rho^2 v^2_S\: +\: M^2_{\widetilde  N} & \rho A_\rho
v_S\: +\: \rho\lambda v_u v_d\\
\rho A^*_\rho v_S\: +\: \rho\lambda v_u v_d &
\rho^2 v^2_S\: +\: M^2_{\widetilde  N}
\end{array} \!\right)\; ,
\end{equation}
where $v_S =  \langle S \rangle$, $v_{u,d} =  \langle H_{u,d} \rangle$
and  $M^2_{\widetilde N}$  is the  soft SUSY-breaking  mass parameters
associated with the sneutrino fields. The sneutrino spectrum will then
consist of 3 heavy (light) right-handed sneutrinos of mass
$$\rho^2 v^2_S\:  +\: M^2_{\widetilde N}\  +\, (-)\ \Big(  \rho A_\rho
v_S\: +\: \rho\lambda v_u v_d\Big)\, .$$ Hence, the 3 light sneutrinos
can   act   as   LSPs,   which   we  collectively   denote   them   by
$\widetilde{N}_{\rm LSP}$.

Recently, the possibility that right-handed sneutrinos are the CDM was
considered  in~\cite{GGP}.  This recent  analysis showed  that thermal
right-handed  sneutrinos have  rather high  relic abundances  and will
generally overclose the Universe  in a supersymmetric extension of the
MSSM with  right-handed neutrino superfields  $\widehat{N}_i$ and bare
Majorana   masses  $(m_M)_{ij}   \widehat{N}_i   \widehat{N}_j$.   The
underlying  reason  is  that  because  of  the  small  Yukawa-neutrino
couplings $h^\nu_{ij}$, the  self- and co-annihilation interactions of
the  sneutrino LSP  with itself  and other  MSSM particles  are rather
weak.  These  weak processes  do not allow  the sneutrino LSP  to stay
long enough in thermal  equilibrium before its freeze-out temperature,
such that its  number density gets reduced to  a level compatible with
the CMB  data, i.e.~$\Omega_{\rm DM} h^2 \approx  0.15$.  Instead, the
predicted values turn  out to be many orders  of magnitude larger than
1.

In the  $F_D$-term hybrid model,  however, there is a  new interaction
that can make the right-handed sneutrinos annihilate more efficiently.
This is  the quartic coupling~\footnote{The implications  of a generic
quartic coupling of the same  form for the CDM abundance and detection
was studied earlier  in~\cite{mcdonald,pospelov} within the context of
a simple non-SUSY model. These studies will not be directly applicable
to  our  more  elaborate   case  of  a  supersymmetric  scenario  with
right-handed sneutrinos.  However, we have used their results to check
our qualitative estimates for the CDM abundance.}
\begin{equation}
  \label{Llsp}
{\cal L}^{\rm LSP}_{\rm int}\ =\
\frac{1}{2}\,\lambda \rho\,  \widetilde{N}^*_i \widetilde{N}^*_i H_u
  H_d\quad  +\quad  {\rm H.c.}
\end{equation}
It  results  from  the  $F$-term  of the  inflaton  field:  $F_S  \sim
\frac{1}{2}\,\rho    \widehat{N}_i    \widehat{N}_i\:   +\:    \lambda
\widehat{H}_u  \widehat{H}_d$.   To  assess  the significance  of  the
interaction~(\ref{Llsp}),   we   estimate   the   relic   density   of
$\widetilde{N}_{\rm LSP}$ in different kinematic regions.

We   first  consider   the   self-annihilation  off-resonant   process
$\widetilde{N}_{\rm   LSP}   \widetilde{N}_{\rm   LSP}   \to   \langle
H_u\rangle\, H_d\to W^+W^-$,  which occurs when $m_{\widetilde{N}_{\rm
LSP}} > M_W$. A simple estimate yields
\begin{equation}
\Omega_{\rm DM}\, h^2\ \sim\
\Bigg(\frac{10^{-4}}{\rho^2\lambda^2}\Bigg)\;
\Bigg(\frac{\tan\beta\, M_H}{g_w\, M_W}\Bigg)^2\, .
\end{equation}
To obtain an acceptable relic density, we need relatively large $\rho$
and  $\lambda$ couplings, i.e.~$\rho,\,  \lambda \stackrel{>}{{}_\sim}
0.1$\footnote{An  upper bound on  the product  $\rho\lambda$, although
somewhat model-dependent,  can be derived from  experimental limits on
the  flux  of  energetic  upward  muons that  occur  in  the  possible
detection  of  CDM  using  neutrino telescopes  \cite{pospelov}.   Our
initial estimates indicate that it should be $\rho\lambda\lesssim0.03$
for $m_{\widetilde N_{\rm LSP}}\sim 50~{\rm GeV}$, which is not a very
rectrictive bound.}.  Such values  go in opposite direction with those
obtained   by  requiring  successful   inflation  with   a  red-tilted
spectrum. Therefore,  as far as  inflation is concerned,  they signify
the necessity of going well beyond the minimal K\"ahler potential.

The  above situation may  slightly improve  for $m_{\widetilde{N}_{\rm
LSP}} <  M_W$, in large $\tan\beta$ scenarios  with light Higgs bosons
that couple appreciably to $b$-quarks~\cite{Sabine}. In particular, in
the kinematic region  $M_{H_d} \approx 2 m_{\widetilde{N}_{\rm LSP}}$,
the     self-annihilation     process     $\widetilde{N}_{\rm     LSP}
\widetilde{N}_{\rm  LSP}  \to  \langle H_u\rangle\,  H_d\to  b\bar{b}$
becomes resonant, and the above estimate modifies to
\begin{equation}
\Omega_{\rm   DM}\,   h^2\    \sim\
10^{-4}\times   B^{-1}(H_d   \to \widetilde{N}_{\rm LSP}
 \widetilde{N}_{\rm LSP})\, \times\,
\Bigg(\frac{M_H}{100~{\rm GeV}}\Bigg)^2\, .
\end{equation}
Consequently, if the  couplings $\lambda ,\, \rho$ are  not too small,
e.g.~$\lambda,\,   \rho   \stackrel{>}{{}_\sim}   10^{-2}$,  the   LSP
right-handed  sneutrinos  $\widetilde{N}_{\rm  LSP}$  can  efficiently
annihilate via  a Higgs  resonance $H_d$ into  pairs of  $b$-quarks in
this kinematic region, thus  obtaining a relic density compatible with
the CMB data.  A~detailed study of the  thermal right-handed sneutrino
as CDM could be given elsewhere.

\setcounter{equation}{0}
\section{Conclusions}\label{conclusions}

We  have analyzed the  cosmological implications  of a  novel $F$-term
hybrid  inflationary  model, in  which  the  inflaton  and the  gauged
waterfall sectors  respect an  approximate discrete symmetry  which we
called here $D$-parity.  The approximate breaking of $D$-parity occurs
explicitly either through the presence of a subdominant FI $D$-term or
through  non-renormalizable operators in  the K\"ahler  potential. For
brevity,  this  scenario of  inflation  was  termed $F_D$-term  hybrid
inflation.  One of the most  interesting features of the model is that
the VEV of  the inflaton field closely relates  the $\mu$-parameter of
the MSSM to an SO(3)  symmetric Majorana mass $m_N$.  If $\lambda \sim
\rho$,  this implies  that $\mu  \sim m_N$,  so the  $F_D$-term hybrid
model may naturally predict lepton-number violation at the electroweak
scale.

Before summarizing the  cosmological and particle-physics implications
of the  $F_D$-term hybrid model, it  might be interesting  to list our
basic assumptions pertinent to inflation and to the model itself:

\begin{itemize}

\item[ (i)] The standard assumption for successful hybrid inflation is
  that the  inflaton field  $\phi$ should be  displaced from  its true
  minimum at the  start of inflation, whereas all  other scalar fields
  in the  spectrum must have  zero VEVs [c.f.~(\ref{initial})].   In a
  nmSUGRA scenario of hybrid  inflation, however, additional tuning is
  required  beyond the  above standard  assumption.  The  horizon exit
  values of the  inflaton field $\phi_{\rm exit}$ have  to be close to
  the value $\phi_{\rm max}$,  at which the inflationary potential has
  a  maximum.    Nevertheless,  such  a  tuning  is   not  so  strong,
  i.e.~$(\phi_{\max}     -     \phi_{\rm    exit})/\phi_{\rm     exit}
  \stackrel{>}{{}_\sim}     10\%$,      as     long     as     $\kappa
  \stackrel{>}{{}_\sim} 10^{-3}$.

\item[(ii)]  Although  there  may  exist  several  ways  of  breaking
  $D$-parity explicitly, we have  considered here two possibilities to
  motivate  the required  small  amount of  $D$-parity violation.   As
  discussed  in  Appendix  A,  we  first  considered  the  case  where
  $D$-parity is broken by a  subdominant FI $D$-term, which is induced
  radiatively after heavy degrees of freedom have been integrated out.
  Another  minimal  way   would  be  to  introduce  non-renormalizable
  operators   in  the   K\"ahler  potential   that   break  $D$-parity
  explicitly.

\item[  (iii)]  In  order  to  be  able  to  realize  thermal  resonant
  leptogenesis  at a  low scale,  the coupling  matrix  $\rho_{ij}$ is
  assumed  to be  close  to SO(3)  symmetric, i.e.~$\rho_{ij}  \approx
  \rho\, {\bf 1}_3$.

\end{itemize}

The $F_D$-term hybrid model has several cosmological implications that
may be summarized as follows:

\begin{itemize}

\item  The  model  can   accommodate  the  currently  favoured  strong
  red-tilted    spectrum    with    $n_{\rm    s}    -    1    \approx
  -0.05$~\cite{WMAP3,Lyman}, if the  radiative corrections dominate the
  slope of  the inflationary potential and  a next-to-minimal K\"ahler
  potential  is assumed,  where the  parameter $c_H$  is in  the range
  0.05--0.2.   The radiative  corrections  dominate the  slope of  the
  potential,  if  the  superpotential couplings,  $\kappa,\  \lambda,\
  \rho$,  lie in  a certain  interval:  $10^{-4} \stackrel{<}{{}_\sim}
  \kappa,\   \lambda,\   \rho   \stackrel{<}{{}_\sim}  10^{-2}$.    In
  addition, the  actual value of  the power spectrum $P_{\cal  R}$ and
  the required  number of $e$-folds, ${\cal N}_e  \approx 55$, provide
  further constraints on these couplings  and the SSB scale $M$ of the
  waterfall gauge symmetry.   For example, for $M\approx 10^{16}$~GeV,
  one finds the allowed parameter space: $\kappa \stackrel{<}{{}_\sim}
  \lambda,\  \rho  \stackrel{<}{{}_\sim}  4\kappa$, for  $\kappa  \sim
  10^{-3}$--$10^{-2}$    and     $0.05    \stackrel{<}{{}_\sim}    c_H
  \stackrel{<}{{}_\sim} 0.1$.

\item For $F_D$-term hybrid  models with spontaneously broken U(1)$_X$
  gauge symmetry, the non-observation  of a cosmic string contribution
  to the  power spectrum at the  10\% level implies an  upper bound on
  the      superpotential     coupling      $\kappa$,     i.e.~$\kappa
  \stackrel{<}{{}_\sim} 10^{-3}$. This  strict upper bound on $\kappa$
  can be  weakened by  one order  of magnitude in  a nmSUGRA  model of
  $F_D$-term hybrid inflation, with $\kappa = \lambda = \rho$ and $c_H
  =  0.14$.  On the  other hand,  this upper  limit can  be completely
  evaded,  if  the watefall  sector  of  the  $F_D$-term hybrid  model
  realizes an  SU(2)$_X$ local symmetry that breaks  completely to the
  identity ${\bf I}$,  i.e.~SU(2)$_X \to {\bf I}$.  In  this case, not
  only  cosmic  strings  but  any  other topological  defects  can  be
  avoided,  such  as  monopoles  and  textures.   As  we  outlined  in
  Section~\ref{FDmodel}, GUTs, such as  those based on the exceptional
  groups E(6) and E(7),  have breaking patterns that contain SU(2)$_X$
  subgroups  uncharged under the  SM gauge  group and  so are  able to
  realize $F_D$-term  hybrid inflation devoid of  monopoles and cosmic
  strings.

\item To  avoid overproduction  of gravitinos, one  needs to  impose a
  strict upper limit on  the reheat temperature $T_{\rm reh}$ obtained
  from   the   perturbative   inflaton   decays,   i.e.~$T_{\rm   reh}
  \stackrel{<}{{}_\sim} 10^{10}$--$10^7$~GeV. This upper bound depends
  on the  decay properties  of the gravitino  and gives rise  to tight
  constraints on the  basic theoretical parameters $\kappa$, $\lambda$
  and $\rho$,  i.e.~$\kappa,\, \lambda,\, \rho\ \stackrel{<}{{}_\sim}\
  10^{-5}$. However,  these tight limits may  be significantly relaxed
  by considering the late decays of the so-called $g$-sector particles
  which are  induced by small $D$-parity violating  couplings that may
  result from  either a subdominant FI  $D$-term or non-renormalizable
  K\"ahler potential  terms.  These $g$-sector  particles are produced
  during the  preheating epoch,  and if they  are abundant,  they will
  lead  to a  second reheating  phase in  the evolution  of  the early
  Universe, giving  rise to a  rather low reheat temperature,  even as
  low as $0.3$~TeV.   In this case, the enormous  entropy release from
  the  $g$-sector   particles  may  reduce   the  gravitino  abundance
  $Y_{\widetilde{G}}$    below   the    BBN   limits    discussed   in
  Section~\ref{reheat}.

\item After the inflaton $S$ receives a VEV, one ends up with 3 nearly
  degenerate heavy  Majorana neutrinos with masses  at the electroweak
  scale.   As we  discussed in  Section~\ref{BAU}, this  opens  up the
  possibility  to  successfully address  the  BAU  within the  thermal
  electroweak-scale   resonant  leptogenesis   framework,  in   a  way
  independent of any pre-existing lepton- or baryon-number abundance.

\item The  $F_D$-term hybrid model  conserves $R$-parity, in  spite of
  the fact that the lepton number is explicitly broken by the Majorana
  operator    $\frac{1}{2}\,    \rho\,    \widehat{S}    \widehat{N}_i
  \widehat{N}_i$.   This is so,  because all  superpotential couplings
  either  conserve the  $B-L$ number  or break  it by  even  number of
  units. The  aforementioned Majorana operator  breaks explicitly $L$,
  as well as $B-L$, by 2~units.  Consequently, the LSP of the spectrum
  is stable and so qualifies  as candidate to address the CDM problem.
  The  new aspect  of  the  $F_D$-term hybrid  model  is that  thermal
  right-handed  sneutrinos  emerge as  new  candidates  to solve  this
  problem,  by virtue of  the quartic  coupling: $\frac{1}{2}\,\lambda
  \rho\, \widetilde{N}^*_i \widetilde{N}^*_i  H_u H_d\ +\ {\rm H.c.}$.
  This new  quartic coupling results  in the Higgs potential  from the
  $F$-terms of the  inflaton field, and it is not  present in the more
  often-discussed extension  of the MSSM,  where right-handed neutrino
  superfields have bare Majorana  masses.  Provided that the couplings
  $\lambda$  and  $\rho$  are  not too  small,  e.g.~$\lambda,\,  \rho
  \stackrel{>}{{}_\sim}  10^{-2}$,  the  LSP  right-handed  sneutrinos
  $\widetilde{N}_{\rm  LSP}$ can  efficiently annihilate  via  a Higgs
  resonance $H_d$  into pairs of  $b$-quarks, in the  kinematic region
  $M_{H_d} \approx 2  m_{\widetilde{N}_{\rm LSP}}$, and so drastically
  reduce its relic density to values compatible with the CMB data.

\end{itemize}

In  addition to  the above  cosmological implications,  the $F_D$-term
hybrid  model has  a  rich particle-physics  phenomenology. The  main
phenomenological characteristics of the model are:

\begin{itemize}

\item[(a)] It is straightforward  to embed the $F_D$-term hybrid model
  into minimal or next-to-minimal  SUGRA, where the soft SUSY-breaking
  parameters are  constrained at the gauge  coupling unification point
  $M_X$.  Instead, electroweak baryogenesis is not viable in a minimal
  SUGRA scenario  of the MSSM.  It requires an  unconventionally large
  hierarchy   between    the   left-handed   and    right-handed   top
  squarks~\cite{EWBAU},  which  is  difficult  to  obtain  within  the
  framework  of minimal SUGRA.   In addition,  the CP-odd  soft phases
  required  for   successful  electroweak  baryogenesis   face  severe
  constraints from  the absence of observable  2-loop contributions to
  the electron and neutron electric dipole moments~\cite{CKP}.

\item[(b)] As has been  discussed in Section~\ref{BAU}, if one assumes
  that  the  neutrino-Yukawa  couplings  $h^\nu_{ij}$ have  a  certain
  hierarchical  structure controlled  by the  approximate  breaking of
  global  flavour  symmetries, the  model  can  have further  testable
  implications  for low-energy  observables of  lepton  flavour and/or
  number  violation, e.g.~$0\nu\beta\beta$  decay, $\mu  \to e\gamma$,
  $\mu \to eee$,  $\mu \to e$ conversion in  nuclei etc.  In addition,
  electroweak-scale heavy Majorana neutrinos may be copiously produced
  at high-energy colliders,  such as the LHC, the  ILC and $e^-\gamma$
  colliders,  whose  decays give  rise  to  distinctive signatures  of
  lepton-number  violation which are  usually manifested  by like-sign
  dileptons accompanied by hadron jets.

\item[(c)]  Since  successful   inflation  requires  small  couplings,
  i.e.~$\kappa,\,  \lambda,\,  \rho\ \stackrel{<}{{}_\sim}\  10^{-2}$,
  the  inflaton  field   decouples  effectively  from  the  low-energy
  spectrum and the Higgs-sector of  the model becomes identical to the
  one of the  MSSM.  In spite of the  aforementioned decoupling of the
  inflaton,  however, the  $F_D$-term hybrid  model could  still point
  towards particular benchmark scenarios of the MSSM.  For example, if
  $\lambda \gg  \kappa$, the $F_D$-term  hybrid model may  explain the
  origin   of  a   possible  large   value  of   the  $\mu$-parameter.
  Specifically, if  $\lambda = 2\kappa$,  $A_\kappa = -a_S  = 2
  M_{\rm SUSY}$, one gets from~(\ref{mu}) the hierarchy $\mu \approx 4
  M_{\rm SUSY}$,  where $M_{\rm SUSY}$ is a  common soft SUSY-breaking
  scale  of  all   scalar  fermion  fields  in  the   model.   If  one
  additionally requires $A_t  = A_b = 2 M_{\rm  SUSY}$, the low-energy
  limit  of  the $F_D$-term  hybrid  model  becomes  identical to  the
  so-called  CPX benchmark  scenario~\cite{CPX} describing  maximal CP
  violation in  the MSSM  Higgs sector at  low and moderate  values of
  $\tan \beta$.  In the CPX scenario, the lightest neutral Higgs boson
  weighing less than 60~GeV might have escaped detection at LEP. There
  have  been several  strategies to  unravel the  existence of  such a
  light CP-violating Higgs boson~\cite{DP}.

\item[(d)] The possible CDM  scenario with the right-handed sneutrinos
  as  LSPs requires large  $\lambda$ and  $\rho$ couplings  that could
  make  Higgs bosons decay  invisibly, e.g.~$H  \to \widetilde{N}_{\rm
  LSP}\, \widetilde{N}_{\rm LSP}$. Also, right-handed sneutrinos could
  be  present in  the  cascade decays  of  the heavier  supersymmetric
  particles.  The  collider phenomenology of such a  CDM scenario lies
  beyond the scope of the present article.

\end{itemize}

The $F_D$-term hybrid  model studied in this paper  should be regarded
as   a  first   attempt   towards  the   formulation   of  a   minimal
Particle-Physics and Cosmology Standard  Model, which does not involve
excessive fine-tuning in the  fundamental parameters of the theory. As
we outlined  above, it might be  possible to test the  validity of our
model by  a number of laboratory experiments  and further substantiate
it by  future astronomical observations.  The  $F_D$-term hybrid model
is   not   plagued  with   the   usual   gauge-hierarchy  problem   of
non-supersymmetric theories and can,  in principle, be embedded within
an  E(6) or  E(7) GUT,  within the  framework of  SUGRA where  SUSY is
softly broken  at the  TeV scale.   In~the same vein,  we note  that a
possible natural solution to  the famous cosmological constant problem
will   shed  valuable   light   on  the   model-building  aspects   of
cosmologically viable  models.  It  will also open  up new  avenues in
quantitatively  addressing  the   major  energy-density  component  of
today's Universe, the  so-called Dark Energy.  We hope  that all these
insights,  along with  new observational  and experimental  data, will
help  us  to  improve   further  our  present  bottom-up  approach  to
formulating  a more  complete minimal  model of  particle  physics and
cosmology.

\subsection*{Acknowledgements}

We thank Richard Battye,  Zurab Tavartkiladze and Thomas Underwood for
illuminating discussions.  This work is supported in part by the PPARC
research grants: PP/D000157/1 and PP/C504286/1.

\newpage

\def\theequation{\Alph{section}.\arabic{equation}}
\begin{appendix}

\setcounter{equation}{0}
\section{Mechanisms of Explicit {\boldmath $D$}-Parity
Breaking}\label{Dappendix}

Here  we will  present mechanisms  for explicitly  breaking $D$-parity
within the  SUGRA framework, pointing out  their possible implications
for  the  decay rates  of  the  $g$-sector  particles.  We  separately
discuss  the breaking  of $D$-parity  for an  Abelian U(1)$_X$  and an
non-Abelian SU(2)$_X$ waterfall-gauge sector.

\subsection{{\boldmath $D$}-Parity
Breaking in the U(1)$_X$ Waterfall-Gauge Sector}

As already  discussed in  Section~\ref{postinfl}, the simplest  way of
breaking $D$-parity  is to add  a subdominant bare FI  $D$-term ${\cal
L}_{\rm  FI}$  to  the  Lagrangian~[cf.~(\ref{LFI})].   As  was  shown
in~\cite{GP},  however, even  if  such  a term  were  absent from  the
tree-level    Lagrangian,   it   could    still   be    generated   by
quantum-mechanical effects  in an effective  manner, after integrating
out Planck-scale degrees  of freedom. It should be  stressed here that
the radiative generation of an {\em effective} FI $D$-term occurs only
{\em after} the SSB of the U(1)$_X$ gauge symmetry.

To elucidate  this point,  let us consider  a simple extension  of the
$F_D$-term  hybrid  model,  which   includes  a  pair  of  superfields
$\widehat{\overline{X}}_{1,2}$    of    opposite   U(1)$_X$    charge,
i.e.~$Q(\widehat{\overline{X}}_2)  = - Q(\widehat{\overline{X}}_1  ) =
Q(\widehat{X}_1  ) =  - Q(\widehat{X}_2  ) =  1$.  In  this  case, the
superpotential $W_{\rm IW}$ pertinent to the inflaton-waterfall sector
may be extended as follows:
\begin{equation}
  \label{Wdterm}
 W_{\rm IW} \ =\ \kappa\, \widehat{S}\, \Big( \widehat{X}_1
\widehat{X}_2\:  -\: M^2\Big)\ +\ \xi\, m_{\rm Pl}\,
\widehat{\overline{X}}_1\,\widehat{\overline{X}}_2\ +\
\xi_1\, \frac{ ( \widehat{\overline{X}}_1\widehat{X}_1 )^2}{2\,m_{\rm
    Pl}}\ +\ \xi'_1\,
   \frac{ ( \widehat{\overline{X}}_2\widehat{X}_2 )^2}{2\,m_{\rm Pl}}\ \dots
\end{equation}
where the  dots stand for  subleading terms that multiply  the leading
operators      by       extra      powers      of      $(\widehat{X}_1
\widehat{X}_2)^n/m^{2n}_{\rm Pl}$,  with $n \ge  1$.  These subleading
operators are irrelevant  for our discussion here and  can be ignored,
within a  perturbative framework of SUGRA.  The  leading operator form
of  the superpotential~(\ref{Wdterm})  may  be reinforced  by the  $R$
symmetry:
\begin{equation}
  \label{Rsym}
\widehat{S}\  \to\ e^{i\alpha}\,  \widehat{S}\,,\quad
\widehat{\overline{X}}_{1,2}\ \to\ e^{i\alpha/2}\,
\widehat{\overline{X}}_{1,2}\,,\quad
(\widehat{L},\ \widehat{Q})\   \to\   e^{i\alpha}\, (\widehat{L},\
\widehat{Q})\,,
\end{equation}
with $W  \to e^{i\alpha} W$.  As  before, all remaining  fields do not
transform   under   the   $R$  symmetry.\footnote{Observe   that   the
$R$-symmetry~(\ref{Rsym}) allows for  the subleading operator $\kappa'
S (\widehat{X}_1 \widehat{X}_2 )^2/m^2_{\rm Pl}$.  This superpotential
term  can break  the U(1)$_X$  gauge symmetry  along  the inflationary
trajectory,    thereby    inflating    away    unwanted    topological
defects~\cite{JKLS}.   This  scenario   is  known  as  shifted  hybrid
inflation.}

We will now show that a  FI $D$-term, $-\frac{1}{2} g m^2_{\rm FI} D$,
will   be  generated  if   the  ultraheavy   Planck-scale  superfields
$\widehat{\overline{X}}_{1,2}$  are  integrated  out.  As  a  starting
point, we consider the U(1)$_X$ $D$-term Lagrangian
\begin{equation}
  \label{LDX}
{\cal L}_{D}\ =\ \frac{1}{2}\, D^2\: +\:
\frac{g}{2}\, D\, \Big(\, X^*_1 X_1\: -\: X^*_2 X_2\: -\:
\overline{X}^*_1 \overline{X}_1\: +\:
\overline{X}^*_2\overline{X}_2\,\Big)\; .
\end{equation}
In  order to integrate  out the  fields $\overline{X}_{1,2}$,  we need
their  mass  spectrum in  the  post-inflationary  era, where  $\langle
X_{1,2}\rangle = M$ and  $\langle \overline{X}_{1,2} \rangle = \langle
S   \rangle   =  0$.    In   the   weak   basis  $\overline{X}_\pm   =
\frac{1}{\sqrt{2}}\, (  \overline{X}_1 \pm \overline{X}_2  )$, this is
approximately given by the Lagrangian
\begin{equation}
  \label{LXmass}
-\, {\cal L}^{\overline{X}_\pm}_{\rm mass}\ \approx\
( \overline{X}^*_+,\ \overline{X}^*_- )\,
\left(\! \begin{array}{cc}
\xi^2 m^2_{\rm Pl} + \xi\, (\xi_1 + \xi'_1 )\, M^2 & (\xi^2_1 -
\xi'^2_1 )\, \frac{\displaystyle M^4}{\displaystyle 2\, m^2_{\rm Pl}}
\\ (\xi^2_1 - \xi'^2_1 )\,
\frac{\displaystyle M^4}{\displaystyle 2\, m^2_{\rm Pl}} &
\xi^2 m^2_{\rm Pl} - \xi\, (\xi_1 + \xi'_1 )\, M^2
\end{array}\!\right)\,
\left(\! \begin{array}{c} \overline{X}_+ \\
\overline{X}_- \end{array} \!\right)\, .
\end{equation}
The approximate mass eigenstates derived from~(\ref{LXmass}) are
\begin{equation}
\widetilde{\overline{X}}_+\ =\ \overline{X}_+\ +\ s_\theta\,
\overline{X}_-\,,\qquad
\widetilde{\overline{X}}_-\ =\ \overline{X}_-\ -\ s_\theta\,
\overline{X}_+\,,
\end{equation}
where $s_\theta \approx (\xi_1 - \xi'_1 ) M^2/ (4\xi m^2_{\rm Pl})$ is
a mixing  angle which is typically  much smaller than~1.   In terms of
the mass-eigenstates $\widetilde{\overline{X}}_{\pm}$, the part of the
Lagrangian~(\ref{LDX})  linear in  the $D$-terms  associated  with the
Planck-scale degrees of freedom reads:
\begin{eqnarray}
  \label{LoXD}
{\cal L}^{\overline{X}_\pm}_{D} & = &
-\, \frac{g}{2}\, D\, \Big(\,
\overline{X}^*_+ \overline{X}_-\: +\:
\overline{X}^*_- \overline{X}_+\,\Big)\nonumber\\
& = &
-\, \frac{g}{2}\, D\, \Big[\,
\widetilde{\overline{X}}^*_+\, \widetilde{\overline{X}}_-\: +\:
\widetilde{\overline{X}}^*_-\, \widetilde{\overline{X}}_+\ +\
2s_\theta\,
\Big( \widetilde{\overline{X}}_+^*\, \widetilde{\overline{X}}_+\: -\:
\widetilde{\overline{X}}^*_-\, \widetilde{\overline{X}}_- \Big)\: +\:
{\cal O}(s^2_\theta)\, \Big]\; .
\end{eqnarray}

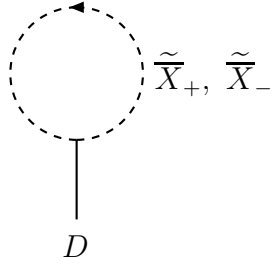
\begin{figure}[t]
\begin{center}
\begin{picture}(200,100)(0,0)
\SetWidth{0.8}

\Line(100,20)(100,50)\DashArrowArc(100,75)(25,-90,270){3}
\Text(100,10)[]{$D$}\Text(130,75)[l]{$\widetilde{\overline{X}}_{+},\
\widetilde{\overline{X}}_{-}$}

\end{picture}
\end{center}
\caption{\sl\small  Radiative  generation   of  an  effective  FI  $D$-term,
  $-\frac{g}{2}\,m^2_{\rm FI}\, D$.}\label{DXtad}
\end{figure}

A FI $D$-tadpole can only be generated from terms linear in $s_\theta$
in  the Lagrangian~(\ref{LoXD}).   This result  should be  expected on
symmetry grounds,  since terms  linear in $s_\theta$  explicitly break
the  $D$-symmetry.   The
$D$-tadpole  $m^2_{\rm FI}$,  calculated  from the  one-loop graph  of
Fig,~\ref{DXtad}, is found to be
\begin{equation}
  \label{FIdterm}
m^2_{\rm FI}\ \approx\ \frac{ \xi^2_1 - \xi'^2_1 }{8\pi^2}\
\frac{M^4}{m^2_{\rm Pl}}\ \ln\left(\frac{m_{\rm Pl}}{M}\right)\ .
\end{equation}
Typically, one gets  $m_{\rm FI}/M \stackrel{<}{{}_\sim} 10^{-6}$, for
$M = 10^{16}$~GeV and $\xi_1,\ \xi'_1 \stackrel{<}{{}_\sim} 10^{-3}$.

\begin{figure}[t]
\begin{center}
\begin{picture}(200,100)(0,0)
\SetWidth{0.8}

\Line(100,20)(100,50)
\DashArrowArc(100,75)(25,-90,90){3}
\DashArrowArc(100,75)(25,90,270){3}
\SetWidth{1}
\Line(96,104)(104,96)
\Line(104,104)(96,96)
\SetWidth{0.5}
\Text(100,10)[]{$D$}
\Text(130,75)[l]{$\overline{X}_{1}$ ($\overline{X}_{2}$)}
\Text(70,75)[r]{$\overline{X}_{1}$ ($\overline{X}_{2}$)}

\Text(100,113)[]{$\cal F$}

\end{picture}
\end{center}
\caption{\sl\small  Diagram pertinent to a short-cut derivation of
the  effective  FI  $D$-term.}\label{DXtad:alt}
\end{figure}
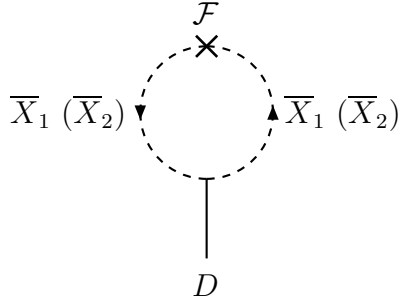

For later use, it is  interesting to outline a short-cut derivation of
the  result~(\ref{FIdterm}),  using the  original  weak  basis of  the
fields,  i.e.~$X_{1,2}$  and  $\overline{X}_{1,2}$.  We  notice  that,
after the  SSB of U(1)$_X$, the $F$-terms  of $\overline{X}_{1,2}$ give
rise to the $D$-odd operator,
\begin{equation}
  \label{FU1}
{\cal F}\ =\ (\xi_1^2-\xi'^2_1)\ \frac{M^4}{2\, m_{\rm Pl}^2}\
\Big(\,\overline X_1^* \overline X_1\:  -\:
\overline X_2^* \overline X_2\, \Big)\; ,
\end{equation}
in the scalar potential of the extended $F_D$-term hybrid model.  This
operator induces,  via the  diagram shown in  Fig.~\ref{DXtad:alt}, an
effective FI $D$-tadpole. Since the scalar fields $\overline{X}_{1,2}$
are degenerate in  mass to leading order, i.e.~$M_{\overline{X}_{1,2}}
\approx \xi  m_{\rm Pl}$, the graph  in~Fig.~\ref{DXtad:alt} is easily
calculated   using   standard    field-theoretic   methods.    It   is
logarithmically divergent,  and in an  effective cut-off theory  it is
given by~(\ref{FIdterm}).   We will use this  short-cut approach below
to calculate  effective $D$-tadpoles  in more elaborate  extensions of
the inflation-waterfall sector.

The  size  of  the FI  $D$-term  may  be  further suppressed,  if  the
Planck-mass chiral  superfields $\widehat{\overline{X}}_{1,2}$ possess
higher U(1)$_X$ charges.  In general, one may assume that the U(1)$_X$
charges         of         $\widehat{\overline{X}}_{1,2}$         are:
$Q(\widehat{\overline{X}}_2)  = -  Q(\widehat{\overline{X}}_1 )  = n$,
where  $n\ge  1$. In  this  case, the  leading  operator  form of  the
inflaton-waterfall superpotential reads:
\begin{equation}
  \label{Wdtermn}
W_{\rm IW} \ =\ \kappa\, \widehat{S}\, \Big( \widehat{X}_1
\widehat{X}_2\: -\: M^2\Big)\ +\ \xi\, m_{\rm Pl}\,
\widehat{\overline{X}}_1\,\widehat{\overline{X}}_2\ +\ \xi_n\, \frac{
(\widehat{\overline{X}}_1)^2\, (\widehat{X}_1)^{n+1}}{2\,m^n_{\rm
Pl}}\ +\ \xi'_n\, \frac{ (
\widehat{\overline{X}}_2)^2\,(\widehat{X}_2)^{n+1}}{2\,m^n_{\rm Pl}}\ .
\end{equation}
Employing the  short-cut method outlined above,  it is straightforward
to compute the loop-induced $D$-term, which is given by
\begin{equation}
  \label{FIdtermn}
m^2_{\rm FI}\ \approx\ \frac{\xi^2_n - \xi'^2_n}{8\pi^2}\
\frac{M^{2(n+1)}}{m^{2n}_{\rm Pl}}\ \ln\left(\frac{m_{\rm Pl}}{M}\right)\ .
\end{equation}
To  obtain a  small ratio  $m_{\rm FI}/M  \sim 10^{-6}$  with $\xi_n,\
\xi'_n \sim 1$, one would simply need $n = 5,\ 6$.  Finally, we should
remark  that the loop-induced  $D$-term does  not lead  to spontaneous
breakdown of global supersymmetry.

\subsection{{\boldmath $D$}-Parity Breaking in the SU(2)$_X$
                                              Waterfall-Gauge Sector}

Here we  outline two possible  mechanisms for explicitly  breaking the
$D$-parities,  $D_1$ and $D_2$  defined in~(\ref{D1})  and (\ref{D2}),
which  govern  the  minimal  inflaton-waterfall  sector  based  on  an
SU(2)$_X$ gauge group.

The  first mechanism utilizes  a non-minimal  K\"ahler waterfall-gauge
sector, where  the two  $D$-parities are broken  by non-renormalizable
operators.  To  be  specific,  a  minimal  $D_{1,2}$-parity  violating
K\"ahler potential of the waterfall-gauge  sector may be cast into the
form:
\begin{eqnarray}
  \label{Kwf}
K_{\rm WF} &=&   \int   d^4\theta\     \Bigg(\,
\widehat{X}^\dagger_1\, {\rm e}^{2g\widehat{V}_X} \widehat{X}_1\
+\  \widehat{X}^T_2\, {\rm e}^{-2g\widehat{V}_X} \widehat{X}_2^*\
+\  \kappa_1\,   \frac{(\widehat{X}^\dagger_1 \, {\rm e}^{2g\widehat{V}_X}
\widehat{X}_1)^2}{2\,m^2_{\rm Pl}} \nonumber\\
&&+\    \kappa_2\, \frac{(\widehat{X}^T_2 \, {\rm e}^{-2g\widehat{V}_X}
\widehat{X}_2^*)^2}{2\,m^2_{\rm   Pl}}\
+\ \frac{
\kappa_1^\prime
(\widehat X_1 ^\dagger {\rm e}^{2 g \widehat V_X} {\rm i} \tau^2 \widehat X_2 )
(\widehat X_1 ^\dagger {\rm e}^{2 g \widehat V_X} \widehat X_1)
+{\rm H.c.}
}{2\,m_{\rm Pl}^2}\nonumber\\
&&
+\ \frac{
\kappa_2^\prime
(\widehat X_1 ^\dagger {\rm e}^{2 g \widehat V_X} {\rm i} \tau^2 \widehat X_2 )
(\widehat X_2 ^T {\rm e}^{-2 g \widehat V_X} \widehat X_2^*)
+{\rm H.c.}
}{2\,m_{\rm Pl}^2}\  +\ \dots\Bigg)\, ,
\end{eqnarray}
where  the ellipses  denote  possible higher-order  non-renormalizable
operators.    The   couplings   $\kappa_{1,2}$   are   real,   whereas
$\kappa^\prime_{1,2}$  can  in  general  be  complex.   Moreover,  the
difference  $\kappa_- =  \kappa_1 -  \kappa_2$  signifies $D_1$-parity
violation, whilst $\kappa_-^\prime=\kappa_1^\prime-\kappa_2^\prime$ is
a parameter of $D_2$-parity  violation.  Hence, non-zero values of the
parameters $\kappa_-$  and $\kappa^\prime_-$ will give  rise to $D_1$-
and $D_2$-parity  violation in the  waterfall-gauge K\"ahler potential
$K_{\rm  WF}$.   Notice that,  as  far  as  $D_1$-parity violation  is
concerned, the present mechanism applies to the Abelian case as well.

There are several sources of $D$-parity violation contained in $K_{\rm
WF}$.  More explicitly, $D$-parity violation will first originate from
the terms  linear in  $D^a$, where $D^a$  are the  auxiliary SU(2)$_X$
components of  the gauge-vector  superfield $\widehat V_X$.   In fact,
these are  the lowest dimensional $D_{1,2}$-odd  operators that emerge
after  the  SSB  of the  SU(2)$_X$  and  are  given by  the  effective
Lagrangian
\begin{equation}
{\cal L}^{D^a-{\rm tad}}_{\rm eff} \ =\
                           \frac{g}{2}\ \frac{M^4}{m_{\rm Pl}^2}\
\Big(\, {\rm Re}\kappa^\prime_-\, D^1\: -\:
{\rm Im}\kappa^\prime_-\, D^2\: +\: \kappa_-\, D^3 \Big)\; .
\end{equation}
These  effective FI  $D$-terms can  be included  in the  Lagrangian by
adding the constants $\frac{g}{2}\,  (m^a_{\rm FI})^2$ to the on-shell
constrained $D^a$ terms, according to the scheme: $D^a \to D^a + \frac
g2 ({m_{\rm FI}^a})^2$, where
\begin{equation}
  \label{mFI123}
(m^1_{\rm FI})^2\ =\ \frac{M^4}{m_{\rm Pl}^2}\
                         {\rm Re}\kappa^\prime_-\,,\qquad
(m^2_{\rm FI})^2\ =\ -\ \frac{M^4}{m_{\rm Pl}^2}\
                         {\rm Im}\kappa^\prime_-\,,\qquad
(m^3_{\rm FI})^2\ =\ \frac{M^4}{m_{\rm Pl}^2}\ \kappa_-\; .
\end{equation}
One may obtain a fair estimate of the $g$-sector particle decay rates,
using  the formula~(\ref{GammaR})  and identifying  $m_{\rm  FI}$ with
$m^a_{\rm FI}$. In this way, we obtain
\begin{equation}
  \label{GgD}
\Gamma [^-\!R_-\,,\ ^-\!I_+\,,\ ^+\!R_- ] \ \sim\
[\,\kappa_-^2\,,\ {\rm Re}^2(\kappa_-^\prime)\,,\ {\rm Im}^2
(\kappa_-^\prime )\,]\
\frac {g^3}{128 \pi}\;  \frac{M^5}{m_{\rm Pl}^4}\ .
\end{equation}
In   addition  to   the  effective   $D$-tadpoles,  higher-dimensional
operators  will  also  break   the  $D$-parities  and  so  render  the
$g$-sector  particles  unstable.   For  example, after  expanding  the
superfields  $\widehat{X}_{1,2}$  about  their  VEVs in  the  K\"ahler
potential~(\ref{Kwf}),  we   find  the  non-renormalizable  $D$-parity
violating interactions described by the Lagrangian
\begin{equation}
  \label{Lnonren}
{\cal L}_{\rm non-ren} \ =\
-\; \frac{M}{2 m_{\rm Pl}^2} \kappa_-\, ^+\!R_-\,
|\partial_\mu {}^+\!X_+|^2\ +\ \frac{M}{4 \sqrt{2} m_{\rm Pl}^2}\;
\Big(\, \kappa^\prime_-\, ^-\!X_-\ +\ {\rm H.c.}\,\Big)\,
|\partial_\mu {}^+\!X_+|^2\; .
\end{equation}
With the aid of~(\ref{Lnonren}), an order of magnitude estimate of the
$g$-sector  particle decay  rates gives:  $\Gamma_g  \sim (\kappa_-^2,
|\kappa_-^\prime|^2)\,  g^3 M^5/m_{Pl}^4$.   These  are of  comparable
order  to the  ones obtained  earlier in~(\ref{GgD}).   For  a typical
inflationary scale, $M  = 10^{16}$~GeV (with $g \sim  1$), we find the
decay   width  $\Gamma_g   \sim   (\kappa_-^2,  |\kappa_-^\prime|^2)\,
10^7$~GeV.  The latter should be compared with the bounds: $3.8 \times
10^{-13}~{\rm        GeV}        \stackrel{<}{{}_\sim}        \Gamma_g
\stackrel{<}{{}_\sim}  4.3~{\rm GeV}$, corresponding  to an  upper and
lower  limit on the  second reheat  temperature $T_g$  of cosmological
interest:      $0.3~{\rm      TeV}     \stackrel{<}{{}_\sim}      T_g\
\stackrel{<}{{}_\sim}  10^9~{\rm GeV}$.  Consequently,  values ranging
from  $10^{-9}$  to  $10^{-2}$  for the  couplings  $\kappa_-$  and/or
$\kappa^\prime_-$ are required  for successful coupled reheating.  The
lower end values  of order $10^{-9}$ may possibly be  seen as a strong
tuning of the  parameters.  One way to explain  the smallness of these
parameters  is  to  contemplate  K\"ahler  manifolds  that  break  the
$D$-parities   by  even  higher-order   non-renormalizable  operators,
e.g.~of  order $\sim  1/m^4_{\rm Pl}$.   In this  case,  the couplings
$\kappa_-$  and $\kappa'_-$  will be  multiplied by  extra  factors of
$M^2/m^2_{\rm Pl} \sim 10^{-4}$, so these couplings can have values of
order~1  and  still predict  a  second  reheat  temperature $T_g  \sim
0.3$~TeV.

We now describe a second mechanism of $D$-parity violation which might
be useful  to obtain small $D$-parity violating  interactions.  Let us
therefore   assume   that   the   K\"ahler  potential   respects   the
$D$-parities. In this  case, we may invoke a  mechanism very analogous
to  the   Abelian  case.    We  extend  the   field  content   of  the
inflaton-waterfall  sector  by adding  a  pair  of Planck-mass  chiral
superfields $\widehat{\overline{X}}_1$ and $\widehat{\overline{X}}_2$,
which belong  to the anti-fundamental  and fundamental representations
of SU(2)$_X$,  respectively.  As in the U(1)$_X$  case, the superheavy
superfields  $\widehat{\overline{X}}_{1,2}$   are  charged  under  the
continuous $R$-symmetry given in~(\ref{Rsym}).  With this restriction,
the leading operator form  of the inflaton-waterfall superpotential is
given by
\begin{eqnarray}
  \label{WIFSU2}
W_{\rm IW} &=& \kappa\, \widehat{S}\, \Big( \widehat{X}_1{}\!^T
\widehat{X}_2\:  -\: M^2\Big)\ +\
\xi\, m_{\rm Pl}\,
\widehat{\overline{X}}_1{}\!^T \,\widehat{\overline{X}}_2\ +\
\theta_1\; \frac{ ( \widehat{\overline{X}}_1{}\!^T
\widehat{X}_1 )\, ( \widehat{\overline{X}}_2{}\!^T \widehat{X}_2)}
{m_{\rm Pl}}\nonumber\\
&&+\; \theta_2\; \frac{ ( \widehat{\overline{X}}_1{}\!^T {\rm i}\tau^2
\widehat{X}_2)\, ( \widehat{\overline{X}}_2{}\!^T {\rm i} \tau^2
\widehat{X}_1)}{m_{\rm    Pl}}\ +\ \zeta_1\;
\frac{(\widehat{\overline{X}}_1{}\!^T {\rm i} \tau^2 \widehat{X}_2)\,
(\widehat{\overline{X}}_2{}\!^T \widehat{X}_2)}{m_{\rm Pl}}\nonumber\\
&&+\ \zeta_2\;
\frac{(\widehat{\overline{X}}_1{}\!^T  \widehat{X}_1)\,
(\widehat{\overline{X}}_2{}\!^T {\rm i} \tau^2 \widehat{X}_1)}
{m_{\rm Pl}}\quad + \quad \dots\; ,
\end{eqnarray}
where  the dots stand  for additional  operators that  turn out  to be
irrelevant  for  the   generation  of  effective  $D^a$-tadpoles,  and
especially  for those  related  to the  $D^1$-  and $D^2$-terms.   The
presence of  these operators is  only important to lift  an accidental
global  U(1)$_{\overline{X}}$ symmetry  that  governs this  restricted
part of  the superpotential  $W_{\rm IW}$ under  consideration.  Here,
all non-renormalizable couplings  $\theta_{1,2}$ and $\zeta_{1,2}$ can
in general be complex.  Extending  the notion of $D_{1,2}$ parities to
the Planck-mass superfields $\widehat{\overline{X}}_{1,2}$, we observe
that  the  operators related  to  the  couplings  $\kappa$, $\xi$  and
$\theta_{1,2}$ are  even under $D_1$ and $D_2$,  whereas those related
to   the  couplings  $\zeta_{1,2}$   are  $D_2$-odd.    Moreover,  the
superpotential operators proportional to the couplings $\zeta_{+(-)} =
\zeta_1 +(-)\ \zeta_2$ are $D_1$-even ($D_1$-odd).

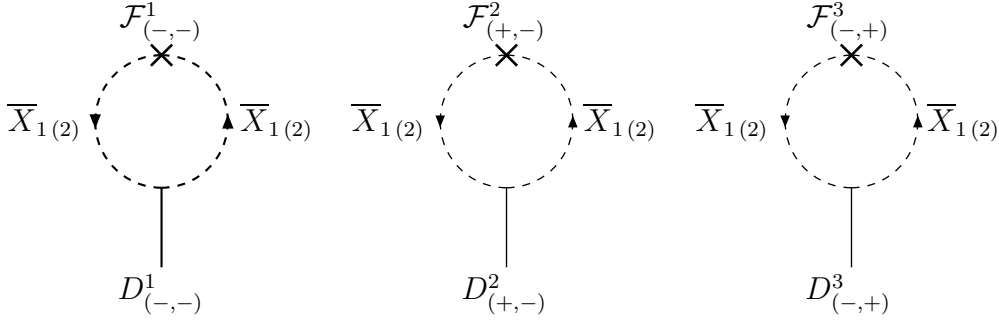
\begin{figure}[t]
\begin{center}
\begin{picture}(600,100)(0,0)
\SetWidth{0.8}

\Line(100,20)(100,50)
\DashArrowArc(100,75)(25,-90,90){3}
\DashArrowArc(100,75)(25,90,270){3}
\SetWidth{1}
\Line(96,104)(104,96)
\Line(104,104)(96,96)
\SetWidth{0.5}
\Text(100,10)[]{$D^1_{(-,-)}$}
\Text(130,75)[l]{$\overline{X}_{1\, (2)}$}
\Text(70,75)[r]{$\overline{X}_{1\, (2)}$}
\Text(100,113)[]{${\cal F}^1_{(-,-)}$}

\Line(230,20)(230,50)
\DashArrowArc(230,75)(25,-90,90){3}
\DashArrowArc(230,75)(25,90,270){3}
\SetWidth{1}
\Line(226,104)(234,96)
\Line(234,104)(226,96)
\SetWidth{0.5}
\Text(230,10)[]{$D^2_{(+,-)}$}
\Text(260,75)[l]{$\overline{X}_{1\, (2)}$}
\Text(200,75)[r]{$\overline{X}_{1\, (2)}$}
\Text(230,113)[]{${\cal F}^2_{(+,-)}$}

\Line(360,20)(360,50)
\DashArrowArc(360,75)(25,-90,90){3}
\DashArrowArc(360,75)(25,90,270){3}
\SetWidth{1}
\Line(356,104)(364,96)
\Line(364,104)(356,96)
\SetWidth{0.5}
\Text(360,10)[]{$D^3_{(-,+)}$}
\Text(390,75)[l]{$\overline{X}_{1\, (2)}$}
\Text(330,75)[r]{$\overline{X}_{1\, (2)}$}
\Text(360,113)[]{${\cal F}^3_{(-,+)}$}

\end{picture}
\end{center}
\caption{\sl\small   Diagrams  responsible for the  generation of  effective
$D^{1,2,3}$-tadpoles  for  the ${\rm  SU}(2)_X$  case,  in the  single
insertion approximation  of the $D$-odd  operators ${\cal F}^{1,2,3}$.
The subscripts  in parentheses label  the $(D_1,D_2)$ parities  of the
respective operator.}\label{DXtad:SU2}
\end{figure}

To calculate  the effective $D^{1,2,3}$-tadpoles after the  SSB of the
SU(2)$_X$ gauge  group, we use the short-cut  approach described above
in     Section~\ref{Dappendix}.1.     Thus,    the     $F$-terms    of
$\widehat{\overline{X}}_{1,2}$  give  rise  to the  following  $D$-odd
contributions to the scalar potential:
\begin{eqnarray}
  \label{F1}
{\cal F}^1_{(-,-)} &=& \theta^*_1\, \zeta_-\
\frac{M^2}{2\,m^2_{\rm Pl}}\
\bigg[\,
\Big( \overline{X}^\dagger_1 \langle X_1^*\rangle \Big)\,
\Big( \overline{X}^T_1\, {\rm i}\tau^2 \langle X_2\rangle \Big)\
-\ (1\leftrightarrow 2)\, \bigg]\nonumber\\
&& -\, \theta^*_2\, \zeta_-\
\frac{M^2}{2\,m^2_{\rm Pl}}\
\bigg[\,
\Big( \overline{X}^\dagger_1\, {\rm i}\tau^2 \langle X^*_2\rangle \Big)\,
\Big( \overline{X}^T_1  \langle X_1 \rangle \Big)\
-\ (1\leftrightarrow 2)\, \bigg]\quad  +\quad {\rm H.c.}\; ,\\[4mm]
  \label{F2}
{\cal F}^2_{(+,-)} &=& \theta^*_1\, \zeta_+\
\frac{M^2}{2\,m^2_{\rm Pl}}\
\bigg[\,
\Big( \overline{X}^\dagger_1 \langle X_1^*\rangle \Big)\,
\Big( \overline{X}^T_1\, {\rm i}\tau^2 \langle X_2\rangle \Big)\
+\ (1\leftrightarrow 2)\, \bigg]\nonumber\\
&& +\, \theta^*_2\, \zeta_+\
\frac{M^2}{2\,m^2_{\rm Pl}}\
\bigg[\,
\Big( \overline{X}^\dagger_1\, {\rm i}\tau^2 \langle X^*_2\rangle \Big)\,
\Big( \overline{X}^T_1  \langle X_1 \rangle \Big)\
+\ (1\leftrightarrow 2)\, \bigg]\quad  +\quad {\rm H.c.}\; ,\\[4mm]
  \label{F3}
{\cal F}^3_{(-,+)} & =&
-\,{\rm Re}\,(\zeta_+\zeta^*_-)\ \frac{M^2}{2\,m^2_{\rm Pl}}\
\bigg[\,  \Big( \overline{X}^\dagger_1\, i\tau^2 \langle X_2^*\rangle \Big)\,
\Big( \langle X^T_2 \rangle\, i\tau^2 \overline{X}_1 \Big)\ +\
\Big( \overline{X}^\dagger_1 \langle X^*_1 \rangle \Big)\,
\Big( \langle X^T_1 \rangle \overline{X}_1 \Big)\nonumber\\
&&-\quad (1 \leftrightarrow 2)\, \bigg]\; ,
\end{eqnarray}
where the subscripts in  parentheses indicate the $(D_1,D_2)$ parities
of  the  above  operators.    Note  that  possible  $D$-odd  operators
proportional to  $\xi \theta_{1,2}$  and $\xi\zeta_\pm$ have  not been
displayed, since they do not contribute to the generation of effective
$D^a$-tadpoles.   To  be  specific,  the  effective  $D$-tadpoles  are
induced radiatively via the graphs shown in~Fig.~\ref{DXtad:SU2}, once
the operators ${\cal F}^{1,2,3}$  are individually contracted with the
$D$-term  operator ${\cal  D}^{1,2,3}_{\overline{X}}$  related to  the
$\overline{X}_{1,2}$ fields:
\begin{equation}
{\cal D}^a_{\overline{X}}\ =\ -\; \frac{g}{2}\ \Big(\,
\overline{X}^T_1\, \tau^a\, \overline{X}^*_1\: -\:
\overline{X}^\dagger_2\, \tau^a\, \overline{X}_2\, \Big)\; .
\end{equation}
These  loop-induced effective  FI  $D$-terms can  be  included in  the
effective Lagrangian by shifting  the on-shell constrained $D^a$ terms
by constants, according to the  above described scheme: $D^a \to D^a +
\frac  g2 ({m_{\rm FI}^a})^2$.   In this  scheme, the  mass parameters
$(m^a_{\rm FI})^2$ are found to be
\begin{eqnarray}
  \label{mFISU2}
\big(m_{\rm FI}^1\big)^2 & = &  -\,
             \frac{{\rm Re}\,(\theta^*_- \zeta_-)}{4\pi^2}\
\frac{M^4}{m_{\rm Pl}^2}\ \ln \left(\frac{m_{\rm Pl}}{M}\right)\; ,\nonumber\\
\big(m_{\rm FI}^2\big)^2 & = &
             \frac{{\rm Im}\,(\theta^*_-\zeta_+)}{4\pi^2}\
\frac{M^4}{m_{\rm Pl}^2}\ \ln \left(\frac{m_{\rm Pl}}{M}\right)\; ,\\
\big( m_{\rm FI}^3 \big)^2 & = & -\,
             \frac{{\rm Re}\,(\zeta_+\zeta^*_-)}{4\pi^2}\
\frac{M^4}{m_{\rm Pl}^2}\ \ln \left(\frac{m_{\rm Pl}}{M}\right)\; ,\nonumber
\end{eqnarray}
where  $\theta_\pm =  \theta_1  \pm \theta_2$.   It  can be  estimated
from~(\ref{mFISU2})  that  for  values $\theta_\pm\,,  \zeta_\pm  \sim
10^{-4}$,   one  gets   $m^{1,2,3}_{\rm   FI}/M  \stackrel{<}{{}_\sim}
10^{-6}$,  leading to  a low  second reheat  temperature  $T_g$, below
1~TeV.   In this  context,  one should  notice  that the  size of  the
effective $D$-tadpoles  is very sensitive to the  cut-off scale, which
we have  chosen here to be  the reduced Planck mass  $m_{\rm Pl}$. For
instance, if a cut-off larger  by one order of magnitude were adopted,
then   values   of   order   $10^{-2}$  for   the   non-renormalizable
superpotential couplings would be sufficient to generate the effective
$D^a$-tadpoles at the required size.

The violation  of $D$-parities will also affect  the particle spectrum
of the  SU(2)$_X$ inflaton-waterfall sector.  This will  depend on the
particular choice of the non-renormalizable part of the superpotential
and K\"ahler potential.  Our intention is not to pursue this issue any
further  here,  by  putting   forward  a  specific  non-minimal  SUGRA
scenario.  Instead,  our goal has  been to explicitly  demonstrate the
existence   of   at   least   two  mechanisms,   which   utilize   the
non-renormalizable part of the K\"ahler potential or superpotential to
break the $D$-parities at the required order of magnitude.

\end{appendix}

\end{document}